\documentclass[11pt,english,onecolumn]{IEEEtran}
\usepackage[T1]{fontenc}
\usepackage[latin9]{inputenc}
\usepackage{array}
\usepackage{float}
\usepackage{url}
\usepackage{enumitem}
\usepackage{multirow}
\usepackage{amsmath}
\usepackage{amsthm}
\usepackage{amssymb}
\usepackage{graphicx}
\usepackage{setspace}
\PassOptionsToPackage{normalem}{ulem}
\usepackage{ulem}
\makeatletter

\providecommand{\tabularnewline}{\\}
\floatstyle{ruled}
\newfloat{algorithm}{tbp}{loa}
\providecommand{\algorithmname}{Algorithm}
\floatname{algorithm}{\protect\algorithmname}

\theoremstyle{plain}
\newtheorem{thm}{\protect\theoremname}
\theoremstyle{plain}
\newtheorem{prop}[thm]{\protect\propositionname}
\theoremstyle{remark}
\newtheorem{rem}[thm]{\protect\remarkname}
\theoremstyle{plain}
\newtheorem{lem}[thm]{\protect\lemmaname}
\theoremstyle{plain}
\newtheorem{cor}[thm]{\protect\corollaryname}

\usepackage{hyperref}
\usepackage{enumitem}
\usepackage{breqn}
\usepackage{bbm} 
\usepackage{cite}

\DeclareMathOperator{\diag}{diag}

\DeclareMathOperator*{\argmax}{arg\,max} \DeclareMathOperator*{\argmin}{arg\,min}
\DeclareMathOperator{\supp}{supp}

\DeclareMathOperator{\Rad}{Rad} 
\DeclareMathOperator{\Hamd}{d_H} 
\DeclareMathOperator{\N}{N} 
\global\long\def\dtv{\mathrm{d_{TV}}}
\global\long\def\dkl{\mathrm{d_{KL}}}
\global\long\def\binent{\mathrm{H_{bin}}}
\global\long\def\dchis{\mathrm{d}_{\chi^2}}

\newcommand{\ind}{\perp\!\!\!\!\perp}

\allowdisplaybreaks

\global\long\def\s[#1]{\textnormal{\scriptsize #1}}
\global\long\def\st[#1]{\textnormal{\tiny #1}}

\global\long\def\P{\mathbb{P}}
\global\long\def\pe{\mathbf{p}}
\global\long\def\peb{\bar{\mathbf{p}}}
\global\long\def\qe{\mathbf{q}}

\global\long\def\E{\mathbb{E}}

\global\long\def\I{\mathbbm{1}}
\global\long\def\v[#1]{\mathbf{#1}} 
\global\long\def\m[#1]{\boldsymbol{#1}} 

\global\long\def\r[#1]{#1}
\global\long\def\d{\mathrm{d}}
\global\long\def\dee{\partial}

\global\long\def\eqd{\stackrel{d}{=}}

\global\long\def\dtv{\mathrm{d_{TV}}}
\global\long\def\dkl{\mathrm{d_{KL}}}
\global\long\def\dchis{\mathrm{d}_{\chi^2}}

\global\long\def\dfn{:=}
\global\long\def\nfd{=:}

\global\long\def\trre[#1,#2]{\overset{{\scriptstyle (#2)}}{#1}} 

\usepackage{algorithm,algpseudocode}

\author{

\IEEEauthorblockN{Nir Weinberger\\}
\IEEEauthorblockA{The Viterbi Faculty of Electrical and Computer Engineering\\
	         Technion - Israel Institute of Technology\\
		Technion City, Haifa 3200004, Israel\\
 	    	nirwein@.technion.ac.il}
}

\makeatother

\usepackage{babel}
\providecommand{\corollaryname}{Corollary}
\providecommand{\lemmaname}{Lemma}
\providecommand{\propositionname}{Proposition}
\providecommand{\remarkname}{Remark}
\providecommand{\theoremname}{Theorem}

\begin{document}
\title{Generalization Bounds and Algorithms for Learning to Communicate over
Additive Noise Channels\thanks{This research was supported by the MIT--Technion fellowship and the Viterbi scholarship, Technion, Israel Institute of Technology. The material in this paper was presented in part at IEEE International Symposium on Information Theory (ISIT) 2020.} }

\maketitle
\renewcommand\[{\begin{equation}}
\renewcommand\]{\end{equation}}
\begin{abstract}
An additive noise channel is considered, in which the distribution
of the noise is nonparametric and unknown. The problem of learning
encoders and decoders based on noise samples is considered. For uncoded
communication systems, the problem of choosing a codebook and possibly
also a generalized minimal distance decoder (which is parameterized
by a covariance matrix) is addressed. High probability generalization
bounds for the error probability loss function, as well as for a hinge-type
surrogate loss function are provided. A stochastic-gradient based
alternating-minimization algorithm for the latter loss function is
proposed. In addition, a Gibbs-based algorithm that gradually expurgates
an initial codebook from codewords in order to obtain a smaller codebook
with improved error probability is proposed, and bounds on its average
empirical error and generalization error, as well as a high probability
generalization bound, are stated. Various experiments demonstrate
the performance of the proposed algorithms. For coded systems, the
problem of maximizing the mutual information between the input and
the output with respect to the input distribution is addressed, and
uniform convergence bounds for two different classes of input distributions
are obtained.

\looseness=-1
\end{abstract}


\section{Introduction}

The additive noise channel is one of the simplest and most widely
applicable models in communication and information theory. The channel
output $Y\in\mathbb{R}^{d}$ in this model is given by 
\begin{equation}
Y=X+Z\label{eq: additive noise channel introduction}
\end{equation}
where $X\in\mathbb{R}^{d}$ is the input (which is almost always restricted
in some way, e.g., to have finite power), and $Z\in\mathbb{R}^{d}$
is the noise, which is assumed to be statistically independent of
the input $X$. As is natural and well-established \cite{cover2012elements,lapidoth2017foundation},
the optimization of a communication system for this channel depends
on the probability distribution of the noise: First, via a proper
choice of the decoder which minimizes the error probability for a
given codebook, and second, via the proper design of the codebook
itself in case of uncoded transmission, or the capacity achieving
input distribution in case of coded transmission. The conventional
model further assumes that $Z$ is Gaussian, which is typically justified
by physical phenomena such as thermal noise at the receiver \cite{proakis2001digital},
as well as the central-limit theorem which implies that the accumulated
noise from a large number of sources tends to be Gaussian. If such
a model is accurate, then the parameters of the noise distribution\footnote{The covariance matrix, or just the variance in case of white noise.}
can be estimated from noise samples obtained by the receiver when
the transmitter is silent, and then used to optimize the system. This
method similarly works whenever the noise density is parametric of
sufficiently low dimension. Most often (e.g. \cite{hassibi2003much}),
this includes the design of a training sequence $\{x_{\text{tr},i}\}_{i=1}^{T_{\text{tr}}}$
which is transmitted over the channel over $T_{\text{tr}}$ uses of
the channel (\ref{eq: additive noise channel introduction}), and
the receiver which knows this sequence uses it to estimate the parameters
of the distribution of $Z$. 

Nonetheless, in more complex scenarios encountered in practice \textendash{}
for example in massive multiple-input multiple-output (MIMO) systems
\cite{marzetta2015massive} or in ultra low-latency \cite{sybis2016channel}
communication \textendash{} the class of possible noise distributions
may be too rich to allow efficient and faithful parameter estimation
under the system constraints, or may be essentially nonparametric
to begin with. Following the \emph{statistical learning} paradigm
\cite{vapnik2013nature,hastie2009elements,shalev2014understanding},
a possible approach is to define a hypothesis class for the encoder
(codebooks) and the decoder, and use the samples of the noise to \emph{select}
a member of the class. The selected decoder in this class is most
assuredly \emph{mismatched }to the noise distribution and is not a
maximum-likelihood decoder for the selected codebook. However, the
great advantage of this approach is that the class of codebooks and
decoders may only include encoders/decoders which are feasible to
implement, by design. 

In this paper, we focus on learning aspects of this problem, and specifically
on the convergence rates of generalization bounds as a function of
the number of noise samples. Nonetheless, we mainly ignore computational
complexity and scalability aspects of the algorithms proposed to perform
such learning. Furthermore, as the noise samples are usually only
available at the decoder side, we also mainly sidestep the problem
of communicating the choice of codebook back to the encoder.\footnote{This problem is inconsequential in offline design, or when a reliable
feedback link of large capacity exists. Analysis of limited feedback
makes the model more complex and is an important topic for further
research.} Rather instead, our goal is to explore the statistical learnability
of coding problems in a distilled and basic form, and highlight basic
properties and challenges. 

\subsection{Contributions and Outline}

Our contributions are of two types \textendash{} for uncoded and coded
systems. For both types we assume that the learner has $n$ independent
samples of the noise $Z$. For uncoded systems, one-shot use of (\ref{eq: additive noise channel introduction})
is assumed, and based on the noise samples, the learner is required
to design a codebook of $m$ codewords, and possibly select a decoder
from a given class in order to minimize a prescribed loss function.
It is tacitly assumed that $m$ is reasonably small, so that the encoder
can memorize all codewords (such codebooks are sometimes referred
to as \emph{multi-dimensional constellations}). For this setup, we
show that: 
\begin{enumerate}
\item Section \ref{sec:Learning error probability}: If the class of decoders
is comprised of all possible minimal distance decoders with respect
to (w.r.t.) the Mahalanobis distance (parameterized by a covariance
matrix), and if the loss function is the standard error probability,
then the generalization error of selecting a codebook of $m$ codewords
scales as $\tilde{O}(m\sqrt{\frac{d}{n}})$ with high probability.
The result is an error bound which is uniformly applicable to all
codebooks and decoders, and thus for any learning algorithm. Specifically,
it holds for the empirical risk minimization (ERM) algorithm. At the
same time, to the best of our knowledge there is no efficient ERM
algorithm for this problem. We also show that the $\frac{1}{\sqrt{n}}$
dependency is tight.
\item Section \ref{sec:Learning surrogate}: If the loss function is replaced
by a hinge-type surrogate loss function, which unlike the standard
error probability loss function is a \emph{continuous} function of
the codebook and decoder covariance matrix, then the generalization
error scales at an improved rate of $\tilde{O}(\sqrt{\frac{(d\vee m)d}{n}})$
with high probability. Moreover, a heuristic alternating optimization
algorithm is proposed for this problem, and a stochastic gradient
descent (SGD) variant of this algorithm is outlined. 
\end{enumerate}
In Section \ref{sec:Learning Expurgation}, we consider a setup in
which a large static codebook of $m_{0}>m$ codewords is given, and
the noise samples are used to dynamically expurgate this codebook
by removing ``bad'' codewords which cause large error probability
(the decoder is assumed to be a standard minimal distance decoder).
Finding the optimal subset of codewords is a combinatorial optimization
optimization problem, which is computationally heavy to solve in general.
We thus propose a randomized \emph{Gibbs algorithm} for this scenario,
which gradually removes codewords with some randomness. The use of
this Gibbs algorithm allows us to obtain a bound on its average generalization
error, which follows directly from an information-theoretic stability
analysis of learning algorithms \cite{raginsky2016information,xu2017information}.
The bound scales as $O(\sqrt{\frac{T\beta}{n}\wedge\frac{T\beta^{2}}{4n^{2}}})$
where $T$ is the number of steps of the algorithm; the error probabilities
conditioned on each codeword are re-computed after each step before
expurgating another subset of $k=\frac{m_{0}-m}{T}$ codewords; and
$\beta$ is an inverse-temperature parameter which controls the balance
between greedy and completely random expurgation. We also state a
high probability generalization bound of order $\tilde{O}(\frac{\sqrt{T}\beta}{n}+\frac{1}{\sqrt{n}})$
which is a direct implication of the \emph{uniform-stability} property
\cite{bousquet2002stability} of the Gibbs algorithm, and the recent
results of \cite{feldman2018generalization,feldman2019high}. The
choice of $T$ and $k$ is dictated by the available computational
power, whereas the value of $\beta$ may be arbitrary optimized.

For coded systems, we focus in Section \ref{sec: Learning MI} on
the problem of finding the input distribution which maximizes the
mutual information $I(X;X+Z)$, or, equivalently, the output differential
entropy $h(X+Z)$. The use of the mutual information as a performance
measure may be justified by the theoretical existence of universal
decoding rules which are able to achieve communications rates arbitrarily
close to the mutual information despite lack of knowledge of the noise
distribution \cite{merhav1993universal} (see also the survey in \cite{lapidoth1998reliable}).\footnote{The exact maximal achievable rate using a fixed-structure mismatched
decoder (``mismatch capacity'') is unknown to date \cite{somekh2015general,scarlett2020information}.
A more complicated alternative to the problem studied here, is to
explore generalization bounds which aim to maximize known lower bounds
on the mismatch capacity (for example, the one known as the LM rate
\cite{csiszar1981graph,hui1983fundamental,merhav1994information}). } Evidently, this problem is closely related to the problem of differential
entropy estimation (surveyed in \cite{wang2009universal}; see also
\cite{han2020optimal} and references therein for a more recent account),
but here the goal is to maximize the entropy, rather than to estimate
it. This can be achieved by showing a empirical convergence of an
estimator $h_{n}(X+Z)$ based on the noise samples which is \emph{uniform}
in allowed class input distributions. The methods used here to achieve
this is to control the difference between differential entropies by
some statistical distance between distributions, and specifically,
between the true and empirical distributions. We thus consider two
types of sets of conditions and show that:
\begin{enumerate}
\item If the set of input distributions is \emph{regular} in the sense defined
in \cite{polyanskiy2016wasserstein}, has finite second moment, and
has absolutely bounded differential entropy, and if the norm of the
noise is sub-Gaussian, then with high probability $|h(X+Z)-h(X+\hat{Z}_{n})|=\tilde{O}(n^{-1/(d\vee4)})$
uniformly for all possible input distributions, where $\hat{Z}_{n}$
is the empirical measure of the noise samples. 
\item If the support of $X$ is fixed to $m$ points $\{x_{1},\ldots,x_{m}\}\subset\mathbb{R}^{d}$
(an atomic distribution), and so the input distribution is determined
by weights $\{a_{j}\}_{j\in[m]}$ such that $a_{j}=\P[X=x_{j}]$,
then a similar uniform high probability bound can be obtained using
a plug-in kernel density estimator (KDE) for the noise density. The
final convergence rates of the uniform error in differential entropy
depend on smoothness assumptions on the noise densities. For example,
for Lipschitz balls \cite{han2020optimal} of smoothness parameter
$s\in(0,2]$ it is given by $O(n^{-s/(s+d)})$. Importantly, this
rate is dominated by the entropy estimation error, and the additional
error due to the requirement for \emph{uniform} convergence scales
as $\tilde{O}(\sqrt{\frac{m}{n}})$ which is typically negligible. 
\end{enumerate}
As evident, the convergence rates in this problem scale exponentially
with the dimension, which stems from the fact that at worst case,
the support of the noise distribution may lie in a $d$-dimensional
space. Possible ways to obtain faster convergence rates are discussed,
along with a short summary and other open problems in Section \ref{sec:Summary}.
Proofs are relegated to Appendix \ref{sec:Proofs}, parameters used
in experiments are summarized in Appendix \ref{sec:Experiments-Details},
and technical details regarding the implementation of the Gibbs algorithm
are provided in Appendix \ref{sec:Memoization-Implementation-of}. 

\subsection{Motivation and Context}

The remarkable practical success of learning methods of deep neural
networks (DNN) \cite{Goodfellow-et-al-2016} architectures in an assortment
of tasks has recently motivated various researchers to consider learning-based
design of communication systems \cite{o2017introduction}. As surveyed
in \cite{ibnkahla2000applications,gruber2017deep}, this idea is not
completely new and was proposed as early as the 1990s by several authors.
The approach to this problem is coarsely categorized into either one
of two types \cite{o2017introduction}. The first type \cite{caid1990neural,wang1996artificial,nachmani2018deep}
is motivated by the expectation to utilize the powerful computational
resources available for implementation of DNN and their efficient
training methods in order to achieve performance competitive to expert-designed
systems. Various recent works have considered this approach, e.g.,
for random and polar codes \cite{gruber2017deep}, for an unfolded
belief propagation algorithm \cite{nachmani2018deep}, for MIMO communication
\cite{samuel2017deep}, for low-latency codes \cite{jiang2019learn,kim2018communication}
motivated by 5th generation standard \cite{sybis2016channel}, for
feedback-based communication \cite{kim2018deepcode}, and for interactive
communication  \cite{sahai2019learning}. 

The second type is closer in spirit to the one taken in this paper,
and follows the general learning-theoretic methodology to refrain
from parametric modeling of the data probability distribution \cite{hastie2009elements,vapnik2013nature,shalev2014understanding}.
Rather instead, a hypothesis class is assumed for the required statistical
inference goal (e.g., classification, regression), and data is used
to select an optimal hypothesis from this class based on the available
data. The application of this approach in communication-theory related
problems is most suitable in scenarios for which channel modeling
is difficult or inaccurate. This includes, e.g., interference, jamming
signals non-linearity \cite{schenk2008rf} finite-resolution quantization
(see \cite{wang2017deep,o2017introduction}). The entire end-to-end
system is considered as a single DNN, encompassing the encoder operation,
the operation and randomness of the channel, and the decoder operation,
and is then typically trained in an unsupervised manner as an autoencoder
\cite{Goodfellow-et-al-2016}. This approach was applied to mutli-user
detection in code-division multiple-access systems \cite{mitra1994neural},
and, more recently, to molecular communication systems \cite{farsad2017detection}.
In \cite{zhang2019regression}, a receiver architecture including
a processing of the input followed by nearest neighbor decoding rule
was proposed. The optimal output processing was identified to be a
regression problem, and it was proposed to use the data in order to
choose the processing function to maximize a generalized mutual information
(under Gaussian codebooks).

In fact, such an approach seems to be also useful in case a parametric
modeling of the channel is reasonably accurate, but too complex to
be utilized, as the dimension of the parameter vector is exceedingly
high (e.g., in massive MIMO systems \cite{marzetta2015massive}).
As mentioned above, parameter estimation is a common practice to provide
channel state information (i.e., noise distribution) to the decoder,
and is usually carried by transmitting a designated training sequence.
However, for high-dimensional parameters, insufficient training time
will lead to large variance of the estimator, and coping this by dimensionality
reduction may lead to a large bias. Thus a good variance-bias trade-off
cannot be achieved. Moreover, the estimation step is only secondary
to the ultimate goal of reliable decoding of the data with low error
probability. In other words, decoder selection based on parameter
estimation is a specific, indirect, way of selecting a decoder based
on noise samples. A more general and direct way is to allow any selection
rule of the decoder based on the noise samples. 

Nonetheless, communication theory enjoys a rich expert knowledge in
optimizing its main components, and many practical communication channels
do have reasonably faithful parametric representation, which is widely
and successfully used in practice (e.g., \cite{proakis2001digital,tse2005fundamentals,barry2012digital,heath2018foundations}).
To overcome this high bar and demonstrate the viability of the learning-based
approach, research has focused on obtaining improved or comparable
performance (with improved complexity) to classical communication-theoretic
methods. To obtain this goal, they utilize sophisticated DNN architectures,
which exhibit both remarkable approximation ability \cite{hornik1989multilayer}
as well as generalization to out-of-sample data. To date, however,
the generalization capabilities and the efficiency of training algorithms
such as SGD of DNNs are not fully understood, and challenge general
principles in statistics and learning (e.g., \cite{belkin2018understand}).
Moreover, even with refined understanding, the application of any
learning methods to a communication problem raises additional questions:
What is the richness of the decoder hypothesis classes? How does the
generalization error is affected by the structure of the code and
its data rate? What are useful and proper loss functions? How to properly
regularize the learning process? In this paper, we focus on a stylized
and basic version of this problem, with the prospect that the opportunities
and challenges associated with using learning-theoretic methods for
communication systems are illuminated by the analysis of this basic
model.

\subsection{Other Related Work}

The analysis of decoders which are not fully optimized to the channel
statistics distribution is typically studied under the framework of
\emph{mismatch decoding}. Sharp characterization of the capacity and
other fundamental limits of such decoders is a notoriously difficult
and extensively studied open problem in information theory \cite{merhav1994information,csiszar1995channel,balakirsky1995converse,lapidoth1996nearest,lapidoth1998reliable,ganti2000mismatched,somekh2014achievable,somekh2015general,huleihel2019gaussian,scarlett2020information}.
In this model, however, the encoder and decoder are assumed to be
completely fixed in advance and are not selected in any way. By contrast,
in this work we focus on the selection of a decoder from a class of
decoders using samples of the noise, and specifically, how well this
can be done in a way which generalizes well to out-of-sample noise. 

While being somewhat new to channel coding problems, empirical design
of encoders is a common practice for lossy source coding problems,
usually via the celebrated Lloyd\textendash Max algorithm \cite{gray1998quantization}.
Learning-theoretic analysis of empirically designed quantizers was
carried out during the 1990's and onward \cite{linder1994rates,linder1997empirical,bartlett1998minimax,linder2000training,linder2002learning,antos2005individual,merhav1997amount}.\footnote{As an exception to this , a channel coding related problem was considered
in \cite{merhav1997many}, which mainly focused on impossibility results.} Among various results, it was established that a distortion redundancy
of $\epsilon$ is achievable with $n=\tilde{O}(\frac{d^{3}m^{1-2/d}}{\epsilon^{2}})$
samples, where here $m$ is the size of quantization codebook (number
of reproduction points). The seemingly peculiar dependency on $m$,
and the optimal minimax tightness of its dependency on $n$ is discussed
and established in \cite[Thm. 2]{bartlett1998minimax} (see also \cite{antos2005improved}).
More recently, this setup was generalized to coding from a general
Hilbert space to a finite-dimensional space, with linear \cite{maurer2010k}
and non-linear \cite{lee2019learning} mappings. Beyond vector quantization
(or clustering), this framework is general enough to also include
principal component analysis, non-negative matrix factorization, dictionary
learning \cite{vainsencher2011sample}, and modern unsupervised representation
learning techniques, such as autoencoders. Given that channel coding
(communication) bears both similarities and differences with source
coding (mainly quantization), understanding learning-theoretic questions
for channel coding seems feasible, but also requires dedicated effort
to handle the unique features of the latter. 

A learning approach that avoids channel modeling and estimation, may
be contrasted with the universal decoding approach prevalent in information
theory \cite{csiszar2011information,lapidoth1998reliable}. Similarly
to the learning-based approach, universal methods avoid explicit estimation
of the channel, and aim to find a single decoding rule which does
not depend on the unknown channel statistics, yet it is simultaneously
nearly optimal for any channel in the class (in a sense which requires
explicit definition, e.g., as in \cite{feder2002universal}). The
universal approach is highly applicable on the theoretical side, but
bares some challenges from the practical side.\footnote{In contrast to the remarkable practical success of universal methods
is lossless data compression \cite{ziv1977universal,ziv1978compression}.} The key problem is that in most problems the universal decoder obtained
is required to compute a metric for all codewords in the codebook,
which is typically infeasible due to the large number of codewords,
e.g., \cite{weinberger2008universal}. This is true even if the code
is structured, e.g., a linear or a convolutional code, such that optimal
or efficient decoding rules, e.g., Viterbi's decoder \cite{viterbi2013principles}
or a belief propagation \cite{richardson2008modern} decoder, can
be efficiently implemented when the channel statistics is available.
Furthermore, it was shown in \cite{merhav2013universal} that a universal
decoder which aims to compete with a family of low complexity decoders
will need to be much more complex than each of those decoders. Loosely
speaking, this is an inherent difficulty for any universal approach
because in some sense, universal decoders \emph{implicitly} learn
the channel statistics and decode the message at the same time. Hence,
they cannot rely on likelihood information that is obtained separately
from each channel output, not even when the channel is known to be
memoryless. Similar situation occurs in lossy source coding, for which
\cite[Sec. V]{wyner1998role} states that ``low computational complexity
is achievable only at the expense of yielding a non-optimal distortion''
(mainly regarding such practical methods which are based on approximate
string matching). By contrast, in a learning-based approach, the class
of decoders may be restricted in advance to feasible decoders, and
the available data may then be used to choose the best one from the
given class. 

\subsection{Main Notation Conventions}

We mainly use standard notation or define it before its first use,
and here only highlight main conventions. The standard Euclidean norm
for $x\in\mathbb{R}^{d}$ is denoted by $\|x\|$. The operator norm
for a matrix $S\in\mathbb{R}^{d\times d}$, viewed as an operator
from $\mathbb{R}^{d}$ to $\mathbb{R}^{d}$ (with both spaces equipped
with the Euclidean norm), is denoted by $\|S\|_{\text{op}}$. The
maximal eigenvalue (resp. minimal) of a symmetric matrix $S$ is denoted
by $\zeta_{\text{max}}(S)$ (resp. $\zeta_{\text{min}}(S)$). The
$d$-dimensional diagonal matrix whose diagonal entries are given
by $v\in\mathbb{R}^{d}$ is denoted by $\diag(v)\in\mathbb{R}^{d\times d}$.
The zero-centered Euclidean ball of radius $r$ is denoted by $\mathbb{B}^{d}(r)\dfn\{x\in\mathbb{R}^{d}\colon\|x\|\leq r\}$
and the unit sphere is denoted by $\mathbb{S}^{d-1}\dfn\{x\in\mathbb{R}^{d}\colon\|x\|=1\}$.
For $n\in\mathbb{N}_{+}$, the set $\{1,2,\ldots,n\}$ is denoted
by $[n]$. The cardinality of a finite set $A$ is denoted by $|A|$.
The Hamming distance between $x=\{x_{1},\ldots,x_{d}\},y=\{y_{1},\ldots,y_{d}\}\in\mathbb{R}^{d}$
is denoted by $\Hamd(x,y)\dfn\left|\{i\in[d]\colon x_{i}\neq y_{i}\}\right|$.
The minimum (resp. maximum) of $a,b\in\mathbb{R}$ is denoted by $a\wedge b$
(resp. $a\vee b$). The $(m-1)$-dimensional probability simplex is
denoted by $\mathbb{A}^{m-1}\dfn\{\boldsymbol{a}=(a_{1},\ldots,a_{m})\colon\sum_{j}a_{j}=1,\;a_{j}\geq0\}$.
The indicator of an event ${\cal A}$ is denoted by $\I\{{\cal A}\}$.
As a general rule, random variables or vectors will be denoted by
capital letters and specific values they may take will be denoted
by the corresponding lower case letters. Statistical independence
between random variables $X,Y$ is denoted by $X\ind Y$. Equality
of the probability distributions of $X,Y$ is denoted by $X\eqd Y$.
The Kullback\textendash Leibler (KL) divergence between the probability
distributions $\mu$ and $\tilde{\mu}$ is denoted by $\dkl(\mu||\tilde{\mu})$.
Integrals of probability densities $f$ (which absolutely continuous
w.r.t. Lebesgue measure $\lambda$) are taken w.r.t. the Lebesgue
measure $\lambda$, and this is abbreviated as $\int f\dfn\int f\d\lambda$.
The set of Borel probability measures on $\mathbb{R}^{d}$ is denoted
by ${\cal P}(\mathbb{R}^{d})$. For $\mu_{U},\mu_{V}\in{\cal P}(\mathbb{R}^{d})$,
the $p$th order Wasserstein distance is denoted by $W_{p}(\mu_{U},\mu_{V})\dfn\inf\E^{1/p}(\|U-V\|^{p})$
where $U\sim\mu_{U}$ and $V\sim\mu_{V}$, and where the infimum is
taken over all couplings $\mu_{UV}$ (i.e., joint distributions which
agree with the marginal measures of $U$ and $V$). A real random
variable $X$ is called $\sigma$-sub-Gaussian whenever
\[
\sigma=\|X\|_{\psi_{2}}\dfn\inf\left\{ t>0\colon\E\exp(X^{2}/t^{2})\leq2\right\} <\infty.
\]

\section{Learning under an Error Probability Loss Function \label{sec:Learning error probability}}

\subsection{Problem Formulation}

Consider again the additive noise channel in (\ref{eq: additive noise channel introduction}):
The input is denoted by $X\in\mathbb{R}^{d}$, the output by $Y=X+Z\in\mathbb{R}^{d}$
where $X\ind Z\in\mathbb{R}^{d}$, and where for the purpose of matrix
multiplication are all taken as column vectors. The distribution of
$Z$ is denoted by $\mu$, and it is assumed that it is completely
unknown to the designer of the communication system. Let $\|x-y\|_{S}\dfn\sqrt{(x-y)^{T}S(x-y)}$
be the Mahalanobis distance between $x,y\in\mathbb{R}^{d}$ which
is parameterized by an inverse covariance matrix $S\in\mathbb{\mathbb{S}}_{+}^{d}$
(where $\mathbb{\mathbb{S}}_{+}^{d}$ is the positive-definite cone).
The transmitter chooses a codebook $C=\{x_{j}\}_{j\in[m]}$ of size
$|C|=m$ where $x_{j}\in\mathbb{R}^{d}$ for all $j\in[m]$, and at
each use of the channel (\ref{eq: additive noise channel introduction}),
chooses a codeword from $C$ uniformly at random to send over the
channel. The decoder employs a nearest neighbor decoder w.r.t. $\|\cdot\|_{S}$,
i.e., given a channel output $y$, the index of the decoded codeword
is chosen as: 
\begin{equation}
\hat{j}(y)\in\argmin_{j\in[m]}\|x_{j}-y\|_{S}.\label{eq: generalized minimum distance}
\end{equation}
Our choice of decoding rule of the form (\ref{eq: generalized minimum distance})
is motivated by the fact that this is the optimal (maximum-likelihood)
rule when $Z\sim N(0,S^{-1})$, that is, whenever the noise distribution
is Gaussian with covariance matrix $S^{-1}$. We stress, however,
that we do not assume that the noise $Z$ is Gaussian, and so (\ref{eq: generalized minimum distance})
is chosen for its simplicity, and, in general, will not be a maximum
likelihood decoder. The decision region of the $j$th codeword is
$\{y\colon\|x_{j}-y\|_{S}\leq\min_{j'\in[m]\backslash\{j\}}\|x_{j'}-y\|_{S}\}$
and the boundary of these decision regions belong to the set of hyperplanes
defined by $\|x_{j}-y\|_{S}=\|x_{j'}-y\|_{S}$ for all $j\neq j'$.

The \emph{expected error probability }(over the noise distribution)
given that the $j$th codeword was transmitted is then given by 
\begin{align}
\pe_{\mu}(C,S\mid j) & \dfn\E_{Z\sim\mu}\left[\I\left\{ \min_{j'\in[m]\backslash\{j\}}\|x_{j}+Z-x_{j'}\|_{S}<\|Z\|_{S}\right\} \right]\\
 & =\E_{Z\sim\mu}\left[\I\left\{ \min_{j'\in[m]\backslash\{j\}}\|x_{j}-x_{j'}\|_{S}^{2}+2(x_{j}-x_{j'})^{T}SZ<0\right\} \right],
\end{align}
and the \emph{expected average error probability} by 
\[
\pe_{\mu}(C,S)\dfn\frac{1}{m}\sum_{j=1}^{m}\pe_{\mu}(C,S\mid j)=\E_{Z\sim\mu}\left[\frac{1}{m}\sum_{j=1}^{m}\I\left\{ \min_{j'\in[m]\backslash\{j\}}\|x_{j}-x_{j'}\|_{S}^{2}+2(x_{j}-x_{j'})^{T}SZ<0\right\} \right].
\]
The learning goal related to this formulation is the use of noise
samples to design either a codebook $C\in{\cal C}\subset(\mathbb{R}^{d})^{m}$,
or a decoder inverse covariance matrix $S\in{\cal S}\subset\mathbb{S}_{+}^{d}$,
or both, with minimal expected error probability. Here, the set ${\cal C}$,
for example, could represent an average power constraint ${\cal C}=\{C\colon\frac{1}{m}\sum_{j=1}^{m}\|x_{i}\|^{2}\leq r\}$
for some $r>0$, and the set ${\cal S}$ may represent a restriction
on $\|S\|$ for some matrix norm. 

Assuming that the codebook/decoder designer is provided with $n$
samples of the noise $\boldsymbol{Z}=\{Z_{i}\}_{i=1}^{n}\stackrel{\tiny\mathrm{i.i.d.}}{\sim}\mu$,
the \emph{empirical average error probability }(over the $n$ samples
$\{Z_{i}\}$ of the noise) is given by 
\[
\pe_{\boldsymbol{Z}}(C,S)=\frac{1}{n}\sum_{i=1}^{n}\frac{1}{m}\sum_{j=1}^{m}\I\left\{ \min_{j'\in[m]\backslash\{j\}}\|x_{j}-x_{j'}\|_{S}^{2}+2(x_{j}-x_{j'})^{T}SZ_{i}<0\right\} .
\]
Note that we may write the empirical average error probability as
\begin{equation}
\pe_{\boldsymbol{Z}}(C,S)=\frac{1}{n}\sum_{i=1}^{n}\ell(C,S,Z_{i})\label{eq: empirical error probability as a sum of losses over samples}
\end{equation}
where the loss function is given by 
\begin{equation}
\ell(C,S,z)\dfn\frac{1}{m}\sum_{j=1}^{m}\ell_{j}(C,S,z)\label{eq: error probability loss as averaging over codewords}
\end{equation}
 with
\[
\ell_{j}(C,S,z)\dfn\I\left[\min_{j'\in[m],j'\neq j}\|x_{j}-x_{j'}\|_{S}^{2}+2(x_{j}-x_{j'})^{T}Sz<0\right].
\]

\paragraph*{Comparison to multiclass classification}

Evidently, the problem here resembles multiclass classification, but
as we next discuss, the problems are not equivalent. According to
our formulation, the data of the learner are $n$ noise samples $\{z_{i}\}_{i=1}^{n}$.
In multiclass classification, each data sample $z_{i}$ should be
equipped with a label $j\in[m]$, and the goal of the learner is to
output a classifier, whose performance is measured under the $0-1$
loss function, which equals $1$ if and only if the label is erroneously
classified. By contrast, in the channel learning problem, the noise
samples $\{z_{i}\}_{i=1}^{n}$ are not labeled at all. The learner
is required to output a decoder, which is indeed a multiclass classifier
(to one of $m$ classes), but is also required to output a codebook
$C$. The codebook is an additional requirement from the learner,
which is not a part of the multiclass classification problem. Furthermore,
even if $C$ is fixed in advance and is not required to be learned,
the problem of learning a channel decoder and learning a multiclass
classifier are still not equivalent. To see this, note that indeed
one can synthesize a multiclass classification problem from $\{z_{i}\}_{i=1}^{n}$
and a codebook $C$ by considering an augmented data set $\{x_{j}+z_{i}\}_{j\in[m],\;i\in[n]}$
of $mn$ data samples. Then, as zero loss is incurred when $x_{j}+z_{i}$
is decoded as $j\in[m]$, one may synthetically attribute the label
$j$ to all the points $\{x_{j}+z_{i}\}_{i\in[n]}$. This indeed results
a multiclass classification problem for the labeled data set $\{x_{j}+z_{i},j\}_{j\in[m],\;i\in[n]}$,
and the loss function $\ell(C,S,z)$ in (\ref{eq: error probability loss as averaging over codewords})
is equivalent to the $0-1$ loss function in multiclass classification.
However, in general multiclass classification problems, a data set
of $mn$ points will not have any specific structure, whereas here
it is comprised of $m$ translations of the same $n$ noise samples
$\{z_{i}\}_{i=1}^{n}$. The difference in the structure of the data
set then distinguishes between the two problems. Finally, while the
decoding regions resulting from the decoding rule (\ref{eq: generalized minimum distance})
are indeed hyperplanes, there is no guarantee that a multiclass classifier
which separates classes in $\mathbb{R}^{d}$ with hyperplanes can
be synthesized as a decoding rule in the form of (\ref{eq: generalized minimum distance})
for some given $S$. This is simply because $S$ is parameterized
by at most $O(d^{2})$ parameters, whereas a general such classifier
is parameterized by $O(m^{2}d)$ such parameters ($d$ parameters
for each hyperplane, and a total of $O(m^{2})$ hyperplanes, one for
each pair of classes). The above differences between the multiclass
classification problem and the channel coding problem lead to differences
in generalization bounds (c.f. the next Theorem \ref{thm:Generalization NN}
with the generalization bounds for standard multiclass classification
\cite[Thm. 29.3]{shalev2014understanding}).

\subsection{Generalization Error Bounds}

As a cornerstone for learning-theoretic analysis of channel codes,
we first establish the following upper bound on the empirical error
which is uniform over $(C,S)$:
\begin{thm}
\label{thm:Generalization NN}Assume that $n\geq d+1$. With probability
of at least $1-\delta$, for all $C\subset(\mathbb{R}^{d})^{m}$ with
$|C|=m$ and $S\in\mathbb{S}_{+}^{d}$
\begin{equation}
\left|\pe_{\mu}(C,S)-\pe_{\boldsymbol{Z}}(C,S)\right|\leq4m\sqrt{\frac{2(d+1)\log\left(\frac{en}{d+1}\right)}{n}}+\sqrt{\frac{2\log(2/\delta)}{n}}.\label{eq: generalization NN bound}
\end{equation}
\end{thm}

\paragraph*{Learning algorithms and ERM}

A learning algorithm $A\colon(\mathbb{R}^{d})^{n}{\cal \mapsto C}\times{\cal S}$
for this problem obtains data samples $\boldsymbol{Z}$ as input and
outputs $A_{\boldsymbol{Z}}\equiv(C_{\boldsymbol{Z}},S_{\boldsymbol{Z}})$.
Since the right-hand side (r.h.s.) of Theorem \ref{thm:Generalization NN}
uniformly bounds $|\pe_{\mu}(C,S)-\pe_{\boldsymbol{Z}}(C,S)|$, it
also bounds the generalization error for any learning algorithm. For
example, a generic learning algorithm is ERM which chooses $(C,S)_{\s[ERM]}$
that is $\epsilon$-close to $\inf_{C\in{\cal C},S\in{\cal S}}\pe_{\boldsymbol{Z}}(C,S)$
for some given $\epsilon>0$. The generalization error $|\pe_{\mu}((C,S)_{\s[ERM]})-\pe_{\boldsymbol{Z}}((C,S)_{\s[ERM]})|$
is then bounded by Theorem \ref{thm:Generalization NN}. Denote by
$(C_{*},S_{*})$ a pair which $\epsilon$-achieve $\inf_{C,S}\pe_{\mu}(C,S)$.
Then, given $n=\tilde{O}(\frac{m^{2}d+\log(1/\delta)}{\epsilon^{2}})$
samples, it holds with probability larger than $1-\delta$ that
\begin{align}
\pe_{\mu}((C_{\boldsymbol{Z}},S_{\boldsymbol{Z}})_{\s[ERM]}) & \leq\pe_{\boldsymbol{Z}}((C_{\boldsymbol{Z}},S_{\boldsymbol{Z}})_{\s[ERM]})+\epsilon\\
 & \leq\pe_{\boldsymbol{Z}}((C_{*},S_{*}))+\epsilon\\
 & \leq\inf_{C,S}\pe_{\boldsymbol{Z}}((C,S))+2\epsilon\\
 & \leq\inf_{C,S}\pe_{\mu}(C,S)+3\epsilon.\label{eq: risk of ERM}
\end{align}
In general, however, an efficient algorithm to find or approximate
the ERM for this problem seems to be out of reach. In the next section,
we will demonstrate that one of the reasons for that is that the loss
function is discontinuous function of $(C,S)$, and that by properly
modifying the loss function, one can propose an algorithm of reasonable
complexity.

\paragraph*{Proof outline and interpretation of the bound}

The proof uses the known relation between uniform error bounds and
Rademacher complexity \cite{bartlett2002rademacher} of the loss class
induced by the samples and the hypothesis class (codebook and decoder
inverse covariance matrix). In accordance, the first term in \ref{eq: generalization NN bound}
is an upper bound on the expected Rademacher complexity, while the
second term is a high probability bound on the deviation of the expected
Rademacher complexity from its expected value. This form is similar
to other generalization bounds ,e.g, for binary classification \cite[Theorem 6.8]{shalev2014understanding}.
In turn, the expected Rademacher complexity is bounded by analyzing
the growth function of the loss class via the Sauer\textendash Shelah
lemma, utilizing the fact that the decoding regions are polyhedral,
and the separating regions between any pair of codewords is a $d$-dimensional
hyperplane, whose Vapnik\textendash Chervonenkis (VC) dimension is
bounded by $d+1$. In fact, any other decoding rule for which the
pairwise decision rule is chosen from a class of binary classifiers
of VC dimension $d+1$ would lead to the same bound. A common and
slightly different approach to bounding the growth function is to
define a proper combinatorial dimension which captures the behavior
of the growth function of the loss class induced by $\pe_{\boldsymbol{Z}}(C,S)$.
This can be done via an analogous result to the regular Sauer\textendash Shelah
lemma (which is aimed for binary classification and uses the VC dimension).
The resulting proof, however, seems to be more complicated than necessary
and the result seems to only be (slightly) worse. Nonetheless, this
approach could be useful in other scenarios as we discuss below. 

\paragraph*{Codebook structure}

Theorem \ref{thm:Generalization NN} does not make any assumptions
regarding the structure of the codebook, and its generalization bound
depends linearly on $m$. We will show in the next section that the
modification of the loss function mentioned above also improves the
dependency on the number of codewords to nearly square-root. However,
if the dimension is high $d\gg1$, then useful codebooks typically
have $m=2^{dR}\gg1$ codewords (where $R>0$ is the rate per dimension).
Therefore, beyond the standard utilization of structured codebooks
for efficient encoding and decoding, the learning-based approach also
requires to utilize the codebook structure for refined generalization
bounds which have better dependency on the number of codewords. Specifically,
it is plausible that for some classes of codes, the combinatorial
dimension mentioned above could be much lower than its value for general,
unstructured, codebooks, and this will lead to generalization bounds
which have reasonable dependency on the codebook size. 

\paragraph*{Learning error exponents}

The bound in Theorem \ref{thm:Generalization NN} represents an \emph{additive}
deviation from the true error probability. However, the required error
probability for various communication could be very low, and in these
cases the interest is shifted to \emph{error exponents}, i.e., $\log\pe_{\mu}(C,S)$.
In these cases, a \emph{multiplicative} deviation bound is of more
importance. However, analysis of the generalization error in this
case seems challenging as $\log\pe_{\boldsymbol{Z}}(C,S)-\log\pe_{\mu}(C,S)$
is not an additive function of the losses over $z_{i}$ (see (\ref{eq: empirical error probability as a sum of losses over samples})),
and $\log(\cdot)$ is not a Lipschitz function, so that $|\log\pe_{\boldsymbol{Z}}(C,S)-\log\pe_{\mu}(C,S)|$
cannot be directly bounded by $|\pe_{\boldsymbol{Z}}(C,S)-\pe_{\mu}(C,S)|$. 

The analysis of learning of structured codes and error exponents remains
an open problem. The bound in Theorem \ref{thm:Generalization NN}
is comprised of a term related to the complexity of the hypothesis
class ($\tilde{O}(m\sqrt{\frac{d}{n}}))$ and a term related only
to the required reliability ($O(\sqrt{\frac{\log(1/\delta)}{n}})$).
We next show that the $O(\frac{1}{\sqrt{n}})$ dependency cannot be
improved, even for a rather basic setting. 
\begin{prop}
\label{prop:Generalization NN minimax rates}Let $A_{\boldsymbol{Z}}\equiv(C_{\boldsymbol{Z}},S_{\boldsymbol{Z}})$
be an arbitrary learning algorithm for a codebook of two codewords
$m=2$ over ${\cal C}=\mathbb{B}^{d}(1)$ and ${\cal S}=\{S\colon\zeta_{\text{\emph{min}}}(S)=1,\;\zeta_{\text{\emph{max}}}(S)\leq r_{s}\}$
where $r_{s}\leq2$, where $\zeta_{\text{\emph{max}}}(S)$ (resp.
$\zeta_{\text{\emph{min}}}(S)$) is the maximal eigenvalue (resp.
minimal) of $S$.\footnote{Since both $\pe_{\mu}(C,S)$ and $\pe_{\boldsymbol{z}}(C,S)$ are
insensitive to scaling of $S$, the condition on $S$ implied by ${\cal S}$
is equivalent to a condition on the condition number of $S$. The
requirement $r_{s}\leq2$ is merely made in order to simplify the
exposition, and any other upper bound on $r_{s}$ would lead to a
different constant in the r.h.s. of the inequality defining the event
whose probability is bounded below.} Let $\delta\in(0,1/4)$ be given. Then, 
\[
\sup_{\mu}\P_{\boldsymbol{Z}\sim\mu^{\otimes n}}\left[\pe_{\mu}(C_{\boldsymbol{Z}},S_{\boldsymbol{Z}})-\inf_{C,S}\pe_{\mu}(C,S)>\frac{27}{100\cdot r_{s}}\cdot\sqrt{\frac{1}{n}\log\left(\frac{1}{4\delta}\right)}\right]\geq\delta.
\]
\end{prop}

\paragraph*{Proof outline}

The proof such results is typically based on a reduction to binary
hypothesis testing, obtained by establishing that if a learning algorithm
does not implicitly distinguish between the true distribution and
an alternative distribution then the resulting loss is large (even
though such identification is not a requirement from the learner).
In standard classification \cite[Sec. 28.2]{shalev2014understanding}
and estimation \cite{tsybakov2008introduction} problems, one postulates
a prior probability on the two distributions, and then the specific
structure of the loss function allows to show that the optimal learning
algorithm should operate as if the true distribution is known to be
the one more likely out of the two, given the noise samples. However,
in the channel code learning problem, given a posterior distribution
on a pair of noise distributions, the optimal codebook/covariance
matrix pair may not match exactly to neither of the distributions,
but rather should provide a ``compromise'' between the two. In the
proof of Proposition \ref{prop:Generalization NN minimax rates} we
use a pair of Gaussian distributions, for which the above phenomenon
occurs due to the fact \textbf{$Q(\frac{1}{\sqrt{t}})$ }(where $Q(\cdot)$
is the tail distribution function of the standard normal distribution)
is neither convex nor concave on $\mathbb{R}_{+}$. To ensure convexity,
it is required that the variance of the equivalent noise induced by
$C,S$ (which moves the codeword towards the boundary of the two decoding
regions) will be bounded by some numerical constant for all $C\in{\cal C},S\in{\cal S}$.
This explains the necessity of the perhaps unintuitive condition $\zeta_{\text{max}}(S)\leq r_{s}$
in the proposition. If this condition does not hold, that noise variance
could be arbitrarily large which invalidates the required condition
for convexity (for the sake of intuition, consider the codewords $x_{1}=(1,0)=-x_{2}$
and a decoder which sets the boundary between the decoding regions
to be the horizontal axis). 

\paragraph*{Tightness of the codebook size/dimension in Theorem \ref{thm:Generalization NN}}

For classification problems, the tightness of the hypothesis class
complexity related term (the first term in the bound of Theorem \ref{thm:Generalization NN})
is established by a proper combinatorial dimension (such as VC dimension)
$D$, and then reducing the learning problem to a hypothesis testing
problem involving $2^{D}$ different distributions. Then again, one
needs to show that identifying the correct distribution is necessary
for learning at sufficiently fast rate (e.g., \cite[Lemma 28.1]{shalev2014understanding}).
When $m>2$, even finding the optimal code for a given noise distribution
is difficult, and so using this method to show tightness for the channel
codes learning problem seems challenging. Finding alternative methods
or showing otherwise that the rate in Theorem \ref{thm:Generalization NN}
is not tight w.r.t. $m$ is left for future research. 

\section{Learning under a Surrogate to the Error Probability Loss Function
\label{sec:Learning surrogate}}

\subsection{Problem Formulation}

We next propose an upper bound to the error probability loss function
and consider it to be a surrogate loss function to the error probability
loss function. We state a uniform convergence result along with a
heuristic alternating-minimization algorithm which attempts to minimize
this loss function. The starting point is the trivial inequality $\I(t<0)\leq(1-t)\vee0$,
which leads to the following \emph{hinge-type} upper bound on the
error probability loss function
\begin{align}
\ell_{j}(C,S,z) & \dfn\I\left[\min_{j'\in[m]\backslash\{j\}}\left(\|x_{j}-x_{j'}\|_{S}^{2}+2(x_{j}-x_{j'})^{T}Sz\right)<0\right]\\
 & \leq\left[1-\min_{j'\in[m]\backslash\{j\}}\left(\|x_{j}-x_{j'}\|_{S}^{2}+2(x_{j}-x_{j'})^{T}Sz\right)\right]\vee0\label{eq: surrogate error probability definition}\\
 & \dfn\overline{\ell}_{j}(C,S,z).\label{eq: surrogate error probability notation}
\end{align}
Analogously to (\ref{eq: empirical error probability as a sum of losses over samples})
and (\ref{eq: error probability loss as averaging over codewords}),
we denote $\overline{\ell}(C,S,z)\dfn\frac{1}{m}\sum_{j=1}^{m}\overline{\ell}_{j}(C,S,z)$,
define $\peb_{\boldsymbol{Z}}(C,S)\dfn\frac{1}{n}\sum_{i=1}^{n}\overline{\ell}(C,S,Z_{i})$,
as well as $\peb_{\mu}(C,S)=\E[\overline{\ell}(C,S,Z)]$. 

\paragraph*{Scaling of inverse covariance matrix}

Unlike the error probability loss function, $\overline{\ell}_{j}(C,S,z)$
is not invariant to scaling of $S$, and in fact, the maximal eigenvalue
of $S$ (i.e., its operator norm) limits the ``resolution'' of the
decoder to distinguish between two codewords. For illustration, suppose
that $S=\sigma\cdot I_{d}$ where $\sigma>0$. We can then consider
\[
\tilde{\ell}_{j,\sigma}(C,z)\dfn\overline{\ell}_{j}(C,\sigma I_{d},z)=\left[1-\sigma\cdot\min_{j'\in[m]\backslash\{j\}}\left(\|x_{j}-x_{j'}\|^{2}+2(x_{j}-x_{j'})^{T}z\right)\right]\vee0
\]
as a family of loss functions indexed by $\sigma$. This loss function
can also bound the standard error probability similarly to (\ref{eq: surrogate error probability definition})
using the inequality $\I(t<0)\leq(1-\sigma t)\vee0$ which holds for
any $\sigma>0$. The larger $\sigma$ is, the slope of the straight
line defining this loss function will be larger, and the surrogate
loss function will more severely penalize competing codewords on shorter
distances. As this amounts to merely scaling of $S$, we simply set
$\sigma=1$, and continue with the definition in (\ref{eq: surrogate error probability notation}). 

\subsection{A Generalization Error Bound}

The following theorem provides a uniform convergence bound for the
surrogate loss function:
\begin{thm}
\label{thm:Generalization NN surrogate}Assume that:
\begin{enumerate}
\item $\P_{Z\sim\mu}\{Z\in\mathbb{B}^{d}(1)\}=1$. 
\item ${\cal C}=({\cal C}_{1})^{m}$ where ${\cal C}_{1}=\{x\in\mathbb{R}^{d}\colon\|x\|\leq r_{x}\}$
for some $r_{x}\ge1$. 
\item ${\cal S}=\{S\in\mathbb{\mathbb{S}}_{+}^{d}\colon\zeta_{\text{\emph{max}}}(S)\leq r_{s}\}$
for some $r_{s}\geq1$.
\end{enumerate}
Then with probability of at least $1-\delta$, for all $C\subset{\cal C}$
and $S\in{\cal S}$
\begin{equation}
\left|\peb_{\mu}(C,S)-\peb_{\boldsymbol{Z}}(C,S)\right|\leq112\cdot\sqrt{\frac{(d\vee m)(d+1)\log(31\cdot dr_{s}r_{x})}{n}}+\sqrt{\frac{2r^{2}\log(2/\delta)}{n}},\label{eq: uniform convergence bound for surrogate}
\end{equation}
where $r\dfn1\vee4r_{x}r_{s}(r_{x}+1)$.
\end{thm}

\paragraph*{Empirical risk minimization}

Assume the typical case in which $m\geq d$. Repeating (\ref{eq: risk of ERM}),
$n=\tilde{O}(\frac{dm+\log(1/\delta)}{\epsilon^{2}})$ samples suffice
to obtain with high probability a $3\epsilon$-approximation for the
minimal error probability using ERM. Specifically, the dependency
of the generalization error on the number of codewords has been improved
from linear in $m$ to nearly square-root. This improved dependency
may be attributed to the fact that the loss function $\overline{\ell}(C,S,z)$
is a Lipschitz continuous function of $(C,S)$ (Lemma \ref{lem: Lipschitzness of surrogate loss function}
in Appendix \ref{sec:Proofs}), and thus ``easier'' to be minimized
compared to the discontinuous $\ell(C,S,z)$. This is reflected in
the proof outlined next. We also remark that the proof method used
to prove Proposition \ref{prop:Generalization NN minimax rates} which
shows that $n=\tilde{O}(\frac{\log(1/\delta)}{\epsilon^{2}})$ samples
are necessary to learn the regular error probability loss function
seems also applicable here.\footnote{Namely, lower bounding the loss by taking the worst case of the same
pair of noise distributions defined in the proof, but modifying Lemma
\ref{lem: error probability of m=00003D2 and Gaussian noise} to match
the surrogate loss function.} The details are omitted.

\paragraph*{Proof outline and a comparison to Theorem \ref{thm:Generalization NN}}

Similarly to the bound of Theorem \ref{thm:Generalization NN}, the
bound (\ref{eq: uniform convergence bound for surrogate}) on the
generalization error is comprised of a term which bounds the expected
Rademacher complexity, and a term which is a high probability bound
on the deviation of the Rademacher complexity from its expectation.
The main goal of the proof is upper bounding the expected Rademacher
complexity of the loss class. However, compared to the standard error
probability loss function, here the Lipschitzness of the loss function
(Lemma \ref{lem: Lipschitzness of surrogate loss function} in Appendix
\ref{subsec:Proof-of-Theorem}) makes the Rademacher complexity of
the induced loss class a maxima of a sub-Gaussian process. Consequently,
bounds on the covering numbers of the loss class can be derived (Lemma
\ref{lem: covering number of inverse covariance matrices} in Appendix
\ref{subsec:Proof-of-Theorem}), which, in turn, are used in Dudley's
entropy integral to bound the Rademacher complexity (Lemma \ref{lem: bounding Rad complexity}
in Appendix \ref{subsec:Proof-of-Theorem}, see, e.g., \cite[Ch. 5]{van2016probability}
for a discussion on the \emph{chaining }argument leading to these
bounds). By contrast, in Theorem \ref{thm:Generalization NN}, the
loss function is not Lipschitz, and this enforces to bound the Rademacher
complexity by the \emph{growth function}. In essence, the growth function
is the size of a covering set at scale \emph{zero}, which is only
larger than Dudley's entropy integral, which integrates over covering
sizes at increasing scales (which are naturally smaller). 

\paragraph*{Conditions}

The conditions in Theorem \ref{thm:Generalization NN surrogate} are
made in order to refrain from over-complicating the analysis. Specifically,
analogous conditions to the bounded-noise support and bounded-norm
codewords are common in the statistical-learning literature, and were
also made in quantizer-learning papers (e.g., \cite{linder1994rates}).
They are mainly assumed because for any bounded interval the function
$t\mapsto t^{2}$ is Lipschitz continuous, and this allows the control
the supremum of empirical process via contraction methods and concentration
inequalities (see a discussion \cite[Sec. 1.1]{mendelson2014learning}).
It is plausible that these conditions can be removed using the techniques
of \cite{mendelson2014learning,mendelson2018learning,liang2015learning}. 

\paragraph*{Numerical constants in (\ref{eq: uniform convergence bound for surrogate})}

In the derivation of the theorem and its proof, a simple form of the
numerical constants was favored to tightness, and the conditions $r_{x}\ge1$
and $r_{s}\geq1$ are inconsequential and were only made for this
purpose. In accordance, the constant in Theorem \ref{thm:Generalization NN surrogate}
can be significantly reduced.

\subsection{Alternating optimization algorithm}

Another benefit of using the surrogate loss function $\overline{\ell}(\cdot)$
is that it allows to introduce a simple alternating optimization algorithm
to minimize $\peb_{\boldsymbol{Z}}(C,S)$, utilizing the fact that
the loss function $\overline{\ell}_{j}(C,S,z_{i})$ is continuous
in $(C,S)$. Nonetheless, $\peb_{\boldsymbol{Z}}(C,S)$ is not a convex
function of $C$ \textendash{} not even for a fixed $S$ \textendash{}
due to the minimization over $j'\in[m]\backslash\{j\}$ appearing
in (\ref{eq: surrogate error probability definition}).\footnote{For a fixed $S$, and in case $m=2$, so that there is no need to
minimize over $j'\in[m]\backslash\{j\}$, $\peb_{\boldsymbol{Z}}(C,S)$
is a convex function of $C$ since $\overline{\ell}(C,S,z)$ is a
convex function of $C$ by the composition rules of convex/concave
functions \cite[Ch. 3.2.4]{Boyd}. In general, however, the pointwise
minimum of convex functions is not necessarily convex.} To circumvent this, we propose the following heuristic. Let us introduce
the auxiliary variables $A\dfn\{\alpha_{j,j'}^{(i)}\}_{j,j'\in[m],i\in[n]}$
where $\alpha_{j,j'}^{(i)}\in[0,1]$ and $\alpha_{j,j}^{(i)}\equiv0$.
For any given data sample $z_{i}$ and codeword index $j\in[m]$,
we define, with a slight abuse of notation,
\[
\overline{\ell}_{j}(C,S,A,z_{i})\dfn\sum_{j'\in[m]}\alpha_{j,j'}^{(i)}\left[1-\left(\|x_{j}-x_{j'}\|_{S}^{2}+2(x_{j}-x_{j'})^{T}Sz_{i}\right)\right],
\]
and would ideally like to set $A$ such that $\overline{\ell}_{j}(C,S,A,z_{i})=\overline{\ell}_{j}(C,S,z_{i})$.
To wit, if the clipping operation is active and $\overline{\ell}_{j}(C,S,z_{i})=0$
then $\alpha_{j,j'}^{(i)}=0$ for all $j'$. Otherwise, $\alpha_{j,j'}^{(i)}$
is arbitrarily supported on 
\[
j'\in\argmin_{j'\in[m]\backslash\{j\}}\left(\|x_{j}-x_{j'}\|_{S}^{2}+2(x_{j}-x_{j'})^{T}Sz_{i}\right)
\]
(for example, uniformly), and satisfies $\sum_{j,j'\in[m]}\alpha_{j,j'}^{(i)}=1$.
Clearly, $\{\alpha_{j,j'}^{(i)}\}_{j'\in[m]}$ depend on $(C,S)$,
and ``encode'' the nearest neighbors of $x_{j}$ w.r.t. to the noise
sample $z_{i}$. The idea of the algorithm is to relax this dependency,
and to alternatively update $A$ and $(C,S)$ at each iteration. Given
an initial guess $(C,S)$, in the first part of the iteration the
value of $A$ is determined, and in the second part of the iteration,
given $A$ the value of $(C,S)$ is optimized, or just updated by
a stochastic gradient step. Given the new guess for $(C,S)$ the second
iteration follows in the same manner, and so on. Let us denote $\overline{\ell}(C,S,A,z_{i})=\frac{1}{m}\sum_{j=1}^{m}\overline{\ell}_{j}(C,S,A,z_{i})$.
For $p\in[m]$, the gradient given the sample $z_{i}$ w.r.t. $x_{p}$,
$p\in[m]$ is given by 
\begin{equation}
\frac{\dee\overline{\ell}(C,S,A,z_{i})}{\dee x_{p}}=-\frac{2}{m}\sum_{j\in[m]}(\alpha_{j,p}^{(i)}+\alpha_{p,j}^{(i)})S(x_{p}-x_{j})+(\alpha_{j,p}^{(i)}-\alpha_{p,j}^{(i)})Sz_{i},\label{eq: codeword gradient}
\end{equation}
and w.r.t. $S$ by 
\begin{equation}
\frac{\dee\overline{\ell}(C,S,A,z_{i})}{\dee S}=-\frac{1}{m}\sum_{j,j'\in[m]}\alpha_{j,j'}^{(i)}\left[(x_{j}-x_{j'})(x_{j}-x_{j'})^{T}+2(x_{j}-x_{j'})z_{i}^{T}\right].\label{eq: inverse covariance gradient}
\end{equation}
Note that the last gradient may be an asymmetric matrix and does not
depend on $S$. The \emph{symmetric} matrix $\tilde{Q}$ which maximizes
$\langle\tilde{Q},\frac{\dee}{\dee S}\overline{\ell}(C,S,A,z_{i})\rangle$
is clearly given by $\tilde{Q}_{*}=\tfrac{1}{2}\frac{\dee}{\dee S}\overline{\ell}(C,S,A,z_{i})+\tfrac{1}{2}[\frac{\dee}{\dee S}\overline{\ell}(C,S,A,z_{i})]^{T}$,
and so we choose this to be the update direction. The matrix $S$
is then updated to $\hat{S}=S-\eta\tilde{Q}_{*}$ where $\eta>0$
is a step size. This is a symmetric matrix and can be decomposed as
$\hat{S}=\sum_{q=1}^{d}\zeta_{q}v_{q}v_{q}^{T}$ where $\{\zeta_{q}\}$
are the eigenvalues and $v_{q}$ are the eigenvectors. We then project
$\hat{S}$ to $\tilde{S}\in{\cal S}\subset\mathbb{S}_{+}^{d}$ as
\[
\tilde{S}=\frac{1}{\frac{r_{s}}{\zeta_{\text{max}}}\wedge1}\sum_{q=1}^{d}(\mu_{q}\vee0)\cdot v_{q}v_{q}^{T}
\]
which makes sure that $\tilde{S}$ is a nonnegative definite matrix
and that its maximal eigenvalue is less than $r_{s}$. 

Algorithm \ref{alg:Stochasic-gradient-descent} lists a possible SGD
variant of such algorithm. A SGD algorithm is specifically adequate
for the alternating optimization since a small change in $(C,S)$
is not expected to change most of the $\{\alpha_{j,j'}^{(i)}\}$.
The inputs to the algorithm are the noise samples $\boldsymbol{z}$,
initial values for $(C,S)$, and step sizes for their update. For
each $z_{i}$, first the values of $\{\alpha_{j,j'}^{(i)}\}_{j,j'\in[m]}$
are determined according to the current codebook $C^{(i-1)}=\{x_{j}^{(i-1)}\}_{j\in[m]}$,
and second, they are updated using the gradients in  (\ref{eq: codeword gradient})
and (\ref{eq: inverse covariance gradient}) (computed based on the
single sample $z_{i}$). 

\subsubsection*{Convergence analysis}

A different version of Algorithm \ref{alg:Stochasic-gradient-descent}
of higher computational complexity is a gradient descent algorithm
in which the gradient is averaged over all $n$ samples at each iteration.
The resulting algorithm is then an iterative generalized alternating
optimization algorithm \cite{gunawardana2005convergence}, which is
a \emph{descent} algorithm for the empirical surrogate loss $\peb_{\boldsymbol{Z}}(C,S)$.
While it is guaranteed that the loss decreases at each iteration,
it is not a priori clear that such algorithm globally converges in
the (rather weak) sense that the iterations $\{C^{(i)},S^{(i)}\}_{i=1}^{\infty}$
tend to a limit for any given initialization \cite{zangwill1969convergence}.
Furthermore, it is not \emph{a priori }clear that the sequence of
algorithms which are based on increasing number of samples is consistent
in the sense that such algorithm tends as $n\to\infty$ to the ideal
population version of this algorithm, which is based on the true distribution
of the noise $\mu$ (in a manner that can be made precise, see \cite{sabin1986global}).
Nonetheless, properties of this nature were proved for canonical alternating-minimization
algorithms like the expectation\textendash maximization (EM) algorithm
\cite{wu1983convergence} and its variants \cite{gunawardana2005convergence},
the Lloyd\textendash Max ($k$-means) algorithm \cite{sabin1986global}.
In addition, such properties were recently proved \cite{weinberger2020k}
for a quantizer design problem which utilizes a surrogate loss function
similar in spirit to the one studied here. As the analysis there is
of technical nature and it seems that similar ideas can be applied
to this paper we do not pursue this direction here. Anyhow, if one
has obtained $\peb_{\boldsymbol{Z}}(C^{(i_{*})},S^{(i_{*})})$ below
an acceptable threshold for some $i_{*}\in\mathbb{N}$ using Algorithm
\ref{alg:Stochasic-gradient-descent}, then the expected error probability
$\peb_{\mu}(C^{(i_{*})},S^{(i_{*})})$ is bounded as in Theorem \ref{thm:Generalization NN surrogate}.

\begin{algorithm}
\begin{algorithmic}[1]

\State  \textbf{input }$\boldsymbol{z}=\{z_{i}\}_{i\in[n]}$, $C^{(0)}=\{x_{j}^{(0)}\}_{j\in[m]}\subset({\cal C}_{1})^{m}$,
$S^{(0)}$, $\{\lambda^{(i)}\}_{i=1}^{n}$, $\{\eta^{(i)}\}_{i=1}^{n}$,
$r_{s}$

\State  \textbf{begin }

\State  \textbf{~for all $1\leq i\leq n$ do}

\State  \textbf{~~for all $1\leq j\leq m$ do}

\State  ~~~\textbf{ ${\cal J}'_{*}\leftarrow\argmax_{j'\in[m]\backslash\{j\}}\left[1-\|x_{j}^{(i-1)}-x_{j'}^{(i-1)}\|_{S}^{2}+2(x_{j}^{(i-1)}-x_{j'}^{(i-1)})^{T}S^{(i)}z_{i}\right]$}

\State \textbf{ ~~~$j'_{*}\leftarrow{\cal J}'_{*}\{1\}$}\Comment{
choose an arbitrary member of ${\cal J}_{*}'$ }

\State  \textbf{~~~if $1-\left(\|x_{j}^{(i-1)}-x_{j'_{*}}^{(i-1)}\|_{S}^{2}-2(x_{j}^{(i-1)}-x_{j'_{*}}^{(i-1)})^{T}S^{(i)}z_{i}\right)<0$
}\Comment{ the clipping in (\ref{eq: surrogate error probability definition})
is active }

\State  \textbf{~~~~for all $1\leq j'\leq m$ do }

\State  \textbf{~~~~~$\alpha_{j,j'}^{(i)}\leftarrow0$} 

\State  \textbf{~~~~end for}

\State  \textbf{~~~else}

\State  \textbf{~~~~for all $1\leq j'\leq m$ do }

\State  \textbf{~~~~~$\alpha_{j,j'}^{(i)}\leftarrow\I[j'=j'_{*}]$}

\State  \textbf{~~~~end for}

\State  \textbf{~~~end if }

\State  ~\textbf{~~}$A^{(i)}\leftarrow\{\alpha_{j,j'}^{(i)}\}_{j,j'\in[m]}$ 

\State  ~\textbf{~~}$x_{j}^{(i-1/2)}\leftarrow x_{j}^{(i-1)}-\lambda^{(i)}\frac{\dee\overline{\ell}(C,S,A^{(i)},z_{i})}{\dee x_{j}}$
\Comment{ a gradient update of codeword $x_{p}$}

\State  ~\textbf{~~}$x_{j}^{(i)}\leftarrow\argmin_{\tilde{x}\in{\cal C}_{1}}\|x_{j}^{(i-1/2)}-\tilde{x}\|_{S^{(i)}}$
\Comment{ projection of $x_{p}$ to the feasible set}

\State  \textbf{~~end for}

\State  \textbf{~~}$S^{(i-1/2)}\leftarrow S^{(i-1)}-\tfrac{1}{2}\eta^{(i)}\frac{\dee\overline{\ell}(C,S,A^{(i)},z_{i})}{\dee S}-\tfrac{1}{2}\eta^{(i)}\left[\frac{\dee\overline{\ell}(C,S,A^{(i)},z_{i})}{\dee S}\right]^{T}$\Comment{
a gradient update of $S$}

\State  ~~~ Compute eigenvalue decomposition $\{(\mu_{q},v_{q})\}_{q=1}^{d}$
of $S^{(i-1/2)}$ \Comment{eigenvalue decomposition}

\State  ~\textbf{~~}$S^{(i)}\leftarrow\frac{1}{\frac{r_{s}}{\zeta_{\text{max}}}\wedge1}\sum_{q=1}^{d}(\mu_{q}\vee0)\cdot v_{q}v_{q}^{T}$
\Comment{ projection of $S$ to the feasible set ${\cal S}$}

\State  \textbf{~end for}

\State  \textbf{end}

\State  \textbf{output $(C^{(n)}=\{x_{j}^{(n)}\},S^{(n)})$.}

\end{algorithmic}

\caption{An SGD algorithm for learning a codebook and a decoder for an additive
noise channel \label{alg:Stochasic-gradient-descent}}
\end{algorithm}

\section{Learning by Codebook Expurgation \label{sec:Learning Expurgation}}

\subsection{Problem Formulation and the Gibbs Algorithm}

In general, finding an optimal codebook is a difficult task even when
the noise distribution $\mu$ is known. A simple way to approach this
problem is to select the $m$ codewords of the codebook $C$ from
a larger super-codebook $C_{0}$ of $m_{0}>m$ codewords. Such a process
is amenable to practical implementation, since the super-codebook
can be statically chosen in advance, and can be simple or well-structured
(such as a grid or a lattice \cite{zamir2014lattice}), whereas the
$m$ codewords in the codebook can be chosen dynamically based on
the noise statistics. This approach is akin to both practical coding
methods \cite{forney1989multidimensional}, as well as to the the
common technique used in the proofs of random coding bounds on the
reliability function of channel coding \cite{gallager1965simple},
in which the codebook is \emph{expurgated} from codewords of large
conditional error probability. We assume in this section, that the
decoder inverse covariance matrix is fixed (say $S=I_{d}$), and thus
omit $S$ from the notation.

Nonetheless, even when $\mu$ is known, finding the optimal set of
$m$ codewords is a combinatorial optimization problem, which is computationally
heavy when $m_{0}\gg m$. To see this, consider the simpler combinatorial
optimization problem of finding the codebook $C$ which minimizes
the average pairwise error probability (known as the \emph{union bound
estimate})
\begin{equation}
\argmin_{C=\{x_{1},\ldots,x_{m}\}\subset C_{0}}\frac{1}{m(m-1)}\sum_{j_{1},j_{2}\in[m]\colon j_{1}\neq j_{2}}\pe_{\mu}(x_{j_{1}}\to x_{j_{2}}),\label{eq: minimizing pairwise error probability}
\end{equation}
where $\pe_{\mu}(x_{j_{1}}\to x_{j_{2}})$ is the error probability
of making an error from $x_{j_{1}}$ to $x_{j_{2}}$ when these are
the only two codewords in the codebook. The problem (\ref{eq: minimizing pairwise error probability})
is then equivalent to the\emph{ $k$-cardinality sub-graph problem}
\cite[Sec. 4]{bruglieri2006annotated} as follows: The $m_{0}$ codewords
of $C_{0}$ can be taken as the nodes of a complete directed graph,
such that the weight of each edge is $\pe_{\mu}(x_{j_{1}}\to x_{j_{2}})$.
Then, $\sum_{j_{1},j_{2}\in[m]\colon j_{1}\neq j_{2}}\pe_{\mu}(x_{j_{1}}\to x_{j_{2}})$
is the total edge weight of the sub-graph of cardinality $m$, which
only contains the codewords of $C$ as nodes. The problem (\ref{eq: minimizing pairwise error probability})
is then equivalent to finding a sub-graph of cardinality $m$ with
minimal weight. Hence, any algorithm which solves or approximates
the $k$-cardinality sub-graph problem can be used find the solution
to (\ref{eq: minimizing pairwise error probability}). Nonetheless,
the $k$-cardinality sub-graph problem is NP-hard \cite{ehrgott1992optimization},
and the problem of interest here, of finding 
\begin{equation}
C_{*}=\argmin_{C=\{x_{1},\ldots,x_{m}\}\subset C_{0}}\pe_{\mu}(C)\label{eq: optimal expurgation}
\end{equation}
is only more difficult since the error probability is a more complicated
function of the codebook compared to the average pairwise error probability
in (\ref{eq: minimizing pairwise error probability}). Similar observations
can be made for the empirical error probability. 

A possible greedy relaxation to this optimization problem is to approximate
the optimum by gradually removing codewords from the codebook, say
$k$ of them at each step (as will be evident, $k$ is practically
expected to be chosen as a small integer). For simplicity of the description
we assume henceforth that $T\dfn\frac{m_{0}-m}{k}\in\mathbb{N}_{+}$.
The general meta-algorithm is as follows. Initialize with a codebook
$C_{0}$ of $m_{0}$ codewords. Then, for $t=1,\ldots T$:
\begin{enumerate}
\item Construct candidate codebooks $\{C_{t}^{[l]}\}_{l\in[{m_{t-1} \choose k}]}$
and evaluate the error probability for each candidate $\pe(C_{t}^{[l]})$.
\item Choose an index $l^{*}$ and set $C_{t}\equiv\{C_{t}^{[l_{*}]}\}$
according to a selection rule (which is based on the the error probabilities).
Renumber the codewords in $C_{t}$ by $[m_{t}]$ where $m_{t}=m_{t-1}-k$. 
\end{enumerate}
The error probabilities computed at the first step may be either according
to the true distribution $\mu$ or according to the empirical distribution
induced by $\boldsymbol{Z}$, and the algorithm is termed, respectively,
the population algorithm or empirical algorithm. 

A proper choice of a selection rule is a delicate question. For example,
a possible variant of such algorithm would remove the $k$ codewords
in $C_{t}$ which have the maximal conditional error probability $\pe(C_{t}\mid j)$.
However, this is both greedy as well as naive since the decision to
remove the codeword $\tilde{x}$ from the codebook should be based
on both types of error events \textendash{} from $\tilde{x}$ when
it is the transmitted codeword to a different (competing) codeword,
as well as the opposite case in which $\tilde{x}$ is decoded but
a different codeword was transmitted. Another problem is that such
algorithm depends strongly on the noise samples for the empirical
algorithm, and thus might not generalize well to out-of-sample noise. 

To circumvent this problem, we next propose a \emph{Gibbs algorithm}
which randomly removes codewords from the codebook. Let $C_{t}$ be
a codebook of $m_{t}$ codewords in ${\cal C}$, let $Q\in{\cal P}(\mathbb{R}^{d})$
be a probability reference measure on $\mathbb{R}^{d}$ whose support
includes ${\cal C}$, and let $\beta>0$ be an inverse temperature
parameter. Given $C_{t}$, a Gibbs algorithm chooses to expurgate
the codewords with indices $\{j_{1},\ldots j_{k}\}$ to obtain $C_{t+1}$
with probability 
\begin{equation}
\P\left[C_{t+1}=C_{t}\backslash\{x_{j}\}_{j\in\{j_{1},\ldots j_{k}\}}\mid\boldsymbol{Z},C_{t}\right]\propto Q(C_{t+1})\cdot\exp\left[-\beta\cdot\pe_{\boldsymbol{Z}}(C_{t+1})\right].\label{eq: Gibbs algorithm}
\end{equation}
For $0<\beta<\infty$, the algorithm compromises between the two extremes
of removing codewords at random according to the prior distribution
$Q$ $(\beta\to0)$ versus strong dependence on the noise samples
$(\beta\to\infty)$. The Gibbs algorithm is listed in Algorithm \ref{alg: Gibbs expurgation}.

\begin{algorithm}
\begin{algorithmic}[1]

\State  \textbf{input }$\boldsymbol{z}=\{z_{i}\}_{i\in[n]}$, $C_{0}=\{x_{j}\}_{j\in[m_{0}]}\subset{\cal C}$,
$\beta>0$, $m\leq m_{0}$ , $k\mid|C_{0}|-m$, $Q\in{\cal P}(\mathbb{R}^{d})$

\State  \textbf{$T\leftarrow\frac{m_{o}-m}{k}$}

\State  \textbf{begin }

\State  \textbf{for all $1\leq t\leq T$ do}

\State  \textbf{~$m_{t}\leftarrow m_{t-1}-k$}, $L\leftarrow{m_{t-1} \choose k}$

\State  ~Choose an arbitrary enumeration of the $L$ sets $\{{\cal J}_{l}\}_{l\in[L]}$
such that ${\cal J}_{l}\subset[m_{t-1}]$ and $|{\cal J}_{l}|=m_{t}$
for all $l\in[L]$

\State  \textbf{~for all $1\leq l\leq L$ do}

\State  ~~$C_{t}^{[l]}\leftarrow\{x_{j}\}_{j\in{\cal J}_{l}}$. 

\State  ~~Compute $\{\pe_{\boldsymbol{z}}(C_{t}^{[l]})\}$

\State  \textbf{~end for}

\State  ~ $\Psi\leftarrow\sum_{l=1}^{L}Q(C_{t}^{[l]})\cdot\exp\left[-\beta\cdot\pe_{\boldsymbol{z}}(C_{t}^{[l]})\right]$
\Comment{ normalization factor for the Gibbs distribution}

\State  \textbf{~for all $1\leq l\leq L$ do}

\State  ~~$p_{l}\leftarrow\Psi^{-1}\cdot Q(C_{t}^{[l]})\cdot\exp\left[-\beta\cdot\pe_{\boldsymbol{z}}(C_{t}^{[l]})\right]$

\State  \textbf{~end for}

\State  ~Randomly select $l_{*}\sim(p_{1},\ldots,p_{L})$. \Comment{
random choice of codebook}

\State  ~$C_{t}\leftarrow C_{t}^{[l_{*}]}$, and renumber the codewords
of $C_{t}$ by $[m_{t}]$

\State  \textbf{end for}

\State  \textbf{output $C_{T}$}

\State  \textbf{end}

\end{algorithmic}

\caption{A Gibbs expurgation algorithm for learning a codebook for an additive
noise channel \label{alg: Gibbs expurgation}}
\end{algorithm}

\paragraph*{Computational load}

Note that the computation of $\pe_{\boldsymbol{Z}}(C_{t+1})$ requires
not only removing the $\pe_{\boldsymbol{Z}}(C_{t}\mid j_{p})$, $p\in[k]$
from the averaging operation in the error probability, but should
also take into account that the decoder cannot err to the codewords
$\{x_{j}\}_{j\in\{j_{1},\ldots j_{k}\}}$. This has to be done for
each of the ${m_{t-1} \choose k}$ candidate codebooks, and so the
choice of $k$ significantly determines the complexity via the required
number of candidate codebooks, where the latter is upper bounded by
$m_{0}^{k}$ (see further discussion on efficient implementation in
Section \ref{subsec:Gaussian-Mixture-noise-Gibbs}). However, once
the algorithm's parameters are fixed, so is the running time of the
algorithm. Furthermore, the algorithm actually produces codebooks
of any size $m\leq m'\leq m_{0}$, typically with lower error probability
for smaller codebook, and thus the codebook size can be dynamically
chosen. We also remark that after being learned by the decoder, the
chosen codebook can be sent back to the encoder via a feedback link
using no more than $m\log_{2}m_{0}$ bits. 

To analyze the error of the algorithm, we arbitrarily set $C_{0}$,
and let $\boldsymbol{C}_{\mu}=(C_{0},C_{\mu,1},\ldots,C_{\mu,T})$
(resp. $\boldsymbol{C}_{\boldsymbol{Z}}=(C_{0},C_{\boldsymbol{Z},1},\ldots,C_{\boldsymbol{Z},T})$)
be the sequence of random codebooks generated by the population (resp.
empirical) Gibbs algorithm, when both are initialized with $C_{0}$.
Let $C_{*}\subset C_{0}$ be the codebook obtained by optimal expurgation,
as in (\ref{eq: optimal expurgation}). The average excess error probability
of the empirical Gibbs algorithm can be decomposed as 
\[
\E\left[\pe_{\mu}(C_{\boldsymbol{Z},T})-\pe_{\mu}(C_{*})\right]=\underbrace{\E\left[\pe_{\mu}(C_{\boldsymbol{Z},T})-\pe_{\mu}(C_{\mu,T})\right]}_{\text{empirical error}}+\underbrace{\E\left[\pe_{\mu}(C_{\mu,T})-\pe_{\mu}(C_{*})\right]}_{\text{approximation error }}
\]
where the expectations are taken w.r.t. both the randomness of $\boldsymbol{Z}$
and the Gibbs algorithm. The empirical error is a result of using
the empirical distribution of $\boldsymbol{Z}$ in lieu of the true
distribution $\mu$ in the Gibbs algorithm, and as we shall see is
upper bounded by $\tilde{O}(\beta\sqrt{\frac{T}{n}})$, and vanishes
as $n\to\infty$. By contrast, the approximation error seems to be
an inevitable price to pay for using using a computationally feasible
method (Gibbs algorithm) for (approximately) solving (\ref{eq: optimal expurgation}).
It cannot be reduced by increasing the number of samples, and seems
challenging to quantify due to its intricate dependency on the noise
distribution.

\paragraph*{An alternative algorithm}

As said, finding $C_{*}$ which minimizes $\pe_{\mu}(C)$ in (\ref{eq: optimal expurgation})
is a combinatorial optimization problem, and as such it can be tackled
using the\emph{ simulated annealing }approach \cite{kirkpatrick1983optimization}.
This approach is based on \emph{local optimization} of the codebook.
In its naive form, such algorithm is initialized with a the super-codebook
$C_{0}$ of $m_{0}$ codewords, and a codebook $C^{(1)}\subset C_{0}$
of $m$ codewords. At iteration $t\in\mathbb{N}_{+}$, the codebook
$C^{(t+1)}\subset C_{0}$ is chosen as the codebook which minimizes
the error probability among all codebooks of size $m$ which are different
from $C^{(t)}$ by $k$ codewords, if such exists, and otherwise the
algorithm stops and outputs its current codebook. As well known, the
simulated annealing \cite{johnson1989optimization} replaces the ``hill-descending''
step, with a randomized step which though it might increase the objective
function (error probability), is essential in order to avoid local
minima. The same algorithm can be used to minimize $\pe_{\boldsymbol{Z}}(C)$
in the empirical setting, however, as discussed above, this might
not guarantee generalization to out-of-sample noise. Nonetheless,
it is easy to see that generalization bounds similar to Theorems \ref{thm:Generalization and empirical error for Gibbs}
and \ref{thm: High probability bound on Gibbs algorithm} can be obtained
for an algorithm which chooses $C$ randomly from
\begin{equation}
\P\left[C\mid\boldsymbol{Z}\right]\propto Q(C)\cdot\exp\left[-\beta\cdot\pe_{\boldsymbol{Z}}(C)\right]\cdot\I\left[C\subset C_{0}\colon|C|=m\right].\label{eq: Metropolis-Hastings distribution}
\end{equation}
Sampling directly from (\ref{eq: Metropolis-Hastings distribution})
is difficult (since this distribution is supported on ${m_{0} \choose m}$
different codebooks and so it is costly or impossible to compute the
required normalization factor), however it can be done indirectly
via the Metropolis\textendash Hastings algorithm \cite{metropolis1953equation,hastings1970monte,bolstad2009understanding},
which generates a Markov chain whose stationary distribution approaches
that of (\ref{eq: Metropolis-Hastings distribution}). Nonetheless,
the time required to converge to (\ref{eq: Metropolis-Hastings distribution})
(mixing time) might be large. Analysis of this method, and experiments
which compare its effectiveness with that of Algorithm \ref{alg: Gibbs expurgation}
are left for future research. 

\subsection{Generalization Error and Empirical Error Bounds }

The next theorem states a bound on the average empirical error, and
also provides a bound on the average generalization error $\E[\pe_{\mu}(C_{\boldsymbol{Z},T})-\pe_{\boldsymbol{Z}}(C_{\boldsymbol{Z},T})]$. 
\begin{thm}
\label{thm:Generalization and empirical error for Gibbs}Assume that
$C_{0}$ of size $|C_{0}|$ is chosen in a data-independent way, that
the Gibbs algorithm is used with $T=\frac{m_{0}-m}{k}\in\mathbb{N}_{+}$
steps, and inverse-temperature $\beta>0$. Also assume that $m_{0}\geq2m$
and 
\begin{equation}
\beta^{2}\left(\log n+\left(\frac{m_{0}}{2}+1\right)\log m_{0}-\log k\right)\leq\frac{n}{2},\label{eq: first assumption on beta}
\end{equation}
Then, the average empirical error is bounded as 
\begin{equation}
\E\left[\pe_{\mu}(C_{\boldsymbol{Z},T})-\pe_{\mu}(C_{\mu,T})\right]\leq3\sqrt{\frac{T\beta^{2}\left(\log n+m_{0}\log m_{0}-\log k\right)}{n}}\label{eq: bound on the average empirical error}
\end{equation}
and the average generalization error is upper bounded as
\begin{align}
\E\left[\pe_{\mu}(C_{\boldsymbol{Z},T})-\pe_{\boldsymbol{Z}}(C_{\boldsymbol{Z},T})\right] & \leq\sqrt{T\left(\frac{\beta}{n}\wedge\frac{\beta^{2}}{4n^{2}}\right)}.\label{eq: bound on the average generalization bound}
\end{align}
\end{thm}

\paragraph*{Discussion}

The average empirical error bound is on the order of $O(\sqrt{\frac{T\beta^{2}\log n}{n}})$.
The proof is based on showing that $|\pe_{\boldsymbol{z}}(C)-\pe_{\mu}(C)|\leq\epsilon$
for all possible sub-codebooks $C\subset C_{0}$ with high probability
(the $\sqrt{\log n}$ dependency can be traced in the proof to a union
bound which leads to this result), and using this to bound generalization
error by bounding the KL divergence $\dkl(P_{C_{\boldsymbol{z},1}\cdots C_{\boldsymbol{z},T}}||P_{C_{\mu,1}\cdots C_{\mu,T}})$
where $P_{C_{\mu,1}\cdots C_{\mu,T}}$ (resp. $P_{C_{\boldsymbol{z},1}\cdots C_{\boldsymbol{z},T}}$)
is the the joint probability measure of $\boldsymbol{C}_{\mu}$ (resp.
$\boldsymbol{C}_{\boldsymbol{Z}}$). The average generalization error
is on the order of $O(\sqrt{T\left(\frac{\beta}{n}\wedge\frac{\beta^{2}}{n^{2}}\right)})$,
which, in the standard case $\beta<n$ is $O(\sqrt{T}\frac{\beta}{n})$,
which decays faster than the average empirical error bound. The bound
is proved by establishing an information-theoretic stability of the
Gibbs algorithm, and using the results of \cite{xu2017information}
which bounds the generalization error of stable algorithms. 

The condition $m_{0}\geq2m$ is only made for simplicity of exposition
of the proof. The generalization error can be used to estimate a bound
on $\pe_{\mu}(C_{\boldsymbol{Z},T})$ using $\pe_{\boldsymbol{Z}}(C_{\boldsymbol{Z},T})$,
which in turn can be computed from the data $\boldsymbol{Z}$. It
should be noted, however, that unlike Theorems \ref{thm:Generalization NN}
and \ref{thm:Generalization NN surrogate}, the stated bound is on
the \emph{average} error, and has no strong concentration properties.
This is a result of the proof method which relies on the chain rule
properties of the KL divergence (and mutual information) to bound
the average error. A bound which holds with high probability will
be presented in what follows. Evidently, both upper bounds in Theorem
\ref{thm:Generalization and empirical error for Gibbs} are monotonically
increasing functions of $\beta$, and thus $\beta$ should be as low
as possible in order to minimize the bounds on the empirical and generalization
errors. This can be contrasted with the goal of minimizing $\pe_{\mu}(C_{\mu,T})$
(or $\pe_{\boldsymbol{Z}}(C_{\boldsymbol{Z},T})$) which typically
requires $\beta$ to be as ``large'' as possible.

Assuming that $\frac{\beta}{n}\to0$ as $n\to\infty$, the average
generalization error indicated by Theorem \ref{thm:Generalization and empirical error for Gibbs}
is $O(\frac{\sqrt{T}\beta}{n})$. This bound involves averaging on
both the randomness of the data samples $\boldsymbol{Z}$, as well
as the randomness of the Gibbs algorithm. To obtain high probability
bounds, we rely on a \emph{uniform-stability} property of the Gibbs
algorithm.\footnote{Recently, \cite{esposito2019generalization} proposed to use the R\'{e}nyi
mutual information between the data samples $\boldsymbol{Z}$ and
the algorithm output, but the resulting bounds in this setting are
weaker than the ones stated here.} Specifically, if we assume the stronger condition that $\frac{T\beta}{n}\to0$
as $n\to\infty$ then a high probability of roughly the same order
can also be obtained. Specifically, for a given algorithm $P_{C\mid\boldsymbol{Z},C_{0}}$,
and a single noise sample $\tilde{z}\in\mathbb{R}^{d}$ we let 
\[
\qe_{\tilde{z}}(\boldsymbol{z})\dfn\E\left[\pe_{\tilde{z}}(C)\mid\boldsymbol{Z}=\boldsymbol{z},C_{0}\right]
\]
be the error probability of $\tilde{z}$ when averaged over a random
codebook $C$ that is drawn according to $P_{C\mid\boldsymbol{Z}=\boldsymbol{z},C_{0}}$
(defined via the Gibbs algorithm), and $\qe_{\mu}(\boldsymbol{z})\dfn\E_{\tilde{Z}\sim\mu}[\qe_{\tilde{Z}}(\boldsymbol{z})]$
as well as $\qe_{\tilde{\boldsymbol{Z}}}(\boldsymbol{z})\dfn\frac{1}{n}\sum_{i=1}^{n}\qe_{\tilde{z}_{i}}(\boldsymbol{z})$.
We have the following: 
\begin{thm}
\label{thm: High probability bound on Gibbs algorithm}Assume that
$\frac{T\beta}{n}\to0$ as $n\to\infty$. Then, there exists an absolute
$n_{0}\in\mathbb{N}_{+}$ and an absolute constant $c>0$ such that
for all $n\geq n_{0}$
\begin{equation}
\P\left[\qe_{\mu}(\boldsymbol{Z})-\qe_{\boldsymbol{Z}}(\boldsymbol{Z})>c\left(\frac{\sqrt{T}\beta}{n}+\frac{1}{\sqrt{n}}\right)\cdot\log\left(\frac{n}{\beta\sqrt{T}}\right)\cdot\log n\cdot\log\frac{n}{\delta}\right]\leq\delta.\label{eq: High proabability bound gibbs}
\end{equation}
\end{thm}
Hence, under the condition of Theorem \ref{thm: High probability bound on Gibbs algorithm},
and assuming that $\beta=\Omega(\sqrt{\frac{n}{T}})$, the high probability
bound is $\tilde{O}(\frac{\sqrt{T}\beta}{n})$, and matches, up to
logarithmic factors, the average generalization error bound of Theorem
\ref{thm:Generalization and empirical error for Gibbs}. 

\paragraph*{Discussion}

As discussed, the output of the Gibbs algorithm is random due to both
the randomness of the samples and the randomness of the Gibbs mechanism.
While the generalization bound of Theorem \ref{thm:Generalization and empirical error for Gibbs}
is averaged w.r.t. both type of randomness, the bound of Theorem \ref{thm: High probability bound on Gibbs algorithm}
is a high probability bound w.r.t. the samples, but still averages
the Gibbs mechanism. Nonetheless, assuming $\frac{T\beta}{n}=o(1)$
the decay rate of Theorem \ref{thm:Generalization and empirical error for Gibbs}
is recovered (up to logarithmic terms), and the generalization error
is $\tilde{O}(\frac{\sqrt{T}\beta}{n})$ with high probability, and
not only on the average. The bound \ref{eq: High proabability bound gibbs}
is proved by establishing that the Gibbs algorithm is a \emph{differentially-private
}algorithm \cite{dwork2014algorithmic} (Lemma \ref{lem: Gibbs algorithm is differntialy stable}
in Appendix \ref{subsec:Proof-of-Theorems}), which, in turn, implies
that it is \emph{uniformly stable }learning algorithm \cite{bousquet2002stability}.
Then, the recent high probability bound on the generalization error
of uniformly stable learning algorithms \cite{feldman2019high} is
utilized. 

\section{Learning Input Distributions which Maximize Mutual Information \label{sec: Learning MI}}

\subsection{Problem Formulation}

Communicating at rates which approach the capacity of the additive
noise channel (\ref{eq: additive noise channel introduction}) requires
knowledge of the noise distribution $\mu_{Z}$\footnote{In this section we add the sub-script $Z$ to $\mu$ and explicitly
$\mu_{Z}$, so it will not be confused with the input distribution
$\mu_{X}$.} in order to optimize the input distribution which we denote here
by $\mu_{X}$.\footnote{As was considered in previous sections, the design of the decoder
is also based on the noise distribution, and in uncoded systems will
affect the error probability. However, in coded systems, and especially
in the random coding regime, this can be circumvented in principle
by use of universal decoders which also achieve capacity \cite{lapidoth1998reliable}. } Thus, whenever coding across multiple $d$-dimensional codewords
is possible, it is desired to find the input distribution $\mu_{X}$
which maximizes the mutual information between the input and the output,
or, equivalently, the differential entropy of the channel output.
Typically, to obtain finite mutual information, the set of feasible
input distributions is restricted to some ${\cal P}$ and so it is
required to solve: 
\begin{equation}
\argmax_{\mu_{X}\in{\cal P}}I(X;X+Z)=\argmax_{\mu_{X}\in{\cal P}}h(X+Z)\label{eq: maximizing MI}
\end{equation}
where we assume that the maximum exists (otherwise, it is required
to find an $\epsilon>0$ approximation of the supremum). 

Closed-form solutions to (\ref{eq: maximizing MI}) are rare, and
currently exist only for the simplest classes ${\cal P}$, even when
the noise distribution is completely known. For example, for a known
Gaussian noise $Z\sim N(0,I_{d})$, the optimal input distribution
is Gaussian if ${\cal P}$ represents an average power constraint,
but if an amplitude constraint is also enforced, to wit 
\[
{\cal P}=\left\{ \mu_{X}\colon\P_{\mu_{X}}(\|X\|\leq A_{X})=1,\;\E_{\mu_{X}}\|X\|\leq d\sigma_{X}^{2}\right\} 
\]
then it is only known that the optimal $\mu_{X}$ is supported on
a finite number of concentric shells with isotropic direction (see
\cite{smith1971information,shamai1995capacity,rassouli2016capacity},
and \cite{dytso2019capacity} for an overview and recent advances).
Finding the optimal support, however, is not trivial, and requires
algorithmic efforts, e.g., the cutting-plane iterative algorithm proposed
in \cite{huang2005characterization} for finding a discrete approximation
to the capacity input distribution. Similarly, for the class of input
distributions which are restricted to some $m$-point codebook $C\subset\mathbb{R}^{d}$
whose power is bounded, to wit,
\[
{\cal P}=\left\{ \mu_{X}\colon\E_{\mu_{X}}\|X\|\leq d\sigma_{X}^{2},\;|\supp(\mu_{X})|\leq m\right\} ,
\]
it is only known that for $d=1$ the optimal input distribution weakly
converges to an equi-lattice when $\sigma_{X}\to\infty$ and to a
Gaussian quadrature when $\sigma_{X}\to0$ \cite[Sec. IV]{wu2010impact,ozarow1990capacity}.
Nonetheless, from an algorithmic point of view, if the support is
restricted to a fixed codebook $C=\{x_{j}\}_{j\in[m]}\subset\mathbb{R}^{d}$
and 
\begin{equation}
{\cal P}=\left\{ \mu_{X}=\sum_{j=1}^{m}a_{j}\delta_{x_{j}}\colon\E_{\mu_{X}}\|X\|\leq d\sigma_{X}^{2},\;x_{j}\in\mathbb{R}\;\forall j\in[m],\;\boldsymbol{a}\in\mathbb{A}^{m-1}\right\} ,\label{eq: input distribution with codebook and weights}
\end{equation}
where $\delta_{x_{0}}=\delta(x-x_{0})$ and $\delta(x)$ is Dirac's
delta function, then the problem of maximizing $h(X+Z)$ over the
weights $\boldsymbol{a}$ (in a feasible subset of $\mathbb{A}^{m-1}$
dictated by the power constraint) is a convex optimization problem
and can be solved using the celebrated Blahut\textendash Arimoto algorithm
\cite{blahut1972computation,arimoto1972algorithm}. 

Following the approach previously taken in this paper, we focus on
the statistical-learning aspect of this problem, i.e., the difference
between empirical and population versions of this optimization problem,
when the noise distribution is unknown, and instead $n$ i.i.d. samples
of the noise are available. We consider two classes for input distributions,
and show for each one that an estimator $\hat{h}_{\boldsymbol{Z}}(X+Z)$
to $h(X+Z)$ which is based on the noise samples converges with high
probability to the true value and that the convergence is uniform
over the chosen class ${\cal P}$. As in the previous sections, this
assures that \emph{any} algorithm which attempts to maximize $\hat{h}_{\boldsymbol{Z}}(X+Z)$
(that can be computed from data) will produce $h(X+Z)$ which is not
very far from the computed value, where specific convergence rates
depend on the class of input distributions.

\subsection{A General Class of Input Distributions}

The first class of input distributions we consider is rather general,
but in turn the resulting convergence rate assured is only $\tilde{O}(n^{-1/d})$
(Theorem \ref{thm: learning input distributions}), and as such deteriorates
fast with the dimension $d$. Following the above discussion, there
is also no known efficient algorithm that finds the maximizing distribution
in this class even for a known noise distribution. Nonetheless, the
motivation of deriving such a result is to demonstrate that uniform
convergence is a possible even for quite general classes. The results
itself necessitates further discussion, and this follows its formal
statement. 

Following \cite{polyanskiy2016wasserstein}, we say that a density
$\mu_{X}$ (absolutely continuous w.r.t. Lebesgue measure) is $(\psi_{1},\psi_{2})$-regular
for $\psi_{1}>0$ and $\psi_{2}\geq0$ if 
\begin{equation}
\|\nabla\log\mu_{X}(x)\|\leq\psi_{1}\|x\|+\psi_{2},\;\forall x\in\supp(\mu_{X}).\label{eq: regular density}
\end{equation}
We then consider input densities $\mu_{X}$ which are $(\psi_{1},\psi_{2})$-regular,
have a bounded second moment, and have a absolutely bounded entropy:
\[
{\cal P}^{*}(\eta_{X},\sigma_{X},\psi_{1},\psi_{2})\dfn\left\{ \mu_{X}\text{ is }(\psi_{1},\psi_{2})\text{-regular}\colon\;\E_{\mu_{X}}\|X\|^{2}\leq d\sigma_{X}^{2},\;|h(\mu_{X})|\leq\eta_{X}\right\} .
\]

\begin{thm}
\label{thm: learning input distributions} Assume that $\mu_{Z}$
is a probability density on $\mathbb{R}^{d}$ such that $A\dfn\|Z\|$
is $(\sqrt{d}\sigma_{Z})$-sub-Gaussian where $\sigma_{Z}=\Omega(n^{-(d-2)/(4d)})$.
Let $\boldsymbol{Z}=(Z_{1},\ldots,Z_{n})\stackrel{\tiny\mathrm{i.i.d.}}{\sim}\mu_{Z}$,
and let $\hat{Z}_{n}\sim\frac{1}{n}\sum_{i=1}^{n}\delta_{Z_{i}}$
denote the empirical distribution of $\boldsymbol{Z}$. Then, there
exists $n_{0}\in\mathbb{N}_{+}$ and a constant $c>0$, both which
depend on $(d,\sigma_{X},\eta_{X},\psi_{1},\psi_{2},\sigma_{Z})$
such that for all $n\geq n_{0}$
\begin{equation}
\max_{X\sim\mu_{X}\colon\mu_{X}\in{\cal P}^{*}(\eta_{X},\sigma_{X},\psi_{1},\psi_{2})}\left|h(X+Z)-h(X+\hat{Z}_{n})\right|\leq c\frac{\log^{2}n}{n^{1/(d\vee4)}}\label{eq: uniform convergence on general input distributions}
\end{equation}
with probability larger then $1-\frac{1}{n}$.
\end{thm}
We proceed with the following discussion.

\paragraph*{Validity of the estimator}

The estimator $h(X+\hat{Z}_{n})$ in (\ref{eq: uniform convergence on general input distributions})
is indeed well-defined. This is because whenever $X$ has a density
$\mu_{X}$, so does $X+\hat{Z}_{n}\sim\frac{1}{n}\sum_{i=1}^{n}\mu_{X}*\delta_{Z_{i}}$
which is a mixture of $n$ translations of $\mu_{X}$. 

\paragraph*{Conditions on the noise distribution}

The assumption $\sigma_{Z}=\Omega(n^{-(d-2)/(4d)})$ is rather mild,
and made to simplify the resulting bound (see Lemma \ref{lem: Wasserstein distance of noise}). 

\paragraph*{Conditions on the input distribution}

The regularity condition (\ref{eq: regular density}) defines a smoothness
condition on $\mu_{X}\in{\cal P}^{*}$, and implies, for example,
that the tail of the density cannot decay faster than the tail of
a Gaussian density, because regularity implies
\[
\mu_{X}(x)\geq\exp\left[-|\log\mu_{X}(0)|-\frac{\psi_{1}}{2}\|x\|^{2}-\psi_{2}\|x\|\right].
\]
The finite entropy requirement in ${\cal P}^{*}$ can be assured by
controlling $|\log\mu_{X}(0)|$, since if $\mu_{X}\text{ is }(\psi_{1},\psi_{2})$-regular
and $\E\|X\|^{2}\leq d\sigma_{X}^{2}$ then \cite[Sec. II]{polyanskiy2016wasserstein}
\[
|h(\mu_{X})|\leq|\log\mu_{X}(0)|+\psi_{2}\sqrt{d}\sigma_{X}+\frac{\psi_{1}}{2}d\sigma_{X}^{2},
\]
and this upper bound can be chosen as $\eta_{X}$.

\paragraph*{Non-density input classes}

Suppose that $\tilde{{\cal P}}\dfn\{\tilde{\mu}_{X}\colon\E\|X\|^{2}\leq d\sigma_{X}^{2}\}$
is a set of probability measures (which are not necessarily densities
w.r.t. the Lebesgue measure). Now consider the set of smoothed densities
\[
\overline{\mathcal{P}}\dfn\{\tilde{\mu}_{X}*\varphi_{0,\sigma}\colon\tilde{\mu}_{X}\in\tilde{{\cal P}}\}
\]
where $\varphi_{\eta,\sigma}$ is the Gaussian density with mean $\eta\in\mathbb{R}^{d}$
and covariance matrix $d\sigma^{2}\cdot I_{d}$. It was shown recently
in \cite{goldfeld2020convergence} that estimating the entropy of
a smoothed distribution can be made at a fast rate of $e^{O(d)}\cdot O(\frac{1}{\sqrt{n}})$.
The smoothness operation is also useful here, though not for improving
the error rates but rather to allow for general input distributions.
By \cite[Prop. 2]{polyanskiy2016wasserstein}, any $\overline{\mu}_{X}\in\overline{\mathcal{P}}$
is $(\overline{\psi}_{1},\overline{\psi}_{2})\dfn\frac{\log e}{d\sigma^{2}}(3,4\sqrt{d}\sigma_{X})$-regular.
Furthermore, $|h(\overline{\mu}_{X})|$ is bounded because for $\overline{\mu}_{X}=\tilde{\mu}_{X}*\varphi_{0,\sigma}$
and $\tilde{X}\sim\tilde{\mu}_{X}$, $\tilde{X}\ind W\sim\varphi_{0,\sigma}$
\[
h(\overline{\mu}_{X})=h(\tilde{X}+W)\geq h(W)=\frac{d}{2}\log(2\pi e\sigma^{2})\dfn\overline{\eta}_{X}^{(-)},
\]
and since Gaussian vector maximizes entropy under a variance constraint
\[
h(\overline{\mu}_{X})=h(\tilde{X}+W)\leq\frac{d}{2}\log\left(2\pi e(\sigma^{2}+\sigma_{X}^{2})\right)\dfn\overline{\eta}_{X}^{(+)}.
\]
Thus, any $\overline{\mu}_{X}\in{\cal P}^{*}(\overline{\eta}_{X},\overline{\sigma}_{X},\overline{\psi}_{1},\overline{\psi}_{2})$
where $\overline{\eta}_{X}=|\overline{\eta}_{X}^{(-)}|\vee|\overline{\eta}_{X}^{(+)}|$
and $\overline{\sigma}_{X}\dfn d(\sigma^{2}+\sigma_{X}^{2})$, and
the result of Theorem \ref{thm: learning input distributions} holds
for the smoothed class of input densities $\overline{\mathcal{P}}$.
Nonetheless, it seems difficult to make any claims regarding the loss
in mutual information due to the Gaussian smoothing operation (and
this is actually the motivation for the restriction to regular densities
in Theorem \ref{thm: learning input distributions} to begin with).

\paragraph*{Proof idea}

The absolute difference in differential entropy of a pair of \emph{regular}
densities can be controlled by the second-order Wasserstein distance
\cite[Sec. II]{polyanskiy2016wasserstein}. Specifically, \emph{assume
}that  both $X+Z$ and $X+\hat{Z}_{n}$ are $(\psi_{1},\psi_{2})$-regular
densities, then \cite[Prop. 1]{polyanskiy2016wasserstein}
\begin{equation}
\left|h(X+Z)-h(X+\hat{Z}_{n})\right|\leq\left(\frac{\psi_{1}}{2}\sqrt{\E\|X+Z\|^{2}}+\frac{\psi_{1}}{2}\sqrt{\E\|X+\hat{Z}_{n}\|^{2}}+\psi_{2}\right)\cdot W_{2}(\mu_{X+Z},\mu_{X+\hat{Z}_{n}}).\label{eq: entropy difference as a function of Wasserstein}
\end{equation}
As discussed in \cite{polyanskiy2016wasserstein}, this bound can
be considered a \emph{reversed} version of \emph{transportation-information
inequalities} \cite{marton1996bounding,boucheron2013concentration,raginsky2018concentration}
which upper bound the Wasserstein distance by the KL divergence (where
the latter is related to entropy difference). Anyway, it follows from
(\ref{eq: entropy difference as a function of Wasserstein}) that
if $\E\|X+Z\|^{2}$ is bounded and if $\E\|X+\hat{Z}_{n}\|^{2}$ is
bounded with high probability, then the decay rate of the error in
the entropy follows directly from the decay rate of $W_{2}(\mu_{X+Z},\mu_{X+\hat{Z}_{n}})$.
In turn, the dependence of this upper bound on $\mu_{X}$ can be washed
out since Wasserstein distances are non-increasing under convolution
operations, and so $W_{2}(\mu_{X+Z},\mu_{X+\hat{Z}_{n}})\leq W_{2}(\mu_{Z},\mu_{\hat{Z}_{n}})$
(as any coupling of $(Z^{*},\hat{Z}_{n}^{*})$ defines a coupling
$(X+Z^{*},X+\hat{Z}_{n}^{*})$ for $(X+Z,X+\hat{Z}_{n})$). Given
such a bound, the proof is then completed by using known results \cite{dereich2013constructive}
on $\E[W_{2}(\mu_{Z},\mu_{\hat{Z}_{n}})]$ and establishing concentration
to this expectation. The actual proof follows these lines, and uses
a truncation argument on the norm of the noise $\|Z\|$ to establish
such properties under the milder premise of the theorem. 

\paragraph*{Relation to source coding (quantization)}

From its definition, one can anticipate that the Wasserstein distance
of order $p$ would be useful in bounding $p$th moment of empirical
errors. Indeed, a classic result of Pollard \cite{pollard1982quantization}
relates the error of a quantizer to the minimal Wasserstein distance
between the distribution of the source and any other distribution
supported on a finite number of points equal to the cardinality of
the codebook (see also \cite[Sec. 2.2.1]{lee2019learning}). To wit,
let ${\cal C}(m)\dfn\{C\subset\mathbb{R}^{d}\colon|C|=m\}$ be all
possible codebooks of cardinality $m\in\mathbb{N}$, and let ${\cal N}(m)\dfn\{\nu\in{\cal P}(\mathbb{R}^{d})\colon|\supp\{\nu\}|=m\}$.
Then, for $p\geq1$ it holds that
\begin{equation}
\inf_{C\in{\cal C}(m)}\E\int\min_{x\in C}\|x-Z\|^{p}\cdot\d\mu_{Z}=\inf_{\nu\in{\cal N}(m)}W_{p}^{p}(\nu,\mu_{Z}).\label{eq: quantization error equals Wasserstein distance}
\end{equation}
In \cite{dereich2013constructive}, the convergence rates as a function
of $m$ of the Wasserstein distance to $\mu_{Z}$ were studied for
the empirical measure $\mu_{\hat{Z}_{m}}$ rather than for the optimal
density, i.e., $\E[W_{p}^{p}(\mu_{\hat{Z}_{m}},\mu_{Z})]$ instead
of the r.h.s. of (\ref{eq: quantization error equals Wasserstein distance}).
So, therein the error of the density estimator is controlled by a
Wasserstein distance. Our proof here, exhibits another use of the
empirical Wasserstein distance, as a \emph{uniform} bound on the error
of an entropy estimator. Thus, despite what might have been apparent
from its definition, the role of Wasserstein distance goes beyond
bounds on the $p$th norm.

\subsection{A Finite Support Class of Input Distributions}

The second class of input distributions is similar to (\ref{eq: input distribution with codebook and weights}),
and seeks to only optimize weights. That is, a codebook $C=\{x_{j}\}_{j\in[m]}\subset\mathbb{R}^{d}$
is \emph{chosen} in advance,\emph{ }and the class of input distributions
is 
\[
{\cal P}_{C}^{**}\dfn\left\{ \mu_{X}=\sum_{j=1}^{m}a_{j}\delta_{x_{j}}\colon\boldsymbol{a}\in\mathbb{A}^{m-1}\right\} .
\]
Thus, an input distribution from ${\cal P}_{C}^{**}$ is equivalent
to a probability vector $\boldsymbol{a}\in\mathbb{A}^{m-1}$. The
problem of maximizing $h(X+Z)$ over $\mu_{X}\in{\cal P}_{C}^{**}$
is a concave optimization problem over the convex set $\mathbb{A}^{m-1}$
which, as said, can also be solved efficiently using the Blahut\textendash Arimoto
algorithm \cite{blahut1972computation,arimoto1972algorithm}. As we
show next it can also be approximated when the noise distribution
is unknown. To this end, consider a kernel $\kappa\colon\mathbb{R}^{d}\to\mathbb{R}$,
and denote by $\kappa_{\theta,z}$ its shift by $z\in\mathbb{R}^{d}$
followed by scaling of $\theta\in\mathbb{R}_{+}$, to wit, $\kappa_{\theta,z}(x)\dfn\kappa(\frac{x-z}{\theta})$.
The learning procedure is obtained by maximizing $h(X+\tilde{Z}_{n})$
where $\tilde{Z}_{n}=\hat{Z}_{n}+V$ and $V\sim\kappa_{\theta,0}$.
In other words, $\mu_{\tilde{Z}_{n}}$ is a KDE  of $\mu_{Z}$ of
bandwidth $\theta$, i.e. 
\begin{equation}
\mu_{\tilde{Z}_{n}}(z)=\frac{1}{n\theta^{d}}\sum_{i=1}^{n}\kappa\left(\frac{Z_{i}-z}{\theta}\right).\label{eq: KDE}
\end{equation}
To state the result, we denote the second-order differential R\'{e}nyi
entropy by $h_{2}(f)\dfn-\log\int f^{2}$. The following theorem is
followed by a discussion on its implications:
\begin{thm}
\label{thm: uniform convergence for learning weights}Let $C=\{x_{j}\}_{j\in[m]}\subset\mathbb{R}^{d}$
be given, and let $X_{\boldsymbol{a}}\sim\sum_{j=1}^{m}a_{j}\delta_{x_{j}}$
for $\boldsymbol{a}\in\mathbb{A}^{m-1}$. Let $\boldsymbol{Z}=(Z_{1},\ldots,Z_{n})\stackrel{\tiny\mathrm{i.i.d.}}{\sim}\mu_{Z}$,
where $|h_{2}(\mu_{Z})|\leq A_{Z,2}$. Let $\hat{Z}_{n}\sim\frac{1}{n}\sum_{i=1}^{n}\delta_{Z_{n}}$
and assume that $\mu_{Z}$ is estimated by $\tilde{Z}_{n}\sim\frac{1}{n\theta^{d}}\sum_{i=1}^{n}\kappa_{\theta,Z_{i}}$
where the kernel density satisfies $|h_{2}(\kappa)|\leq A_{\kappa,2}$.
Then, there exists an absolute constant $c>0$ and $n_{0}\in\mathbb{N}$
which depend on $(A_{\kappa,2},A_{Z,2})$ such that for any given
$\delta>0$, 
\begin{equation}
\max_{\boldsymbol{a}\in\mathbb{A}^{m-1}}\left|h(X_{\boldsymbol{a}}+Z)-h(X_{\boldsymbol{a}}+\tilde{Z}_{n})\right|\leq\max_{\boldsymbol{a}\in\mathbb{A}^{m-1}}\E\left[\left|h(X_{\boldsymbol{a}}+Z)-h(X_{\boldsymbol{a}}+\tilde{Z}_{n})\right|\right]+\Delta\label{eq: learning weights theorem}
\end{equation}
with probability larger than $1-\delta$, where 
\[
\Delta\leq c\left(d\log\tfrac{1}{h}+\log n+\log m\right)\sqrt{\frac{\log\frac{1}{\delta}+m\log m+m\log n}{n}}.
\]
\end{thm}

\paragraph*{Convergence rates and choice of bandwidth}

As evident, the bound on the convergence rate in the r.h.s. of (\ref{eq: learning weights theorem})
depends on two terms. The convergence rate of the first term hinges
on the ability to properly estimate $h(X_{\boldsymbol{a}}+Z)$ for
any $\boldsymbol{a}\in\mathbb{A}^{m-1}$, and more importantly, for
any possible noise distribution. This is typically assured by smoothness
assumptions on the possible noise densities, and requires choosing
the bandwidth to be $\theta=n^{-r}$ for some $r>0$ that depends
on the dimension $d$ and smoothness defining parameters. With this
choice, the exponent $r$ only multiplicatively affects the redundancy
term $\Delta$ via $d\log\tfrac{1}{\theta}+\log n=(dr+1)\log n$.
For concreteness, we may consider the family of noise densities defined
by Lipschitz balls, as proposed and analyzed in \cite{han2020optimal}.
Given a smoothness parameter $s\geq0$ and $r=\lceil s\rceil$, a
norm parameter $p\in[2,\infty)$, and the dimension $d$, a Lipschitz
norm is defined as follows:
\[
\|\mu\|_{\text{Lip}}\dfn\|\mu\|_{L_{p}}+\sup_{t>0}t^{-s}\omega_{r}(\mu,t)_{p}
\]
\[
\omega_{r}(\mu,t)_{p}\dfn\sup_{e\in\mathbb{R}^{d}\colon\|e\|\leq1}\|\Delta_{te}^{r}\mu\|_{L_{p}}
\]
\[
\Delta_{\theta}^{r}\mu(z)=\sum_{k=0}^{r}(-1)^{r-k}{r \choose k}\mu\left(z+(k-\tfrac{r}{2})\theta\right),\quad z\in\mathbb{R}^{d}
\]
where $\|\mu\|_{L_{p}}=\E^{1/p}[\|X\|^{p}]$ with $X\sim\mu$ is the
$L_{p}$ norm. The definition $\|\mu\|_{\text{Lip}}$ indeed induces
a norm, and so, it is specifically convex. This can be seen from the
observation is that $\Delta_{\theta}^{r}$ is a linear operator, and
that since $L_{p}$ norm are convex functions, it holds for any mixture
$\mu=q\mu_{1}+(1-q)\mu_{2}$ that 
\[
\omega_{r}(\mu,t)_{p}\leq\sup_{e\in\mathbb{R}^{d}\colon\|e\|\leq1}q\cdot\|\Delta_{te}^{r}\mu_{1}\|_{L_{p}}+(1-q)\|\Delta_{te}^{r}\mu_{1}\|_{L_{p}}\leq q\cdot\omega_{r}(\mu_{1},t)_{p}+(1-q)\omega_{r}(\mu_{2},t)_{p}
\]
and consequently 
\[
\|\mu\|_{\text{Lip}}\leq q\|\mu_{1}\|_{\text{Lip}}+(1-q)\|\mu_{2}\|_{\text{Lip}}\leq\|\mu_{1}\|_{\text{Lip}}\vee\|\mu_{2}\|_{\text{Lip}}.
\]
Generalizing this, we obtain that the density of $X_{\boldsymbol{a}}+Z$
is only smoother than that of $Z$ in the sense that $\|\mu_{X_{\boldsymbol{a}}+Z}\|_{\text{Lip}}\leq\|\mu_{Z}\|_{\text{Lip}}$.
Now, following \cite{han2020optimal}, consider the set of densities
\[
{\cal B}_{s,p,d}(L)\dfn\left\{ \mu\colon\|\mu\|_{\text{Lip}}\leq L,\;\supp(\mu_{Z})\subseteq[0,1]^{d}\right\} .
\]
Given this definition, it was shown in \cite[Thm. 3]{han2020optimal}
that whenever the kernel $\kappa$ satisfies several regularity assumptions
(non-negativity, unit total mass, zero mean, finite second moment,
and compact support ; see \cite[Assumption 1]{han2020optimal}) an
upper bound on the entropy estimator can be obtained as follows. Assuming
that $s\in(0,2]$, and $p\geq2$, there exists a constant $C>0$ independent
of $n,L$ such that if $L\leq n^{s/d}$ and $\theta\asymp(Ln)^{-1/(s+d)}$
then
\begin{equation}
\sup_{\mu\in{\cal B}_{s,p,d}(L)}\E\left[\left|h(\mu)-h(\mu_{\tilde{Z}_{n}})\right|\right]\leq C\left(n^{-s/(s+d)}L^{d/(s+d)}+n^{-1/2}\cdot\log L\right)\label{eq: entropy estimation for Lipschitz balls}
\end{equation}
where $\mu_{\tilde{Z}_{n}}$ is the KDE as in (\ref{eq: KDE}).\footnote{In fact, in \cite[Thm. 3]{han2020optimal}, the expectation in the
left-hand side of (\ref{eq: entropy estimation for Lipschitz balls})
is replaced by $\E^{1/2}[(h(\mu)-h(\tilde{\mu}_{n}))^{2}]$ and the
corresponding statement is stronger. Moreover \cite[Thm. 3]{han2020optimal}
shows that the rate on the r.h.s of (\ref{eq: entropy estimation for Lipschitz balls})
is sub-optimal, and by using a more sophisticated estimator than (\ref{eq: KDE}),
one can improve the $n^{-s/(s+d)}$ term in (\ref{eq: entropy estimation for Lipschitz balls})
to $(n\log n)^{-s/(s+d)}$, and also that this rate is minimax optimal.
However, uniform convergence as in Theorem \ref{thm: uniform convergence for learning weights}
is more difficult to obtain for that estimator.} As discussed above, if $\mu_{Z}\in{\cal B}_{s,p,d}(L)$ then $\mu_{X+Z}\in{\cal B}_{s,p,d}(\tilde{L})$
holds too for some $\tilde{L}$ (with possibly larger support, which
can be re-normalized at the expense of a multiplicative factor in
$L$ that results $\tilde{L}$). Thus, the convergence rates of the
first term on the r.h.s. of (\ref{eq: learning weights theorem})
can be bounded as in (\ref{eq: entropy estimation for Lipschitz balls}).
Consequently, for a fixed $L$, the rate in (\ref{eq: learning weights theorem})
is determined by the first term on the r.h.s., and equals to $O(n^{-s/(s+d)})$.
This is valid for any $d\geq1$ and since $s\in(0,2]$ may amount
to better rates than $\tilde{O}(n^{-1/d})$ obtained in Theorem \ref{thm: learning input distributions}
whenever $s>\frac{d}{d-1}$ (assuming $d\geq5$ so that Theorem \ref{thm: learning input distributions}
is valid). Nonetheless, the rate $O(n^{-s/(s+d)})$ is fastest when
$s=2$ and then $O(n^{-2/(2+d)})$ which, similarly to the $O(n^{-1/d})$
rate of Theorem \ref{thm: learning input distributions} also requires
a number of samples which is exponential in the dimension.

\paragraph*{Proof outline}

The proof is based on a bound on $|h(X_{\boldsymbol{a}}+Z)-h(X_{\boldsymbol{b}}+Z)|$
for $\boldsymbol{a},\boldsymbol{b}\in\mathbb{A}^{m-1}$ in terms of
the total variation and chi-square divergence between $\boldsymbol{a}$
and $\boldsymbol{b}$ (Lemma \ref{lem: Entropy difference wrt TV and Chi2}).
This bound is used to show a bounded-difference inequality for $|h(X_{\boldsymbol{a}}+Z)-h(X_{\boldsymbol{a}}+\tilde{Z}_{n})|$
when one of the samples in $\boldsymbol{Z}$ is changed and consequently
establishes the concentration of this quantity to its mean for a given
$\boldsymbol{a}\in\mathbb{A}^{m-1}$ via McDiarmid's inequality. Then,
uniform concentration over $\mathbb{A}^{m-1}$ is established by a
covering argument of the simplex w.r.t. the total variation and chi-square
divergence. 

\section{Experiments}

\subsection{Alternating Optimization for Surrogate Error Probability \label{subsec:Experiments Alternating-Minimization- }}

Let us denote the projection operator $\Psi_{r}^{d}\colon\mathbb{R}^{d}\mapsto{\cal B}^{d}(r)$
by $\Psi_{r}^{d}(x)=\frac{(\|x\|\wedge r)}{\|x\|}\cdot x$. For brevity,
the subscript $r$ will be omitted when $r=1$. This operator will
be used to truncate Gaussian noise to have bounded norm so that Theorem
\ref{thm:Generalization NN surrogate} will be valid, but with the
expectation that it has little affect on the results. For purpose
of illustration we begin with a simple example for $d=2$, and then
experiment more extensively with in a more complicated setting.\footnote{Matlab code which implements the learning algorithms proposed in the
paper and use in the experiments is available at \url{http://drive.google.com/open?id=1YSX0cnac7zpviQxjCx_cdjT_zW6zVx3e},
and documented at the end of this manuscript. }

\subsubsection{Two-dimensional Gaussian noise\label{subsec:Two-dimensional-Gaussian-noise}}

\begin{figure}
\centering{}\includegraphics[scale=0.3]{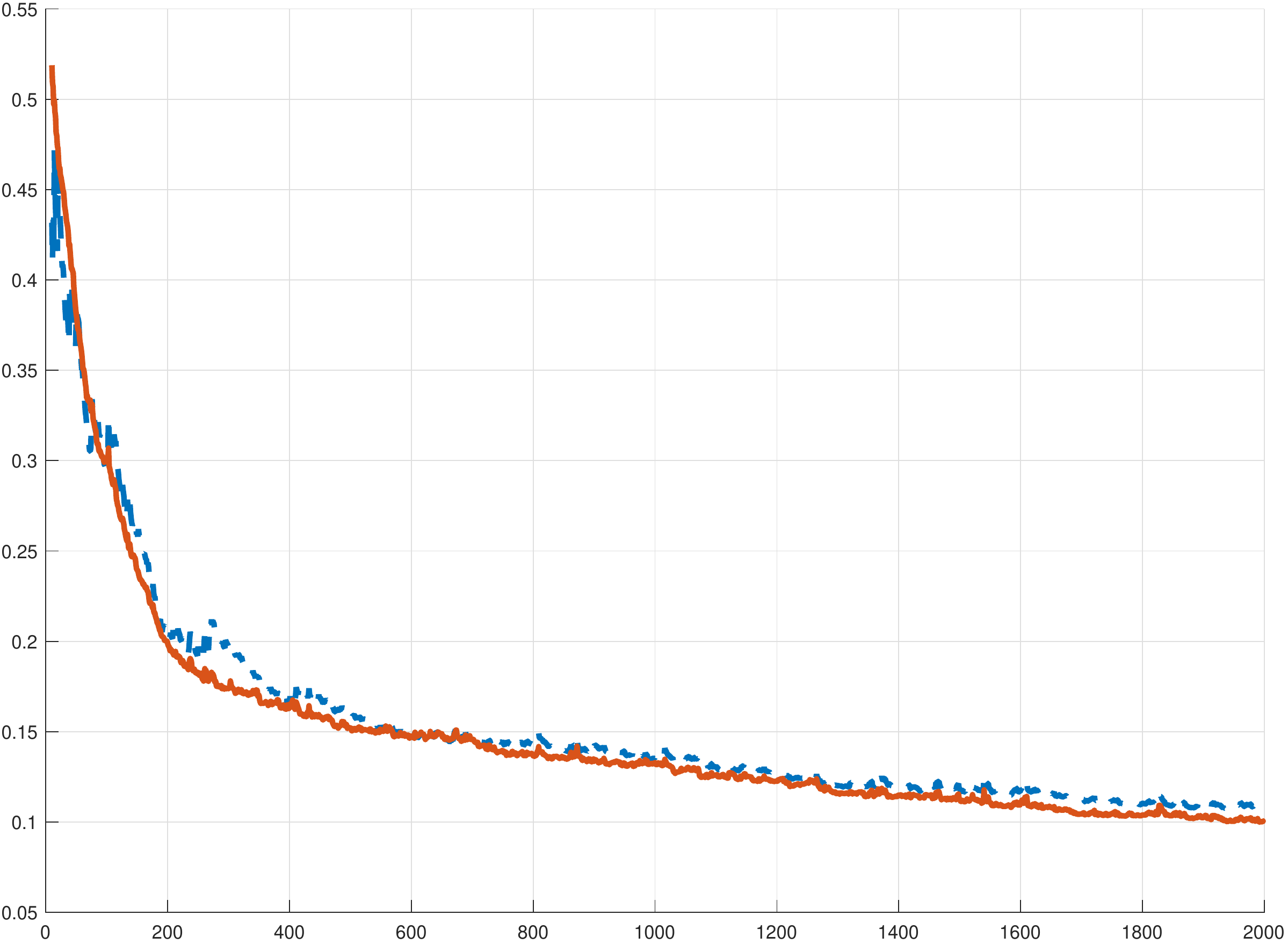}\includegraphics[scale=0.3]{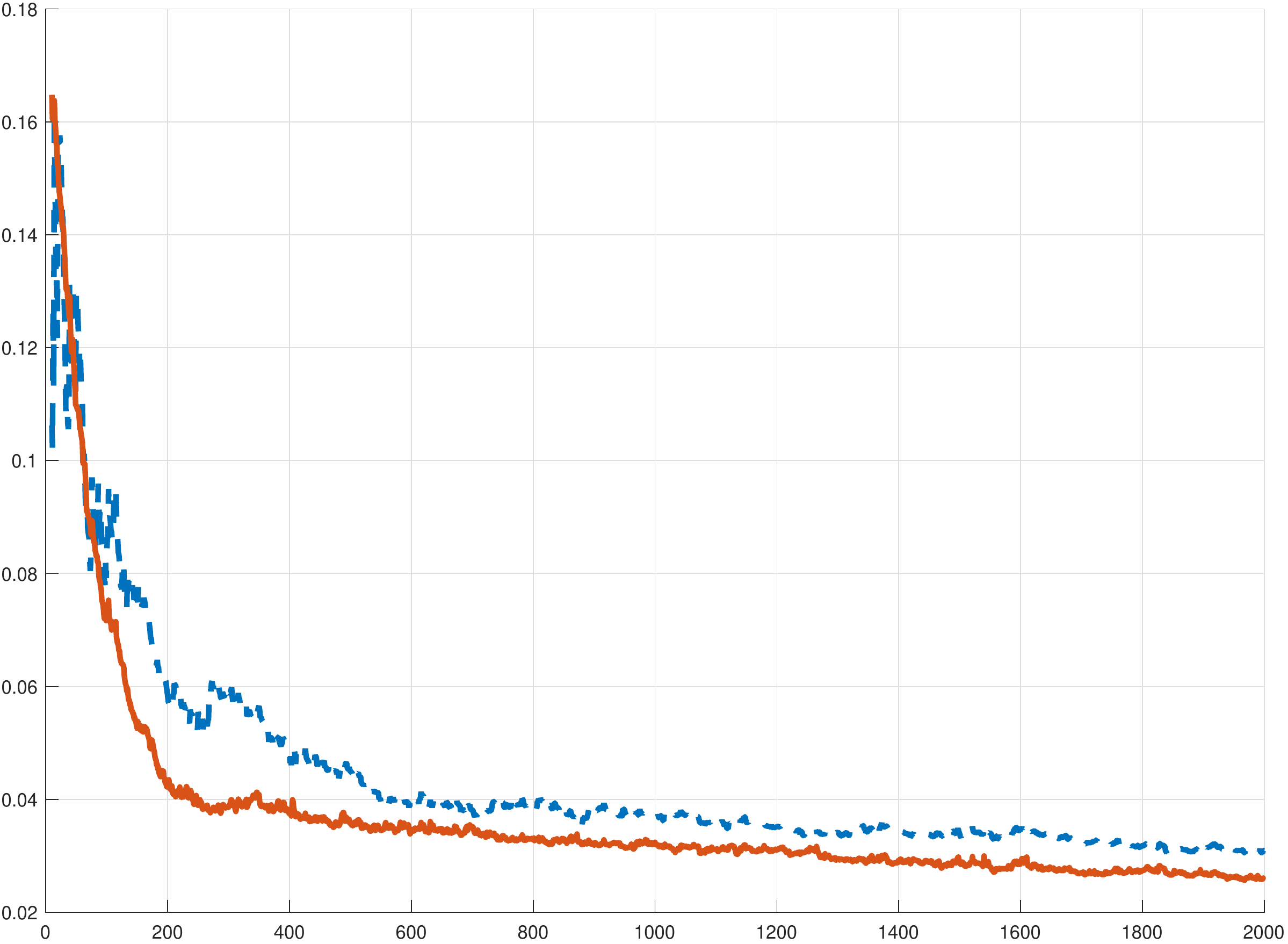}\caption{Left: surrogate error probability $\peb_{\boldsymbol{Z}^{(i)}}(C^{(i)},S^{(i)})$
\textendash{} blue dashed, $\peb_{\tilde{\boldsymbol{Z}}}(C^{(i)},S^{(i)})$
\textendash{} red solid. Right: error probability $\pe_{\boldsymbol{Z}^{(i)}}(C^{(i)},S^{(i)})$
\textendash{} blue dashed, $\pe_{\tilde{\boldsymbol{Z}}}(C^{(i)},S^{(i)})$
\textendash{} red solid \label{fig: two-dimensional convergence iteration}}
\end{figure}

We assume the noise is (projected) Gaussian $Z\sim\Psi^{d}(N(0,K))$,
with covariance matrix $K\in\mathbb{S}_{+}^{d}$ unknown to the learner.
It is assumed that $K$ is chosen such that the projection operation
only affects $Z$ with low probability and thus rather negligible.
We further assume that the codewords of the initial codebook $C^{(0)}=(X_{1}^{(0)},\ldots,X_{m}^{(0)})$
are randomly drawn $X_{j}^{(0)}\sim\Psi_{r_{x}}^{d}\left(N(0,\frac{r_{x}^{2}}{\phi_{x}d}\cdot I_{d})\right)$
for $j\in[m]$ mutually independently, where $r_{x}>0$ is the amplitude
constraint, and $\phi_{x}$ is similarly chosen such that projection
occur with low probability and does not significantly affect the random
codebook. With this choice, the mutual information between $X\in\mathbb{R}^{d}$
and $Y=X+Z\in\mathbb{R}^{d}$ is approximated by
\begin{equation}
I(r_{x})\dfn I(X;Y)=\frac{1}{2}\log\frac{\det\left(\frac{r_{x}^{2}}{\phi_{x}d}\cdot I_{d}+K\right)}{\det(K)}.\label{eq: capacity of Gaussian vector channel with white input}
\end{equation}
The expression in (\ref{eq: capacity of Gaussian vector channel with white input})
is only approximation since the input distribution is not exactly
Gaussian due to the projection operation, but it is only used to roughly
gauge the required power for various codebook sizes. For a codebook
of size $m$, the minimal required power to obtain negligible error
probability in case coding across multiple realizations are allowed
is (approximately) $r_{\text{min}}\dfn\min\{r>0\colon I(r)>\log m\}$.
Since we are considering only a single use of the $d$-dimensional
channel, we choose the input power to be $r_{x}=\sqrt{\Gamma}\cdot r_{\text{min}}$
where $\Gamma>0$ is the so-called \emph{gap-to-capacity }\cite[Ch. 4]{forney2005principles}.
The inverse covariance matrix is initialized as the inverse of the
empirical covariance matrix over the first $d$ samples, to wit, $S^{(0)}=\cdots=S^{(d)}=(\sum_{i=1}^{d}z_{i}z_{i}^{T})^{-1}$.
The parameters used in the experiment are detailed in Table \ref{tab: SGD Experiments-parameters}
in Appendix \ref{sec:Experiments-Details}. Fig. \ref{fig: two-dimensional scatter plot}
displays the output codebook $C^{(n)}$ of a run of Algorithm \ref{alg:Stochasic-gradient-descent}
on a realization of $n$ noise samples, as well as the Voronoi regions
determined by this codebook, and the minimal Mahalanobis distance
rule w.r.t. the output inverse covariance matrix $S^{(n)}$. In addition,
the figure also displays the noise samples used for training superimposed
on each of the codewords, i.e., $X_{j}+Z_{i}$ for all $j\in[m]$
and $i\in[n]$ (with slightly different color tone for any $j\in[m]$).
Fig. \ref{fig: two-dimensional convergence iteration} displays the
evolution of the empirical error $\peb_{\boldsymbol{Z}^{(i)}}(C^{(i)},S^{(i)})$
where $\boldsymbol{Z}^{(i)}=(Z_{1},\ldots,Z_{i})$ are the noise samples
used up to iteration $i\in[n]$, and the evolution of $\peb_{\tilde{\boldsymbol{Z}}}(C^{(i)},S^{(i)})$
where $\tilde{\boldsymbol{Z}}=(\tilde{Z}_{1},\ldots,\tilde{Z}_{\tilde{n}})$
are validation samples. The samples $\tilde{\boldsymbol{Z}}$ are
drawn independently of the training samples $\boldsymbol{Z}$, and
for simplicity only drawn once for all iterations. The validation
average error probability serves as a proxy to the statistical average
loss $\peb_{\tilde{\boldsymbol{Z}}}(C^{(i)},S^{(i)})\approx\peb_{\mu}(C^{(i)},S^{(i)})$.
The same evolution is shown for the standard error probability loss
function. The results displayed in those figures are for a single
``successful'', yet representative, run of the algorithm, in which
the error probability on the validation samples has decreased by a
factor of $4$ after about $n=200$ iterations and samples.

\begin{figure}
\centering{}\includegraphics[scale=0.5]{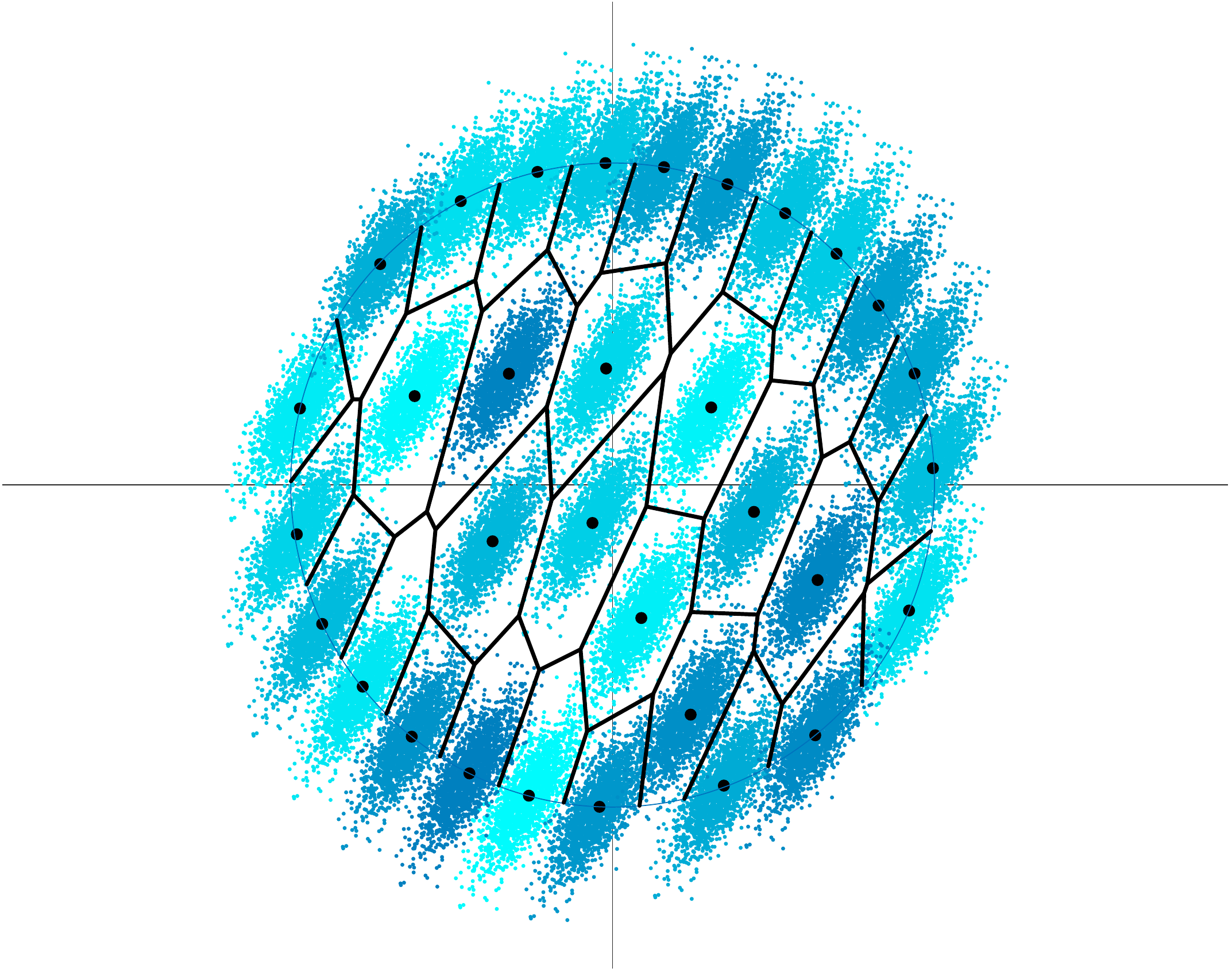}\caption{A scatter plot of the final codebook $C^{(n)}$ (black filled circles),
Voronoi regions w.r.t. the Mahalanobis distance $\|\cdot\|_{S^{(n)}}$
, and noise training samples superimposed on the final codebook. \label{fig: two-dimensional scatter plot}}
\end{figure}

\subsubsection{Gaussian Mixture noise\label{subsec:Gaussian-Mixture-noise}}

We consider next the scenario in which the noise $Z$ is comprised
of interference from other digital communication transmitters, in
addition to standard Gaussian noise. A common strategy is to avoid
special processing of this interference and treat such interference
as noise. It is then also common to assume that the noise distribution
is Gaussian (perhaps with an unknown covariance matrix), which is
justified via the central-limit theorem (in case of large number of
interferers), as well as the extremal properties of the Gaussian distribution
(such as minimizing capacity under covariance constraint, e.g. \cite{ihara1978capacity,pinsker1956calculation},
\cite[Lemma II.2]{diggavi2001worst}). The learning-based framework
proposed here provides a compromise between a worst-case assumption
of Gaussian distribution, and a detailed parametric characterization
of the noise. 

To experiment with this approach, we consider $k$ interfering transmitters,
where the $l$th transmitter sends a binary signal $R_{l}\sim\text{Uniform}\{-1,1\}$,
i.i.d. for $l\in[s]$, such that the noiseless interference from transmitted
$l$ is received as $R_{l}\cdot v_{l}$ where $v_{l}\in\mathbb{R}^{d}$.
The noise distribution thus assumed to satisfy
\[
Z\eqd\Psi^{d}\left[\sqrt{\alpha}\left(\sum_{l=1}^{s}R_{l}\cdot v_{l}+W\right)\right]
\]
where $W\sim N(0,K_{W})\ind(R_{1},\ldots,R_{s})$ with $K_{W}\in\mathbb{S}_{+}^{d}$
being the covariance matrix of the pure Gaussian noise. The covariance
matrix of the noise before projection is then given by $K=\alpha\sum v_{l}\cdot v_{l}^{T}+\alpha K_{W}$,
and $\alpha$ is determined so that the maximal eigenvector of $K$
is at most $\frac{1}{\chi_{Z}}$ for some parameter $\chi_{Z}>0$
(so that the projection does not significantly affect the noise distribution).

\begin{figure}
\centering{}\includegraphics[scale=0.2]{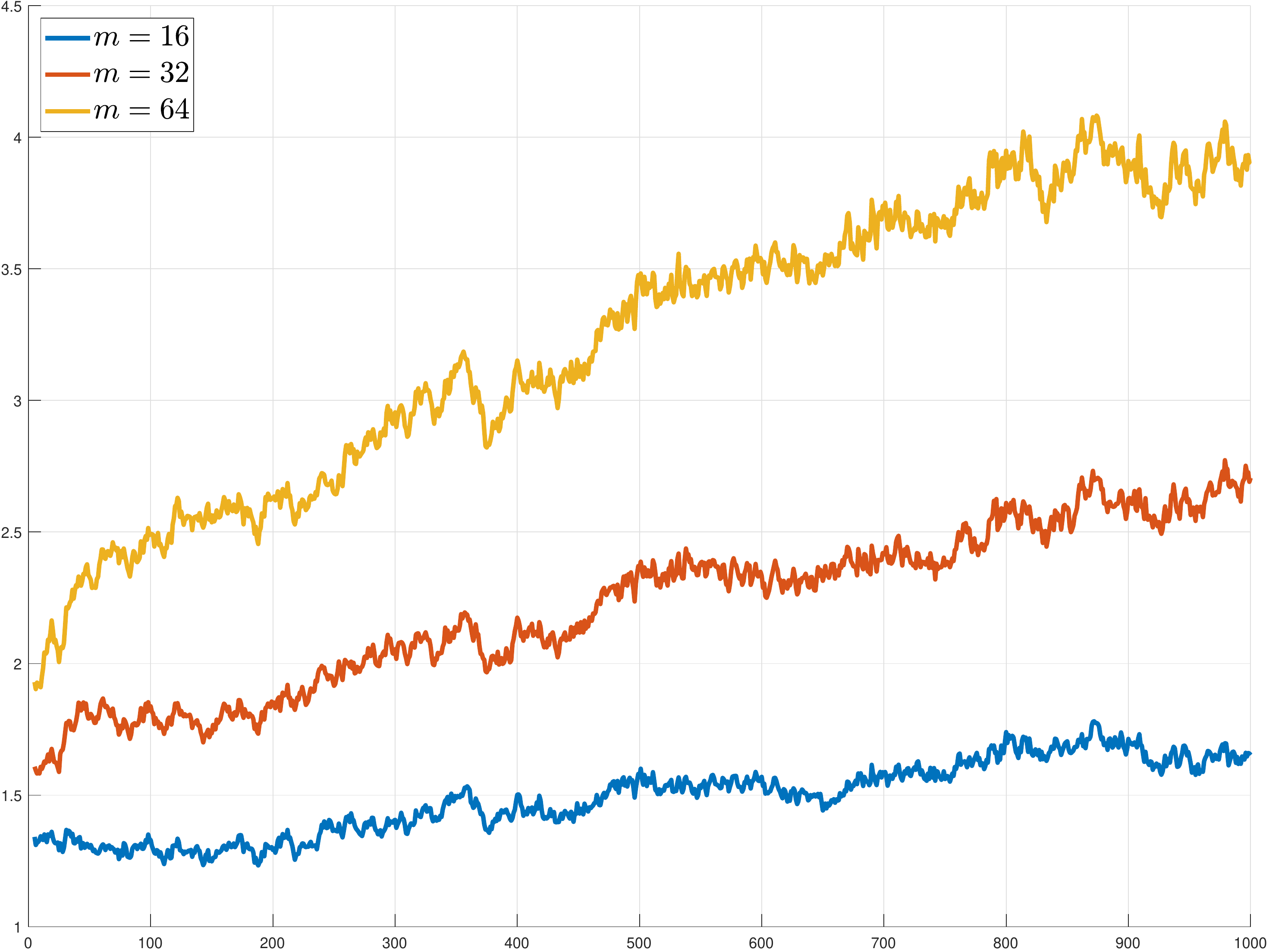}\includegraphics[scale=0.2]{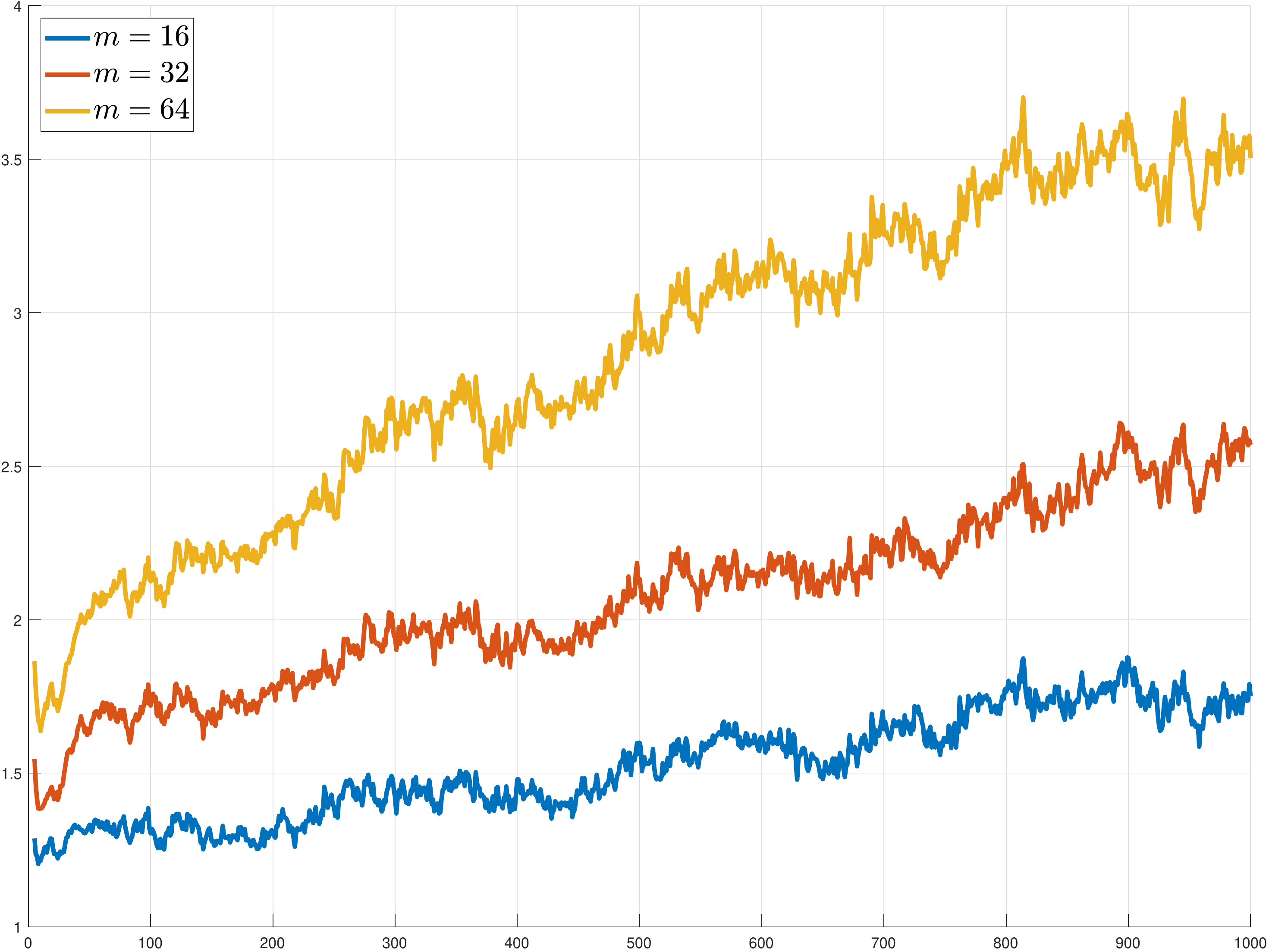}\caption{Surrogate $\overline{g}_{m}^{(i)}/\overline{g}_{8}^{(i)}$ (left)
and standard $g_{m}^{(i)}/g_{8}^{(i)}$ (right) generalization error
ratio, averaged over $10^{3}$ experiments. \label{fig: GM generalization}}
\end{figure}

We again assume initial random (projected) Gaussian codebook, where
$I(r)$ gauges the gap-to-capacity as before.\footnote{We use the same metric even though the noise is not Gaussian because
Gaussian noise leads to the minimal capacity under a noise covariance
constraint.} The experiment parameters are detailed in Table \ref{tab: SGD Experiments-parameters}
in Appendix \ref{sec:Experiments-Details}. Fig. \ref{fig: GM convergence iteration}
shows graphs that track the evolution of the loss over iterations,
in a similar fashion to Fig. \ref{fig: two-dimensional convergence iteration}.
However, in Fig. \ref{fig: GM convergence iteration} the losses are
averaged over multiple noise distributions and multiple runs in the
following way. First, a noise distribution is chosen where $\{v_{l}\}$
are random and chosen as $V_{l}\sim N(0,I_{d})$, and $K_{W}$ is
randomly chosen from the Wishart distribution, i.e., $K_{W}=QQ^{T}$
where the the $d^{2}$ entries of $Q\in\mathbb{S}^{d}$ are independent
$Q_{i_{1}i_{2}}\sim N(0,1)$. Then, for each realization of noise
distribution, multiple runs were performed, where in each run a noise
realization (both for training and validation) and a random codebook
is drawn, independent of all other runs. Fig. \ref{fig: GM convergence iteration}
displays an averaging and a $0.8$-quantile of $10^{3}$ runs where
the noise distribution was re-drawn every $10$ runs. It is apparent
that both type of loss functions follow the same trend and the reduction
in the surrogate error probability leads to a reduction in the ordinary
error probability. Convergence is achieved by a few hundred samples.
It can also be observed that the $0.8$-quantile is typically less
than the average. This indicates that the events where no convergence
is achieved lead to high error, and so can easily be detected, after
$100-200$ samples, which marginally increases the convergence time.
We also remark that the step-sizes are chosen to be the same for all
iterations, and were not thoroughly optimized. In general, the convergence
time is longer as the constellation size is bigger. However, the required
number of samples (or iterations) required for convergence does not
seem to be significantly vary from the surrogate and standard error
probability loss functions. To further inspect the dependency on the
codebook size, we denote the generalization error for the surrogate
and the standard loss functions, respectively, as explicit functions
of the codebook size $m$: 
\[
\overline{g}_{m}^{(i)}\dfn\left|\peb_{\tilde{\boldsymbol{Z}}}(C^{(i)},S^{(i)}\mid m)-\peb_{\boldsymbol{Z}}(C^{(i)},S^{(i)}\mid m)\right|
\]
and 
\[
g_{m}^{(i)}\dfn\left|\pe_{\tilde{\boldsymbol{Z}}}(C^{(i)},S^{(i)}\mid m)-\pe_{\boldsymbol{Z}}(C^{(i)},S^{(i)}\mid m)\right|.
\]
The dependency on the codebook size can then be found by the ratios
$g_{m}^{(i)}/g_{m_{0}}^{(i)}$ and $\overline{g}_{m}^{(i)}/\overline{g}_{m_{0}}^{(i)}$
for various values of $m$. In Fig. \ref{fig: GM generalization},
these ratios are computed with $m_{0}=8$, and roughly shows that
the dependency of the generalization error on the codebook size is
not different between the surrogate and ordinary error probability.
In fact, it seems that that the dependency is close to square-root
in $m$, as was theoretically obtained for the surrogate error probability
(Theorem \ref{thm:Generalization NN surrogate}), rather than to the
linear dependence theoretically obtained for the standard error probability
(Theorem \ref{thm:Generalization NN}). This can be attributed to
the fact that Theorem \ref{thm:Generalization NN} states a generalization
bound which assumes an arbitrary noise distribution, and the worst
case distribution can be significantly worse than the average-case
distribution, or worse than the worst-case distribution within a family
of \emph{structured} distributions such as the Gaussian mixture. 

\begin{figure}
\centering{}\includegraphics[scale=0.2]{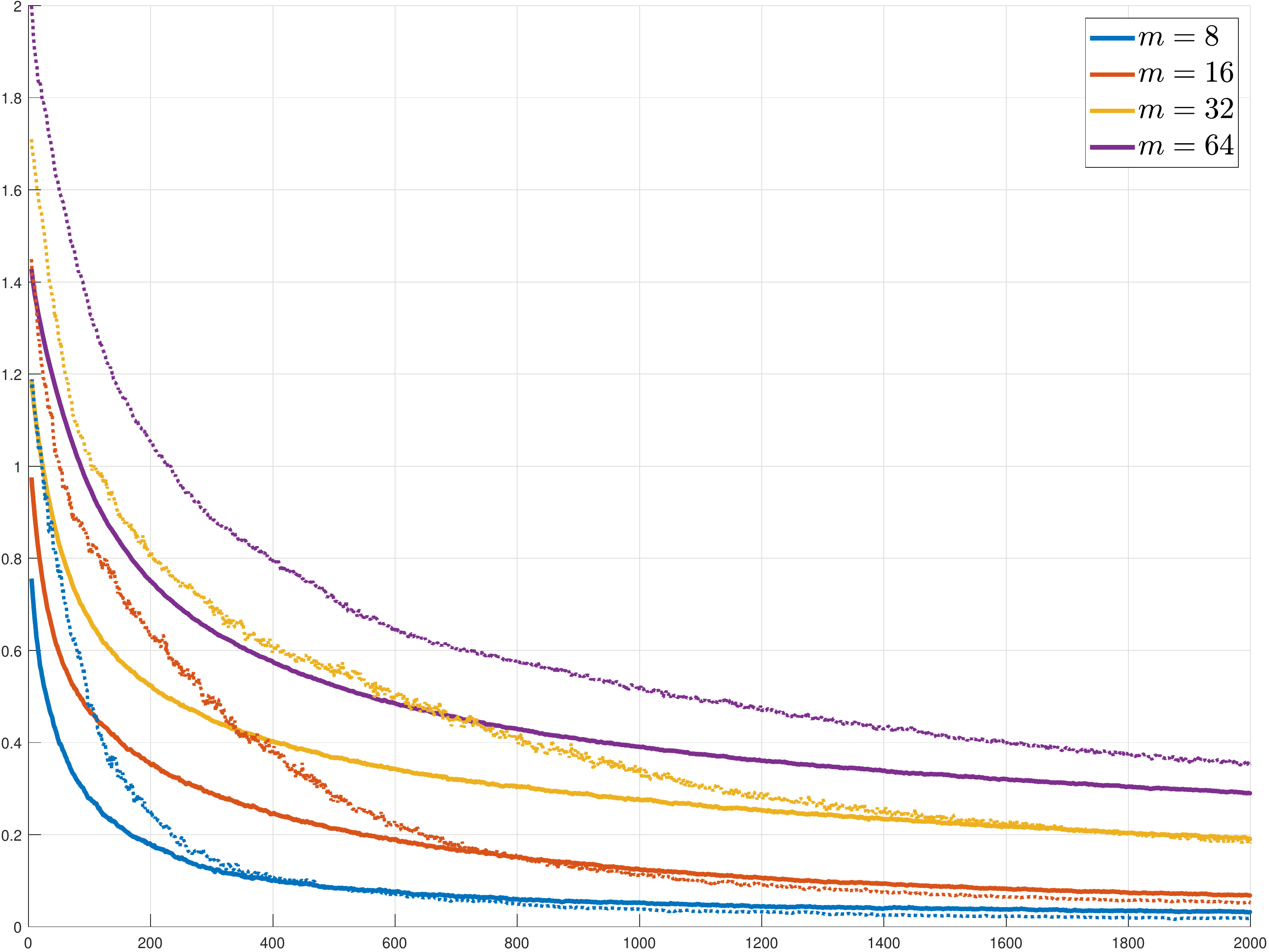}\includegraphics[scale=0.2]{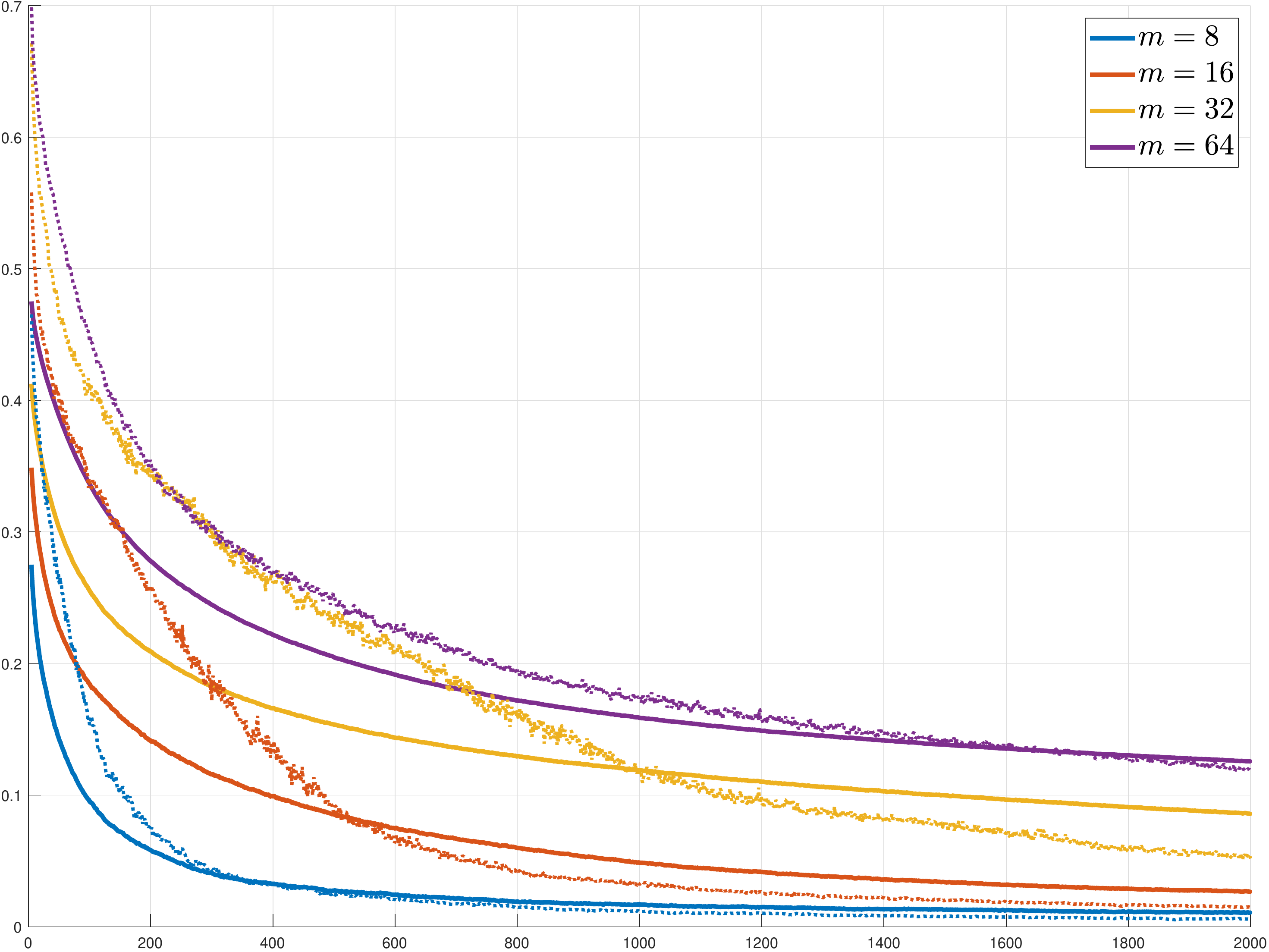}\caption{Surrogate $\peb_{\tilde{\boldsymbol{Z}}}(C^{(i)},S^{(i)})$ (left)
and standard $\pe_{\tilde{\boldsymbol{Z}}}(C^{(i)},S^{(i)})$ (right)
error probability. Averaged over $10^{3}$ experiments \textendash{}
solid, $0.8$-quantile \textendash{} dotted. \label{fig: GM convergence iteration}}
\end{figure}

\subsection{Gibbs Algorithm \label{subsec:Gaussian-Mixture-noise-Gibbs}}

\begin{figure}
\centering{}\includegraphics[scale=0.3]{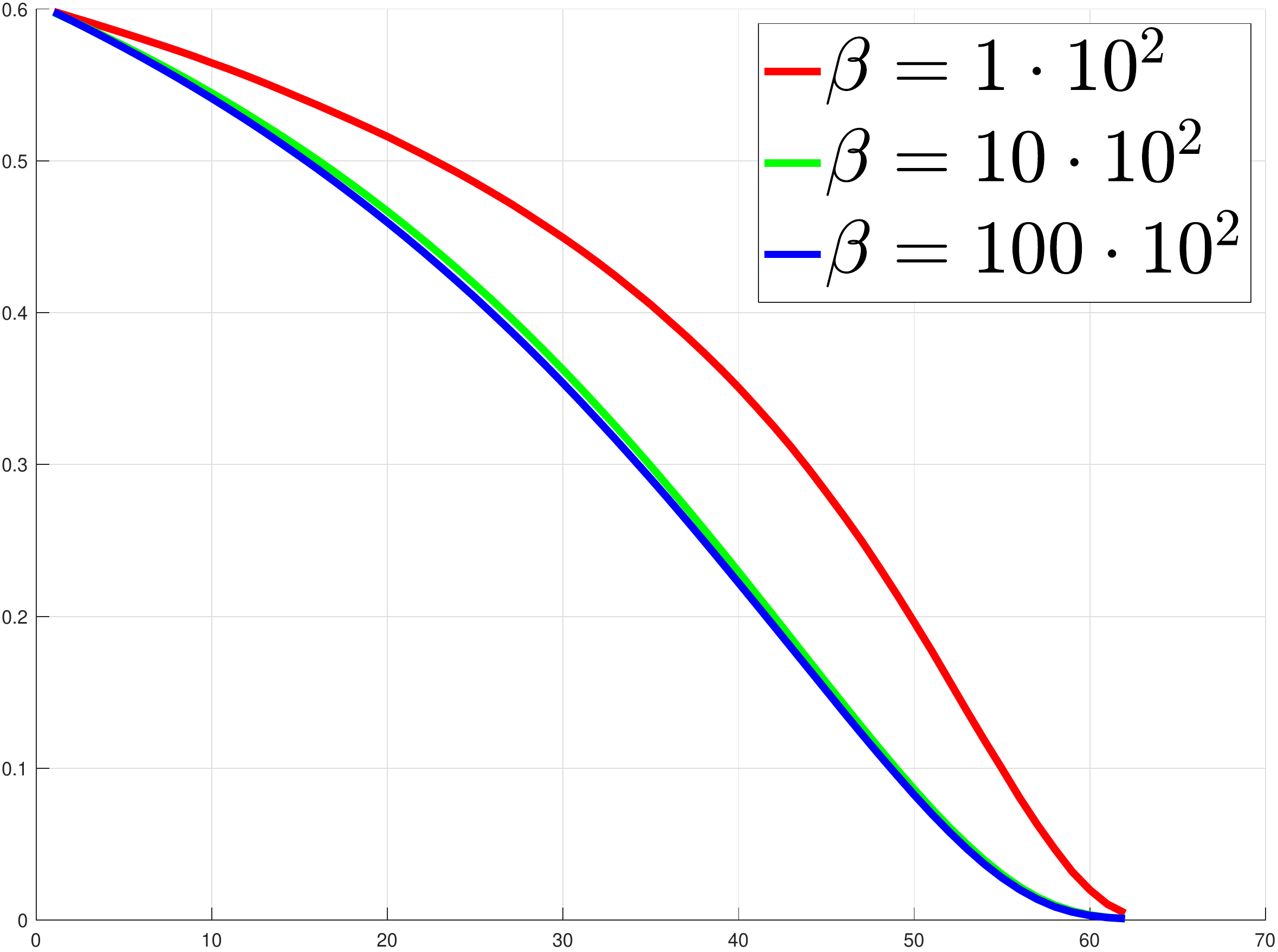}\includegraphics[scale=0.3]{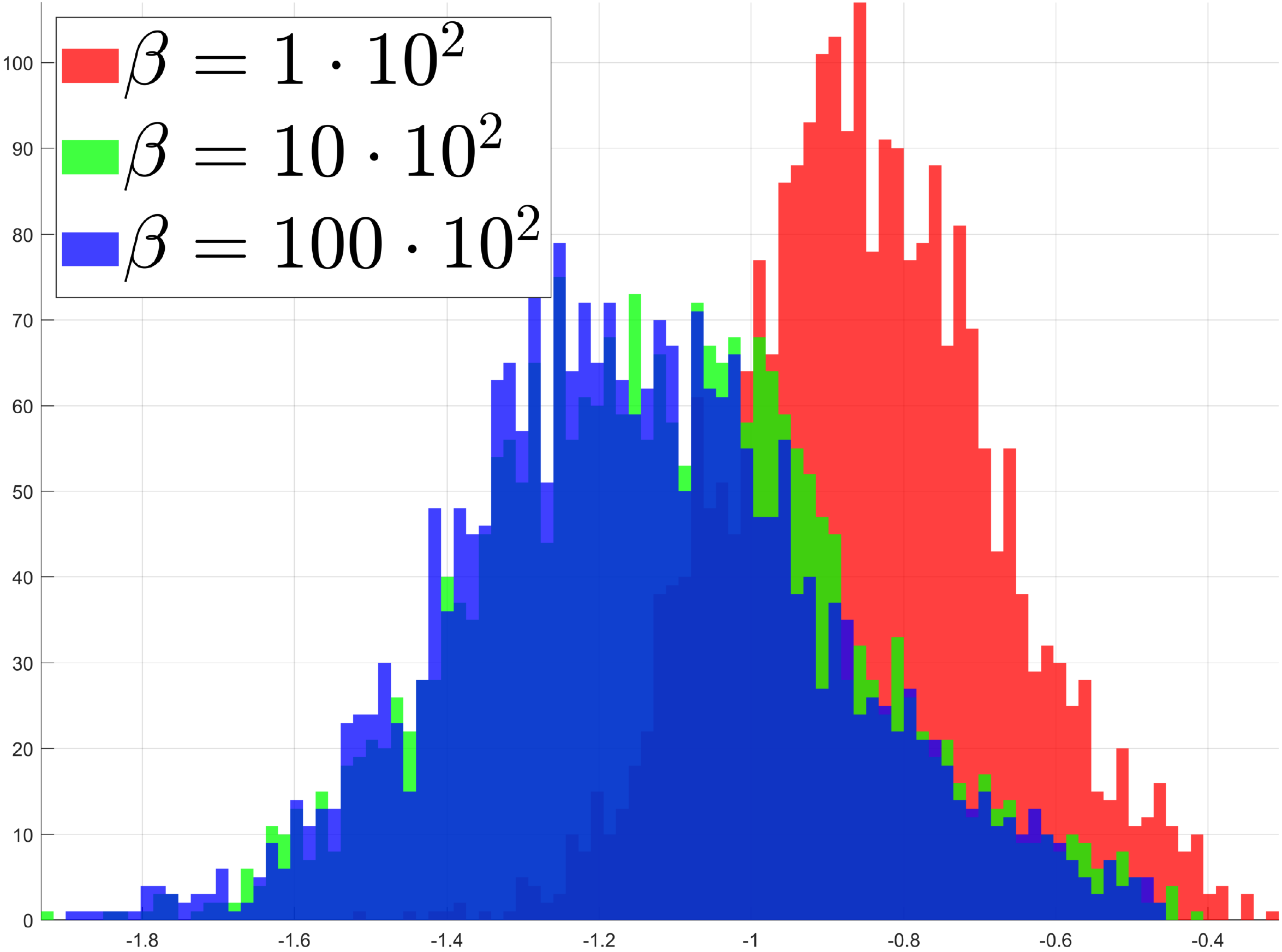}\caption{Left: error probability on the validation data $\pe_{\tilde{\boldsymbol{Z}}}(C^{(i)})$
per iteration \textendash{} averaged over $2.5\cdot10^{3}$ trials.
Right: log error probability on the validation data $\log\pe_{\tilde{\boldsymbol{Z}}}(C^{(i)})$
\textendash{} histogram over $2.5\cdot10^{3}$ trials for codebooks
of size $m=32$. \label{fig: Gibbs GM convergence iteration}}
\end{figure}

We repeat the Gaussian mixture noise experiment from Section \ref{subsec:Gaussian-Mixture-noise}.
Here there is no need to constraint the noise to ${\cal B}^{d}(1)$
and the codebook to ${\cal B}^{d}(r_{x})$ and so we omit the projection
operation. We choose the reference measure $Q$ to be the standard
Lebesgue measure, so that the codebook at each stage is only chosen
based on its error probability and not affected by other factors such
as its average power. The decoder is the standard minimum Euclidean
decoder. To facilitate the computational load of the Gibbs algorithm
we take a memoization-based approach which is detailed in Appendix
\ref{sec:Memoization-Implementation-of}. As evident from Fig. \ref{fig: Gibbs GM convergence iteration},
practically there is no reason to increase $\beta$ beyond $\beta=1000$,
and the loss of reducing $\beta$ beyond that by a factor of $10$
is rather mild. The question whether using finite $\beta$ is merely
a theoretical tool to prove generalization bounds or useful in practice
remains open. In principle, the theoretical bounds do not depend on
any property of the noise distribution, but rather on the algorithmic
stability of the Gibbs algorithm. Thus, the results are applicable
to \emph{arbitrary }noise distribution, no matter how complex. In
various other experiments we have conducted with Gaussian mixture,
there were cases in which reducing the value of $\beta$ has improved
performance, yet not in a very consistent or statistically significant
way. A noise distribution which is more intricate than Gaussian mixture
might leads to finite values of $\beta$ obtaining better performance,
and a finite value of $\beta$ universally ``protects'' against
any possible noise distribution.

\section{Summary and Future Research \label{sec:Summary}}

We have considered the problem of empirical design of an encoder and
a decoder of a communication system operating over an additive noise
channel, given samples of the noise. We summarize here the main findings
and various open problems for future research.

A bound on the generalization error for the regular error probability
loss function was stated, which shows a $O(\frac{1}{\sqrt{n}})$ dependency
on the number of samples. No structure is assumed on the codebook
and in turn the generalization bound depends linearly on the codebook
size $m$. A goal for future research is to settle the dependency
of the generalization bound on the allowed codebook structure and
size, via a possibly refined generalization bound and a proper converse
result. Another possibility of obtaining convergence rates faster
than $O(\frac{1}{\sqrt{n}})$ is to make assumptions on the noise
distribution. Such rates were established for empirical design of
quantizers in \cite{antos2005individual,levrard2013fast,levrard2015nonasymptotic}.
Future research could derive analogous conditions and bounds for the
problem studied in this paper.

It was further shown that the use of a surrogate loss function to
the standard error probability loss function improves the dependency
of the generalization error bound on the number of codewords from
linear to (nearly) square root. As evident from the proof, this property
essentially follows from the continuity of the loss function in the
design parameters \textendash{} the codebook and the inverse covariance
matrix. Furthermore, an alternating optimization algorithm was proposed
to minimize the empirical loss. At each iteration, a ``local'' approximation
of the loss function\footnote{Which is also a provably upper bound to the loss function.}
is obtained by ``freezing'' the nearest neighbors for each codeword
in the current codebook and decoder, and the iteration is completed
by optimizing this upper bound. We expect that similar techniques
involving surrogate loss functions and their local approximation could
be useful (perhaps with some innovation) to obtain generalization
bounds and empirical loss minimization algorithms for much more complicated
scenarios, for example, decoders which are based on DNN. A goal for
future research is to develop and analyze such algorithms. Furthermore,
we have not provided a dedicated analysis of the alternating-minimization
SGD algorithm, but rather rely on the general uniform convergence
bound for the surrogate loss function. The uniform convergence bound
is oblivious to the way the algorithm explore the possible input distributions
(based on the noise samples), and thus could be pessimistic when applied
to the SGD algorithm. In this case, various stability properties (e.g.,
\cite{bousquet2002stability,raginsky2020information}) could be used
to obtain sharper generalization bounds, as was done, e.g., in \cite{hardt2016train}. 

Another possibility for future research is the analysis of different
loss functions. For example, the cross-entropy loss function is a
commonly used surrogate in the practical learning-based design of
end-to-end communication systems (e.g., \cite{farsad2017detection,wang2017deep,nachmani2018deep}).
Similarly, to the surrogate loss function studied here, it also directly
bounds the standard error probability loss function via Pinsker's
inequality. However, its analysis would differ from the analysis in
this paper mainly as it requires considering a different class of
decoders \textendash{} those which output a posterior probability
distribution on the messages given the channel output. In addition,
the cross entropy is not a bounded loss function, and so analyzing
it it requires more elaborate methods compared to the ones used in
this paper. 

We have then considered a Gibbs algorithm that expurgates large codebooks
in order to obtain smaller codebooks with improved error probability.
The randomness in this algorithm, as manifested by the inverse temperature
parameter $\beta$, was mainly introduced in order to provably bound
the generalization error of such algorithms. In practice, however,
most of our experiments have shown that backing $\beta$ from infinity
does not improve performance. An open problem is an analytical characterization
of the optimal value of $\beta$, and identifying cases in which $\beta<\infty$
improves performance in relevant practical scenarios. Furthermore,
finding efficient combinatorial algorithms with theoretical guarantees
for the problem of finding an optimal codebook withing a super-codebook
is an interesting avenue for future research. 

Finally, we have considered the problem of maximizing the mutual information
$I(X;X+Z)$ over the distribution of $X$. We have seen that uniform
convergence empirical error rates of $\tilde{O}(n^{-1/(d\vee4)})$
are possible, under rather general regularity conditions on the distribution
of $Z$, and the set of possible distributions for $X$. Some improvement
in the rate is possible when the input distribution is constrained
to a given support of finite cardinality, and smoothness assumptions
on the distribution of $Z$. However, the required number of samples
is still exponential in the dimension $d$. Obtaining fast uniform
convergence rates seems possible only if the family of input distributions
is restricted or the noise distribution is supported on a low-dimensional
manifold. This is left for further research. From a different angle,
improved rates can be obtained by considering a specific algorithm.
As was discussed, finding the optimal input distribution could be
a difficult problem even for a known noise distribution. Thus, typically
a specific algorithm is used to approximate the optimal input distribution.
It is possible that stability-based analysis would be useful to this
setting too.

\section*{Acknowledgment}

The support of N. Merhav and the comments of the anonymous reviewers
are acknowledged with gratitude. 

\appendices{}

\section{Proofs\label{sec:Proofs}}

\subsection{A Preliminary \textendash{} Uniform Convergence by Rademacher Complexity}

We denote the \emph{empirical Rademacher complexity} of a set ${\cal L}_{n}\subset\mathbb{R}^{n}$
by
\begin{equation}
\Rad(\mathcal{L}_{n})\dfn\frac{1}{n}\E\left[\sup_{l^{n}\in\mathcal{L}_{n}}\sum_{i=1}^{n}R_{i}l_{i}\right],\label{eq: empircal Rademacher complexity}
\end{equation}
where $l^{n}\dfn(l_{1},\ldots l_{n})\in\mathbb{R}^{n}$ and $R^{n}\dfn(R_{1},\ldots R_{n})\in\{\pm1\}^{n}$
are Rademacher random variables (i.e., $R_{i}\sim\text{Uniform}\{-1,1\}$,
i.i.d.). 
\begin{prop}
\label{prop:Rademacher complexity uniform convergence bound}Let ${\cal Z}$
be a data samples space and let ${\cal H}$ be a hypothesis class.
If the loss function $\ell\colon({\cal H},{\cal Z})\mapsto\mathbb{R}$
is absolutely bounded $|\ell(h,z)|\leq r$ then
\begin{equation}
\P_{\boldsymbol{Z}\stackrel{\tiny\mathrm{i.i.d.}}{\sim}\mu}\left[\bigcap_{h\in{\cal H}}\left\{ \left|\E_{Z\sim\mu}\left[\ell(h,Z)\right]-\frac{1}{n}\sum_{i=1}^{n}\ell(h,Z_{i})\right|\leq A_{n}(\delta)\right\} \right]\geq1-\delta,\label{eq: uniform bound Rad complexity}
\end{equation}
where 
\[
A_{n}(\delta)\dfn4\cdot\E\left[\Rad({\cal L}_{\boldsymbol{Z}})\right]+r\sqrt{\frac{2\ln(2/\delta)}{n}},
\]
and 
\[
{\cal L}_{\boldsymbol{Z}}=\left\{ (\ell(h,Z_{1}),\ldots,\ell(h,Z_{n}))\colon h\in{\cal H}\right\} .
\]
\end{prop}
\begin{IEEEproof}
It is well-established that Rademacher complexity uniformly bounds
convergence the deviation of empirical averages from the statistical
average \cite{bartlett2002rademacher}, and the statement in (\ref{eq: uniform bound Rad complexity})
was made in \cite[Thm. 26.5]{shalev2014understanding}, albeit without
the absolute value inside the probability term, and where the first
additive term in $A_{n}(\delta)$ is $2\E[\Rad({\cal L}_{\boldsymbol{Z}})]$.
By repeating the same arguments in \cite[Ch. 26]{shalev2014understanding}
that lead to that theorem, while replacing $\E_{Z\sim\mu}\left[\ell(h,Z)\right]-\frac{1}{n}\sum_{i=1}^{n}\ell(h,Z_{i})$
therein with its absolute value, one can obtain that the probability
bound in (\ref{eq: uniform bound Rad complexity}) is valid when is
$A_{n}(\delta)$ replaced with 
\[
\tilde{A}_{n}(\delta)\dfn\frac{2}{n}\cdot\E\left[\sup_{l^{n}\in{\cal L}_{\boldsymbol{Z}}}\left|\sum_{i=1}^{n}R_{i}l_{i}\right|\right]+r\cdot\sqrt{\frac{2\ln(2/\delta)}{n}}.
\]
But, using the notation $-{\cal L}_{\boldsymbol{Z}}\dfn\left\{ (-\ell(h,Z_{1}),\ldots,-\ell(h,Z_{n}))\colon h\in{\cal H}\right\} $
we have that $\tilde{A}_{n}(\delta)\leq A_{n}(\delta)$ since for
any given $\boldsymbol{Z}\in(\mathbb{R}^{d})^{n}$
\begin{align}
\E\left[\sup_{l^{n}\in{\cal L}_{\boldsymbol{Z}}}\left|\sum_{i=1}^{n}R_{i}l_{i}\right|\right] & =\E\left[\left\{ \sup_{l^{n}\in{\cal L}_{\boldsymbol{Z}}\colon\sum_{i=1}^{n}R_{i}l_{i}>0}\sum_{i=1}^{n}R_{i}l_{i}\right\} \vee\left\{ \sup_{l^{n}\in{\cal L}_{\boldsymbol{Z}}\colon\sum_{i=1}^{n}R_{i}l_{i}<0}-\sum_{i=1}^{n}R_{i}l_{i}\right\} \right]\\
 & \leq\E\left[\sup_{l^{n}\in{\cal L}_{\boldsymbol{Z}}\colon\sum_{i=1}^{n}R_{i}l_{i}>0}\sum_{i=1}^{n}R_{i}l_{i}+\sup_{l^{n}\in-{\cal L}_{\boldsymbol{Z}}\colon\sum_{i=1}^{n}R_{i}l_{i}>0}\sum_{i=1}^{n}R_{i}l_{i}\right]\\
 & \leq\E\left[\sup_{l^{n}\in{\cal L}_{\boldsymbol{Z}}}\sum_{i=1}^{n}R_{i}l_{i}+\sup_{l^{n}\in-{\cal L}_{\boldsymbol{Z}}}\sum_{i=1}^{n}R_{i}l_{i}\right]\\
 & =\E\left[\sup_{l^{n}\in{\cal L}_{\boldsymbol{Z}}}\sum_{i=1}^{n}R_{i}l_{i}\right]+\E\left[\sup_{l^{n}\in{\cal L}_{\boldsymbol{Z}}}\sum_{i=1}^{n}(-R_{i})l_{i}\right]\\
 & =2\Rad({\cal L}_{\boldsymbol{Z}}),
\end{align}
where in the last equality we have used $R_{i}\eqd-R_{i}$. 
\end{IEEEproof}

\subsection{The Proofs of Theorem \ref{thm:Generalization NN} and Proposition
\ref{prop:Generalization NN minimax rates}}
\begin{IEEEproof}[Proof of Theorem \ref{thm:Generalization NN}]
 For a given training set $\{z_{i}\}_{i=1}^{n}$, we define the loss
vector as 
\[
l^{n}(C,S)\dfn(\ell(C,S,z_{1}),\ldots,\ell(C,S,z_{n}))
\]
and the loss class as ${\cal L}_{n}(m)\dfn\{l^{n}(C,S)\colon C\in(\mathbb{R}^{d})^{m},\;S\in\mathbb{S}_{+}^{d}\}$.
We next bound the Rademacher complexity (\ref{eq: empircal Rademacher complexity})
of ${\cal L}_{n}(m)$ and then use Prop. \ref{prop:Rademacher complexity uniform convergence bound}.
As $|{\cal L}_{n}(m)|<\infty$, Massart's lemma \cite[Thm. 26.8]{shalev2014understanding}
implies that 
\begin{equation}
\Rad({\cal L}_{n}(m))\leq\max_{l^{n}\in{\cal L}_{n}(m)}\|l^{n}-\overline{l^{n}}\|\frac{\sqrt{2\log|{\cal L}_{n}(m)|}}{n}\leq\sqrt{\frac{2\log|{\cal L}_{n}(m)|}{n}},\label{eq: bound on Rad via Massarat}
\end{equation}
where $\overline{l^{n}}=\frac{1}{|{\cal L}_{n}(m)|}\sum_{l^{n}\in{\cal L}_{n}(m)}l^{n}$,
and the last inequality holds since as $l^{n}\in[0,1]^{n}$. We further
bound $|{\cal L}_{n}(m)|$ as follows. A loss vector $l^{n}(C,S)$
is unequivocally determined by $m$ loss vectors 
\[
l_{j}^{n}(C,S)\dfn(\ell_{j}(C,S,z_{1}),\ldots,\ell_{j}(C,S,z_{n}))\in\{0,1\}^{n}
\]
for $j\in[m]$. In turn, each loss vector $l_{j}^{n}(C,S)$ is unequivocally
determined by the $m-1$ binary classification vectors
\[
b_{j'\mid j}^{n}(C,S)=(b_{j'\mid j}(C,S,z_{1}),\ldots,b_{j'\mid j}(C,S,z_{n}))
\]
for $j'\neq j$, with 
\[
b_{j'\mid j}(C,S,z)=\I\left\{ \|x_{j}-x_{j'}\|_{S}^{2}+2(x_{j}-x_{j'})^{T}Sz<0\right\} .
\]
Hence, $b_{j'\mid j}^{n}(C,S)$ is the result of the binary classifier
induced by $\{x_{j},x_{j'}\}$ and the nearest neighbor decoding rule.
Since this classifier is in fact an affine hyperplane, its VC dimension
is upper bounded by $d+1$ \cite[Thm. 9.3]{shalev2014understanding}.
By the Sauer\textendash Shelah lemma \cite[Lemma 6.10]{shalev2014understanding},
for $n\geq d+1$ 
\[
\left|\{b_{j'\mid j}^{n}(C,S)\colon C\in(\mathbb{R}^{d})^{m},\;S\in\mathbb{S}_{+}^{d}\}\right|\leq\left(\frac{en}{d+1}\right)^{d+1}.
\]
Accounting for all pairs $j,j'\in[m]$ we thus obtain
\begin{equation}
|{\cal L}_{n}(m)|\leq\left(\frac{en}{d+1}\right)^{m^{2}(d+1)}.\label{eq: bound on size of loss class for error probability loss}
\end{equation}
Inserting the bound (\ref{eq: bound on size of loss class for error probability loss})
to (\ref{eq: bound on Rad via Massarat}) and using Prop. \ref{prop:Rademacher complexity uniform convergence bound}
completes the proof.
\end{IEEEproof}
\begin{rem}
Despite (\ref{eq: bound on size of loss class for error probability loss})
being a crude estimate, other bounding techniques such as using Natarajan
dimension for multiclass classification do not yield improved bounds
(at least not using using direct arguments). Furthermore, a seemingly
natural way of defining a class of classifiers which determine $\ell_{j}(C,S,z)$
is as an intersection of $m$ hyperplanes. For this class, the classic
paper \cite{blumer1989learnability} (see also \cite{csikos2018optimal})
gives a bound of $O((d+1)m\log m)$ which leads to a worse bound than
the one obtained here. 
\end{rem}
We now turn to the proof of Proposition \ref{prop:Generalization NN minimax rates}
which requires a few lemmas. Let $Q(t)\dfn\P[W>t]$ where $W\sim N(0,1)$
be the Gaussian tail distribution function. The next lemma states
the error probability of a codebook of two antipodal codewords. 
\begin{lem}
\label{lem: error probability of m=00003D2 and Gaussian noise}Let
$d=2$ and consider a codebook of $m=2$ codewords $C=(x_{1},x_{2})$
given by $x_{1}=\left(\cos\alpha,\sin\alpha\right)^{T}=-x_{2}$ for
some $\alpha\in[0,\pi]$. Parameterize the decoder inverse covariance
matrix by
\begin{equation}
S=\left[\begin{array}{cc}
\cos(\beta) & \sin(\beta)\\
-\sin(\beta) & \cos(\beta)
\end{array}\right]\left[\begin{array}{cc}
s(1) & 0\\
0 & s(2)
\end{array}\right]\left[\begin{array}{cc}
\cos(\beta) & -\sin(\beta)\\
\sin(\beta) & \cos(\beta)
\end{array}\right]\label{eq: S eigendecomposition d=00003D2}
\end{equation}
with $s(1),s(2)\in\mathbb{R_{+}}$ and $\beta\in[0,\pi]$. If $Z\sim N(\boldsymbol{0},\diag(\sigma^{2}(1),\sigma^{2}(2)))$
then,
\begin{equation}
\pe_{\mu}(C,S)=Q\left(\frac{\cos^{2}(\alpha+\beta)s(1)+\sin^{2}(\alpha+\beta)s(2)}{\tau}\right),\label{eq: error probability of general two codewords}
\end{equation}
where 
\begin{align}
\tau^{2}\left(\sigma^{2}(1),\sigma^{2}(2),s(1),s(2),\alpha,\beta\right) & \dfn\sigma^{2}(1)\cdot\left[s(1)\cos(\beta)\cos(\alpha+\beta)+s(2)\sin(\beta)\sin(\alpha+\beta)\right]^{2}\nonumber \\
 & \hphantom{=}+\sigma^{2}(2)\cdot\left[-s(1)\sin(\beta)\cos(\alpha+\beta)+s(2)\cos(\beta)\sin(\alpha+\beta)\right]^{2}.\label{eq: definition of tau}
\end{align}
If $\sigma(1)=\sigma(2)\equiv\sigma$ then $\min_{\alpha,\beta,s(1),s(2)}\pe_{\mu}(C,S)=Q\left(\frac{1}{\sigma}\right)$
is achieved with $s(1)=s(2)=1$.
\end{lem}
The proof of Lemma \ref{lem: error probability of m=00003D2 and Gaussian noise}
is a trivial exercise and thus omitted.
\begin{cor}
\label{cor: maximal error probability for m=00003D2} If $s(1)\wedge s(2)\geq1$
then $\pe_{\mu}(C,S)=Q\left(\frac{1}{\tau}\right)$ for some $\tau\leq[\sigma(1)\wedge\sigma(2)]\cdot\sqrt{s(1)\wedge s(2)}$. 
\end{cor}
\begin{IEEEproof}
We consider three cases. First, if $\cos(\alpha+\beta)=0$ and $\sin(\alpha+\beta)=1$,
then by (\ref{eq: error probability of general two codewords}) and
(\ref{eq: definition of tau}), the error probability is 
\begin{align}
\pe_{\mu}(C,S) & =Q\left(\frac{1}{\sqrt{\sigma^{2}(1)\sin^{2}(\beta)+\sigma^{2}(2)\cos^{2}(\beta)}}\right)\\
 & \leq Q\left(\frac{1}{\sigma(1)\vee\sigma(2)}\right)\\
 & \leq Q\left(\frac{1}{[\sigma(1)\vee\sigma(2)]\cdot\sqrt{s(1)\vee s(2)}}\right)
\end{align}
where the last equality follows since $s(1)\wedge s(2)\geq1$ is assumed,
and since $Q(t)$ is monotonic decreasing in $t$. Second, if $\cos(\alpha+\beta)=1$
and $\sin(\alpha+\beta)=0$ then similar analysis leads to the same
result. Third, if both $\cos(\alpha+\beta)\neq0$ and $\sin(\alpha+\beta)\neq0$
then since the bound (\ref{eq: error probability of general two codewords})
is homogeneous w.r.t. $(s(1),s(2))$, we will obtain the same bound
for any $c\cdot(s(1),s(2))$, and thus we may assume that 
\begin{equation}
\cos^{2}(\alpha+\beta)s(1)+\sin^{2}(\alpha+\beta)s(2)\geq1.\label{eq: assumption on s alpha and beta}
\end{equation}
In addition, from (\ref{eq: definition of tau})
\begin{align}
\tau^{2} & \leq\left[\sigma^{2}(1)\vee\sigma^{2}(2)\right]\times\nonumber \\
 & \hphantom{=}\Big[s^{2}(1)\cos^{2}(\beta)\cos^{2}(\alpha+\beta)+s^{2}(2)\sin^{2}(\beta)\sin^{2}(\alpha+\beta)\nonumber \\
 & \hphantom{=}+s^{2}(1)\sin^{2}(\beta)\cos^{2}(\alpha+\beta)+s^{2}(2)\cos^{2}(\beta)\sin^{2}(\alpha+\beta)\Big]\\
 & =\left[\sigma^{2}(1)\vee\sigma^{2}(2)\right]\cdot\left[s^{2}(1)\cos^{2}(\alpha+\beta)+s^{2}(2)\sin^{2}(\alpha+\beta)\right]\\
 & \leq\left[\sigma^{2}(1)\vee\sigma^{2}(2)\right]\left[s(1)\vee s(2)\right]\cdot\left[s(1)\cos^{2}(\alpha+\beta)+s(2)\sin^{2}(\alpha+\beta)\right].
\end{align}
Substituting this bound in (\ref{eq: error probability of general two codewords}),
and utilizing the monotonicity of the $Q$ function and the assumption
(\ref{eq: assumption on s alpha and beta}) results 
\begin{align}
\pe_{\mu}(C,S) & \leq Q\left(\frac{\cos^{2}(\alpha+\beta)s(1)+\sin^{2}(\alpha+\beta)s(2)}{\sqrt{\sigma^{2}(1)\vee\sigma^{2}(2)}\cdot\sqrt{s(1)\vee s(2)}\cdot\sqrt{s(1)\cos^{2}(\alpha+\beta)+s(2)\sin^{2}(\alpha+\beta)}}\right)\\
 & \leq Q\left(\frac{1}{\sqrt{\sigma^{2}(1)\vee\sigma^{2}(2)}\cdot\sqrt{s(1)\vee s(2)}}\right).
\end{align}
\end{IEEEproof}
The next lemma states two\textbf{ }properties related to the $Q(\cdot)$
function.
\begin{lem}
\label{lem: properties of Q(1/sqrt(t))}$Q(1/\sqrt{t})$ is convex
on $[0,1/3]$, and $Q\left(1/\sqrt{t}\right)-Q\left(1/\sqrt{s}\right)\geq0.145\cdot(t-s)$
for $1/6<s<t<1/3$. 
\end{lem}
\begin{IEEEproof}
For the first property, $\frac{\d}{\d t}Q\left(\frac{1}{\sqrt{t}}\right)=\frac{1}{\sqrt{8\pi}}e^{-1/(2t)}t^{-3/2}$
by the Leibniz integral rule, and so
\[
\frac{\d^{2}}{\d t^{2}}Q\left(\frac{1}{\sqrt{t}}\right)=\frac{1}{\sqrt{8\pi}}e^{-1/(2t)}\left[\frac{1}{2t^{7/2}}-\frac{3}{2t^{5/2}}\right]
\]
which is nonnegative for $t\in[0,1/3]$. For the second property,
using the first property 
\begin{align}
Q\left(1/\sqrt{t}\right)-Q\left(1/\sqrt{s}\right) & =\int_{s}^{t}\frac{\d}{\d x}Q\left(\frac{1}{\sqrt{x}}\right)\cdot\d x\\
 & \geq(t-s)\cdot\min_{\tilde{x}\in[1/6,1/3]}\left.\frac{\d}{\d x}Q\left(\frac{1}{\sqrt{x}}\right)\right|_{x=\tilde{x}}\\
 & =(t-s)\cdot\left.\frac{\d}{\d x}Q\left(\frac{1}{\sqrt{x}}\right)\right|_{x=1/6}
\end{align}
for which the bound numerically holds. 
\end{IEEEproof}
The following lemma essentially states a large-deviations property
of central chi-square random variables.
\begin{lem}
\label{lem: large deviations chi square}Let $\{W_{i},\tilde{W}_{i}\}_{i\in[n]}$
be i.i.d., such that $W_{i}\sim N(0,\sigma^{2}),\;\tilde{W}_{i}\sim N(0,\tilde{\sigma}^{2})$
with $\frac{\sigma^{2}}{\tilde{\sigma}^{2}}=1-t$ for some $t\in(0,1)$
and $W_{i}\ind\tilde{W}_{i}$ for all $i\in[n]$. Then
\[
\P\left[\sum_{i=1}^{n}W_{i}^{2}>\sum_{i=1}^{n}\tilde{W}_{i}^{2}\right]>\frac{1}{4}\exp\left[-\frac{3}{4}nt^{2}\right].
\]
\end{lem}
\begin{IEEEproof}
The proof is based on the standard change-of-measure argument. Denote
by $\varphi_{\sigma}$ the Gaussian probability measure on the Borel
sets of $\mathbb{R}$ of mean zero and variance $\sigma^{2}$, and
let $\nu\dfn\varphi_{\sigma}^{\otimes2n}$ and $\tilde{\nu}\dfn\varphi_{\sigma}^{\otimes n}\otimes\varphi_{\tilde{\sigma}}^{\otimes n}$
(which are probability measures on the Borel sets of $\mathbb{R}^{2n}$).
By the tensorization property of the KL divergence and a standard
calculation 
\begin{equation}
\dkl(\nu||\tilde{\nu})=n\dkl(\varphi_{\sigma}||\varphi_{\sigma})+n\dkl(\varphi_{\sigma}||\varphi_{\tilde{\sigma}})=\frac{n}{2}\left(\frac{\sigma^{2}}{\tilde{\sigma}^{2}}-1-\log\left(\frac{\sigma^{2}}{\tilde{\sigma}^{2}}\right)\right).\label{eq: KL between normal}
\end{equation}
Let $(W_{1}\ldots,W_{n},\tilde{W}_{1}\ldots,\tilde{W}_{n})\in\mathbb{R}^{2n}$
be the identity random variable on the space defined above, and define
$I=\I[\sum_{i=1}^{n}W_{i}^{2}>\sum_{i=1}^{n}\tilde{W}_{i}^{2}]\in\{0,1\}$.
Under $\nu$, $I$ is Bernoulli with probability of success $\E_{\nu}[I]=\frac{1}{2}$
(by symmetry), and under $\tilde{\nu}$ it is Bernoulli with probability
of success given by $p\dfn\E_{\tilde{\nu}}[I]$, which is the probability
required to be lower bounded. By the data processing inequality for
the KL divergence \cite[Lemma 4.1]{csiszar2004information}, \cite[Corollary 5.2.2]{gray2011entropy}
\[
\dkl(\nu||\tilde{\nu})\geq\dkl\left(\left.(\frac{1}{2},\frac{1}{2})\right\Vert (p,1-p)\right)\geq\frac{1}{2}\log\frac{1}{p}-\log2
\]
and so $p\geq\frac{1}{4}\exp[-2\dkl(\nu||\tilde{\nu})]$ which, along
with (\ref{eq: KL between normal}), implies that 
\[
\P\left[\sum_{i=1}^{n}W_{i}^{2}(1)>\sum_{i=1}^{n}\tilde{W}_{i}^{2}\right]>\frac{1}{4}\exp\left[-n\left(\frac{\sigma^{2}}{\tilde{\sigma}^{2}}-1-\log\left(\frac{\sigma^{2}}{\tilde{\sigma}^{2}}\right)\right)\right].
\]
The proof is completed by bounding the exponent in the last bound.
Using a second order Taylor approximation of $\log x$ around $x=1$,
with a Lagrange form for the remainder, which is valid for, say, $[\frac{4}{7},1]$,
we have 
\[
\left|\log(x)-\left[(x-1)-\frac{1}{2}(x-1)^{2}\right]\right|\leq\max_{\tilde{x}\in[\frac{4}{7},1]}\left|\left(\frac{1}{3\tilde{x}^{3}}\right)(x-1)^{3}\right|\leq\frac{1}{4}(x-1)^{2}
\]
and so $\log(x)\geq(x-1)-\frac{3}{4}(x-1)^{2}$ holds for $x\in[\frac{4}{7},1]$.
Hence, if $\frac{\sigma^{2}}{\tilde{\sigma}^{2}}=1-t$ then
\[
\frac{\sigma^{2}}{\tilde{\sigma}^{2}}-1-\log\left(\frac{\sigma^{2}}{\tilde{\sigma}^{2}}\right)\leq\frac{3}{4}t^{2}.
\]
\end{IEEEproof}
\begin{IEEEproof}[Proof of Proposition \ref{prop:Generalization NN minimax rates}]
We first assume that $d=2$, and afterwards reduce the $d>2$ case
to the $d=2$ case. The proof follows the standard reduction to binary
hypothesis testing (specifically \cite[Sec. 28.2.1]{shalev2014understanding}).
We consider a pair of distributions $\mu_{\pm}$ for $Z$, and begin
by showing that a single codebook and inverse covariance matrix pair
cannot simultaneously achieve low $\pe_{\mu}(C,S)-\inf_{C',S'}\pe_{\mu}(C',S')$
for both $\mu=\mu_{-}$ and $\mu=\mu_{+}$. 

Let $\eta\in(0,\frac{1}{4})$ to be set later. Under $\mu_{+}$ (resp.
$\mu_{-}$) it is assumed that $Z\sim N(\boldsymbol{0},\diag(\frac{1}{3r_{s}},\frac{1-2\eta}{3r_{s}}))$
(resp. $Z\sim N(\boldsymbol{0},\diag(\frac{1-2\eta}{3r_{s}},\frac{1}{3r_{s}}))$).
Given any $(x_{1},x_{2})$, if the Euclidean distance between $x_{1}$
and $x_{2}$ is increased, while the relative angle between them is
preserved then the error probability decreases under both $\mu_{\pm}$.
Thus, $\inf_{C,S}\pe_{\mu_{\pm}}(C,S)$ is achieved with $x_{1}=(\cos(\alpha_{\pm}),\sin(\alpha_{\pm}))=-x_{2}$
for some $\alpha_{\pm}$, and similarly, it can be assumed that any
optimal algorithm $A$ outputs an codebook of the form $x_{1}=(\cos(\alpha_{\boldsymbol{\boldsymbol{z}}}),\sin(\alpha_{\boldsymbol{\boldsymbol{z}}}))=-x_{2}$
for some $\alpha_{\boldsymbol{\boldsymbol{z}}}$. Now, note that $S$
can be parameterized by $(\beta,s)$ with $s(1)\equiv1$ and $s(2)\equiv s$
in (\ref{eq: S eigendecomposition d=00003D2}). Now, it holds from
symmetry that 
\[
\inf_{C,S}\pe_{\mu_{+}}(C,S)=\inf_{C,S}\pe_{\mu_{-}}(C,S).
\]
Furthermore, under $\mu_{-}$, it is intuitively clear since the noise
variance $\frac{1}{3r_{s}}$ of the second dimension is larger than
then one of the first dimension $\frac{1-2\eta}{3r_{s}}$ , it is
preferable to choose $x_{1}=x_{1}^{-}\dfn(1,0)$ and $x_{2}=x_{2}^{-}\dfn(-1,0)$,
that is $\alpha_{-}=0$. Letting $C_{-}=(x_{1}^{-},x_{2}^{-})$ we
may bound\footnote{In fact, $C_{-}$ is also an optimal choice, but this stronger property
is not required for the bound.}
\begin{align}
\inf_{C,S}\pe_{\mu_{-}}(C,S) & \trre[\leq,a]\inf_{S}\pe_{\mu_{-}}(C_{-},S)\\
 & =\inf_{s\geq1,\beta}\pe_{\mu_{-}}(C_{-},S)\\
 & \trre[=,b]\inf_{s\geq1,\beta}Q\left(\frac{\cos^{2}(\beta)+\sin^{2}(\beta)s}{\sqrt{\frac{(1-2\eta)}{3r_{s}}\cdot\left[\cos^{2}(\beta)+s\sin^{2}(\beta)\right]^{2}+\frac{1}{3r_{s}}\left[-\sin(\beta)\cos(\beta)+s\cos(\beta)\sin(\beta)\right]^{2}}}\right)\\
 & \trre[\leq,c]Q\left(\frac{1}{\sqrt{(1-2\eta)/(3r_{s})}}\right)
\end{align}
where $(a)$ follows from the choice $C=C_{-}$, $(b)$ follows from
Lemma \ref{lem: error probability of m=00003D2 and Gaussian noise}
by setting $\alpha=\alpha_{-}=0$, $s(1)=1$, $s(2)=s$, $\sigma^{2}(1)=\frac{1-2\eta}{3r_{s}}$
and $\sigma^{2}(2)=\frac{1}{3r_{s}}$, and $(c)$ follows from the
choice $s=1$. Hence, 
\begin{equation}
\inf_{C,S}\pe_{\mu_{-}}(C,S)=\inf_{C,S}\pe_{\mu_{+}}(C,S)\leq Q\left(\frac{1}{\sqrt{(1-2\eta)/(3r_{s})}}\right).\label{eq: bound on a optimal codebook and decoder for d 2}
\end{equation}
In addition, for an arbitrary fixed $(C,S)$ with $x_{1}=(\cos(\alpha),\sin(\alpha))=-x_{2}$,
and $S$ parameterized by $(\beta,s)$, Lemma \ref{lem: error probability of m=00003D2 and Gaussian noise},
together with (\ref{eq: bound on a optimal codebook and decoder for d 2}),
further imply that 
\begin{align}
 & \max_{\mu\in\{\mu_{\pm}\}}\left[\pe_{\mu}(C,S)-\inf_{C',S'}\pe_{\mu}(C',S')\right]\nonumber \\
 & \geq\frac{1}{2}\left[\pe_{\mu_{-}}(C,S)-\inf_{C',S'}\pe_{\mu_{-}}(C',S')\right]+\frac{1}{2}\left[\pe_{\mu_{+}}(C_{\boldsymbol{Z}},S_{\boldsymbol{Z}})-\inf_{C',S'}\pe_{\mu_{-}}(C',S')\right]\\
 & =\frac{1}{2}Q\left(\frac{\cos^{2}(\alpha+\beta)+\sin^{2}(\alpha+\beta)s}{\sigma_{-}(\alpha,\beta,s)}\right)+\frac{1}{2}Q\left(\frac{\cos^{2}(\alpha+\beta)+\sin^{2}(\alpha+\beta)s}{\sigma_{+}(\alpha,\beta,s)}\right)-Q\left(\frac{1}{\sqrt{(1-2\eta)/(3r_{s})}}\right),\label{eq:lower bound before convexity}
\end{align}
where 
\begin{align}
\sigma_{-}^{2}(\alpha,\beta,s) & \equiv\tau^{2}\left(\frac{1}{3r_{s}},\frac{1-2\eta}{3r_{s}},\alpha,\beta,s\right),\\
\sigma_{+}^{2}(\alpha,\beta,s) & \equiv\tau^{2}\left(\frac{1-2\eta}{3r_{s}},\frac{1}{3r_{s}},\alpha,\beta,s\right),
\end{align}
with $\tau$ defined in (\ref{eq: definition of tau}). By Corollary
\ref{cor: maximal error probability for m=00003D2}, it is assured
that the arguments of the three $Q$ functions in (\ref{eq:lower bound before convexity})
are at least $\sqrt{3}$, and so Lemma \ref{lem: properties of Q(1/sqrt(t))}
assures that $t\mapsto Q(1/\sqrt{t})$ is convex. By that convexity
property

\begin{align}
 & \frac{1}{2}Q\left(\frac{\cos^{2}(\alpha+\beta)+\sin^{2}(\alpha+\beta)s}{\sigma_{-}(\alpha,\beta,s)}\right)+\frac{1}{2}Q\left(\frac{\cos^{2}(\alpha+\beta)+\sin^{2}(\alpha+\beta)s}{\sigma_{+}(\alpha,\beta,s)}\right)\nonumber \\
 & \geq Q\left(\frac{\cos^{2}(\alpha+\beta)+\sin^{2}(\alpha+\beta)s}{\sqrt{\frac{1}{2}\sigma_{-}^{2}(\alpha,\beta,s)+\frac{1}{2}\sigma_{+}^{2}(\alpha,\beta,s)}}\right)\\
 & =Q\left(\frac{\cos^{2}(\alpha+\beta)+\sin^{2}(\alpha+\beta)s}{\sigma(\alpha,\beta,s)}\right)\\
 & \geq Q\left(\frac{1}{\sqrt{(1-\eta)/(3r_{s})}}\right)\label{eq: convexity bounds on two Gaussian distributions}
\end{align}
where 
\[
\sigma^{2}(\alpha,\beta,s)\equiv\tau^{2}\left(\frac{1-\eta}{3r_{s}},\frac{1-\eta}{3r_{s}},\alpha,\beta,s\right),
\]
and the last inequality follows from the last statement of Lemma \ref{lem: error probability of m=00003D2 and Gaussian noise}.
Combining (\ref{eq:lower bound before convexity}), (\ref{eq: convexity bounds on two Gaussian distributions}),
and the second property in Lemma \ref{lem: properties of Q(1/sqrt(t))}
leads to 
\[
\max_{\mu\in\{\mu_{\pm}\}}\left[\pe_{\mu}(C_{\boldsymbol{Z}},S_{\boldsymbol{Z}})-\inf_{C,S}\pe_{\mu}(C,S)\right]\geq Q\left(\frac{1}{\sqrt{(1-\eta)/(3r_{s})}}\right)-Q\left(\frac{1}{\sqrt{(1-2\eta)/(3r_{s})}}\right)\geq0.145\cdot\frac{\eta}{3r_{s}}\equiv\epsilon,
\]
where the last inequality holds since $\eta\in(0,\frac{1}{4})$ and
$r_{s}\leq2$ is assumed. Hence, if we denote $L_{\pm}\dfn\pe_{\mu_{\pm}}(C,S)-\inf_{C',S'}\pe_{\mu_{\pm}}(C',S')$
then the last display implies that if $L_{+}<\epsilon$ then $L_{-}\geq\epsilon$
and vice-versa. Consequently,
\begin{equation}
\I\{L_{+}\geq\epsilon\}+\I\{L_{-}\geq\epsilon\}=\I\{L_{+}>\epsilon\}+1-\I\{L_{-}<\epsilon\}>1.\label{eq: dichotomy}
\end{equation}
Thus, a single codebook cannot be simultaneously ``good'' for both
$\mu_{\pm}$.

Now, let $A$ be an arbitrary algorithm which outputs $C_{\boldsymbol{z}},S_{\boldsymbol{z}}$
with $C_{\boldsymbol{z}}$ of the form $x_{1}=(\cos(\alpha_{\boldsymbol{\boldsymbol{z}}}),\sin(\alpha_{\boldsymbol{\boldsymbol{z}}}))=-x_{2}$
for some $\alpha_{\boldsymbol{z}}$, which can be assumed without
loss of generality (w.l.o.g.) for an optimal algorithm. Denote by
$f_{\pm}(\boldsymbol{z})$ the density of $\mu_{\pm}^{\otimes n}$,
and denote $z_{i}\equiv(z_{i}(1),z_{i}(2))$. Further denote the event
$A\dfn\{\sum_{i=1}^{n}z_{i}^{2}(1)\geq\sum_{i=1}^{n}z_{i}^{2}(2)\}$
and its complement by $A^{c}$. Then,
\begin{align}
 & \sup_{\mu}\P_{\boldsymbol{Z}\sim\mu^{\otimes n}}\left[\pe_{\mu}(C_{\boldsymbol{Z}},S_{\boldsymbol{Z}})-\inf_{C',S'}\pe_{\mu}(C',S')>\epsilon\right]\nonumber \\
 & \geq\max_{\mu\in\{\mu_{\pm}\}}\P_{\boldsymbol{Z}\sim\mu^{\otimes n}}\left[\pe_{\mu}(C_{\boldsymbol{Z}},S_{\boldsymbol{Z}})-\inf_{C',S'}\pe_{\mu}(C',S')>\epsilon\right]\\
 & \trre[\geq,a]\frac{1}{2}\P_{\boldsymbol{Z}\sim\mu_{-}^{\otimes n}}\left[L_{-}>\epsilon\right]+\frac{1}{2}\P_{\boldsymbol{Z}\sim\mu_{+}^{\otimes n}}\left[L_{+}>\epsilon\right]\\
 & \trre[=,b]\frac{1}{2}\int f_{-}(\boldsymbol{z})\cdot\I\left[L_{-}>\epsilon\right]\I(A)+f_{+}(\boldsymbol{z})\cdot\I\left[L_{+}>\epsilon\right]\I(A)\cdot\d\boldsymbol{z}\nonumber \\
 & \hphantom{==}+\frac{1}{2}\int f_{-}(\boldsymbol{z})\cdot\I\left[L_{-}>\epsilon\right]\I(A^{c})+f_{+}(\boldsymbol{z})\cdot\I\left[L_{+}>\epsilon\right]\I(A^{c})\cdot\d\boldsymbol{z}\\
 & \trre[\geq,c]\frac{1}{2}\int f_{-}(\boldsymbol{z})\cdot\left(\I\left[L_{-}>\epsilon\right]+\I\left[L_{+}>\epsilon\right]\right)\I(A)\cdot\d\boldsymbol{z}\nonumber \\
 & \hphantom{==}+\frac{1}{2}\int f_{+}(\boldsymbol{z})\cdot\left(\I\left[L_{-}>\epsilon\right]+\I\left[L_{+}>\epsilon\right]\right)\I(A^{c})\cdot\d\boldsymbol{z}\\
 & \trre[\geq,d]\frac{1}{2}\int f_{-}(\boldsymbol{z})\cdot\I(A)\cdot\d\boldsymbol{z}+\frac{1}{2}\int f_{+}(\boldsymbol{z})\cdot\I(A^{c})\cdot\d\boldsymbol{z}\\
 & \trre[=,e]\int f_{-}(\boldsymbol{z})\cdot\I(A)\cdot\d\boldsymbol{z}\\
 & =\P_{\boldsymbol{Z}\sim\mu_{-}^{\otimes n}}\left[\sum_{i=1}^{n}Z_{i}^{2}(1)\geq\sum_{i=1}^{n}Z_{i}^{2}(2)\right]\\
 & \trre[\geq,f]\frac{1}{4}\exp\left[-3n\eta^{2}\right]\\
 & \trre[=,g]\delta,
\end{align}
where $(a)$ utilizes the definition of $L_{\pm}$, $(b)$ utilizes
the definition of the event $A$ and $f_{\pm},$ $(c)$ follows since
$f_{+}(\boldsymbol{z})>f_{-}(\boldsymbol{z})$ if and only if $\boldsymbol{z}\in A$,
$(d)$ follows from (\ref{eq: dichotomy}), $(e)$ follows from symmetry,
$(f)$ follows from Lemma \ref{lem: large deviations chi square},
and $(g)$ is obtained by setting $\eta=\sqrt{\frac{1}{3n}\log\frac{1}{4\delta}}$
while taking $n\geq n_{0}(\delta)$ such that $\eta<\frac{1}{4}$. 

In case $d>2$, one may choose $\mu_{1}$ to be zero-mean Gaussian
of covariance matrix $\diag(\frac{1}{3r_{s}},\frac{1-2\eta}{3r_{s}},\tilde{\sigma},\ldots,\tilde{\sigma})$
(and similarly $\mu_{2}$ with the variances of the first two coordinates
interchanged), and take an arbitrarily large value of $\tilde{\sigma}$
so the optimal algorithm will always choose the codewords to lie in
the two-dimensional subspace spanned by the first two coordinates.
The problem is then reduced to the $d=2$ case. 
\end{IEEEproof}

\subsection{The Proof of Theorem \ref{thm:Generalization NN surrogate}\label{subsec:Proof-of-Theorem}}

We will need several lemmas. The first lemma characterizes the continuity
of the surrogate loss function w.r.t. $(C,S)$. 
\begin{lem}
\label{lem: Lipschitzness of surrogate loss function}Suppose that
$C=\{x_{j}\}_{j\in[m]}\in{\cal C},\;\tilde{C}=\{\tilde{x}_{j}\}_{j\in[m]}\in{\cal C}$,
and $S,\tilde{S}\in{\cal S}$ satisfy that there exists $\gamma_{x},\gamma_{s}\geq0$
such that $\|x_{j}-\tilde{x}_{j}\|\leq\gamma_{x}$ for all $j\in[m]$
and $\|S-\tilde{S}\|_{\text{\emph{op}}}\leq\gamma_{s}$. Then, for
any $z\in\mathbb{B}^{d}(1)$
\begin{equation}
\left|\overline{\ell}(C,S,z)-\overline{\ell}(\tilde{C},\tilde{S},z)\right|\leq\max_{j\in[m]}\left|\overline{\ell}_{j}(C,S,z)-\overline{\ell}_{j}(\tilde{C},\tilde{S},z)\right|\leq8\gamma_{x}r_{s}(r_{x}+1)+\gamma_{s}(\gamma_{x}^{2}+4r_{x}).\label{eq: difference of loss functions vs. difference in codebook/decoder}
\end{equation}
\end{lem}
\begin{IEEEproof}
For any $v,\tilde{v}\in\mathbb{R}^{d}$, it holds that
\begin{align}
\left|v^{T}Sz-\tilde{v}^{T}\tilde{S}z\right| & \leq\left|v^{T}Sz-\tilde{v}^{T}Sz\right|+\left|\tilde{v}^{T}Sz-\tilde{v}^{T}\tilde{S}z\right|\\
 & =\left|(v-\tilde{v})^{T}Sz\right|+\left|\tilde{v}^{T}\left(S-\tilde{S}\right)z\right|\\
 & \leq\|v-\tilde{v}\|\cdot\|S\|_{\text{op}}\cdot\|z\|+\|\tilde{v}\|\cdot\|S-\tilde{S}\|_{\text{op}}\cdot\|z\|,
\end{align}
and
\begin{align}
\left|\|v\|_{S}^{2}-\|\tilde{v}\|_{\tilde{S}}^{2}\right| & \leq\left|\|v\|_{S}^{2}-\|\tilde{v}\|_{S}^{2}\right|+\left|\|\tilde{v}\|_{S}^{2}-\|\tilde{v}\|_{\tilde{S}}^{2}\right|\\
 & =\left|\left(S^{1/2}v\right)^{T}\left(S^{1/2}v\right)-\left(S^{1/2}\tilde{v}\right)^{T}\left(S^{1/2}\tilde{v}\right)\right|+\left|\tilde{v}^{T}(S-\tilde{S})\tilde{v}\right|\\
 & =\left|\left(S^{1/2}v+S^{1/2}\tilde{v}\right)^{T}\left(S^{1/2}v-S^{1/2}\tilde{v}\right)\right|+\left|\tilde{v}^{T}(S-\tilde{S})\tilde{v}\right|\\
 & \leq\|S\|_{\text{op}}\cdot\left(\|v\|+\|\tilde{v}\|\right)\|v-\tilde{v}\|+\|\tilde{v}\|^{2}\cdot\|S-\tilde{S}\|_{\text{op}},\label{eq: bounding quadratic diff}
\end{align}
where $S^{1/2}$ is the symmetric square root of $S$. Setting $v=x_{j}-x_{j'}$
and $\tilde{v}=\tilde{x}_{j}-\tilde{x}_{j'}$ the last two displays
and the triangle inequality imply that 
\begin{equation}
\left|(x_{j}-x_{j'})^{T}Sz-(\tilde{x}_{j}-\tilde{x}_{j'})^{T}\tilde{S}z\right|\leq2\gamma_{x}r_{s}+2r_{x}\gamma_{s}\label{eq: metric diff first type term bound}
\end{equation}
and 
\begin{equation}
\left|\|x_{j}-x_{j'}\|_{S}^{2}-\|\tilde{x}_{j}-\tilde{x}_{j'}\|_{\tilde{S}}^{2}\right|\leq8r_{x}\gamma_{x}r_{s}+\gamma_{x}^{2}\gamma_{s}\label{eq: metric diff second type term bound}
\end{equation}
for all $j,j'\in[m]$. We may now peel the difference between the
loss functions. Let us denote 
\[
a_{j,j'}(C,S,z)\dfn\left(\|x_{j}-x_{j'}\|_{S}^{2}+2(x_{j}-x_{j'})^{T}Sz\right).
\]
Then,
\begin{align}
\left|\overline{\ell}_{j}(C,S,z)-\overline{\ell}_{j}(\tilde{C},\tilde{S},z)\right| & \trre[\leq,a]\left|\min_{j'\in[m]\backslash\{j\}}a_{j,j'}(C,S,z)-\min_{j'\in[m]\backslash\{j\}}a_{j,j'}(\tilde{C},\tilde{S},z)\right|\\
 & \trre[\leq,b]\max_{j'\in[m]\backslash\{j\}}\left|a_{j,j'}(C,S,z)-a_{j,j'}(\tilde{C},\tilde{S},z)\right|\\
 & \leq\max_{j'\in[m]\backslash\{j\}}\left\{ \left|\|x_{j}-x_{j'}\|_{S}^{2}-\|\tilde{x}_{j}-\tilde{x}_{j'}\|_{\tilde{S}}^{2}\right|+\left|2(\tilde{x}_{j}-\tilde{x}_{j'})^{T}\tilde{S}z-2(x_{j}-x_{j'})^{T}Sz\right|\right\} \\
 & \trre[\leq,c]8\gamma_{x}r_{s}(r_{x}+1)+\gamma_{s}(\gamma_{x}^{2}+4r_{x}),
\end{align}
where $(a)$ holds since $t\mapsto[1-t]_{+}$ is a $1$-Lipschitz
function, $(b)$ holds since for any $\{\alpha_{j,j'}\},\{\tilde{\alpha}_{j,j'}\}$
\[
\left|\min_{j'\in[m]\backslash\{j\}}\alpha_{j,j'}-\min_{j'\in[m]\backslash\{j\}}\tilde{\alpha}_{j,j'}\right|\leq\max_{j'\in[m]\backslash\{j\}}\left|\alpha_{j,j'}-\tilde{\alpha}_{j,j'}\right|,
\]
and $(c)$ holds by utilizing (\ref{eq: metric diff first type term bound})
and (\ref{eq: metric diff second type term bound}). 
\end{IEEEproof}
We denote by $\N({\cal S},\|\cdot\|_{\text{op}},\gamma_{s})$ the
covering number (e.g. \cite[Definition 4.2.2]{vershynin2018high})
of ${\cal S}$, for the operator norm and covering radius $\gamma_{s}$. 
\begin{lem}
\label{lem: covering number of inverse covariance matrices}It holds
that 
\begin{equation}
\N({\cal S},\|\cdot\|_{\text{\emph{op}}},\gamma_{s})\leq2\left[\frac{12dr_{s}}{\gamma_{s}}\right]^{d(d+1)}.\label{eq: covering number bound of S}
\end{equation}
\end{lem}
\begin{IEEEproof}
Denote the eigendecomposition of $S\in\mathbb{S}_{+}^{d}$ by $S=U\Lambda U^{T}$,
where $U\in\mathbb{R}^{d\times d}$ is an orthonormal matrix and $\Lambda\in\mathbb{R}_{+}^{d\times d}$
is diagonal. Further let $u_{i}$ be the $i$th eigenvector of $S$
(i.e., the $i$th column of $U$), and $\lambda_{i}\in[0,r_{s}]$
be the $i$th eigenvalue (i.e., the $(i,i)$ element of $\Lambda$).
Then, using analogous notation for $\tilde{S}\in\mathbb{S}_{+}^{d}$,
\begin{align}
\|S-\tilde{S}\|_{\text{op}} & =\|U\Lambda U^{T}-\tilde{U}\tilde{\Lambda}\tilde{U}^{T}\|_{\text{op}}\\
 & \leq\|U\Lambda U^{T}-U\tilde{\Lambda}U^{T}\|_{\text{op}}+\|U\tilde{\Lambda}U^{T}-\tilde{U}\tilde{\Lambda}\tilde{U}^{T}\|_{\text{op}}\\
 & \leq\max_{i\in[d]}|\lambda_{i}-\tilde{\lambda}_{i}|+2d\max_{i\in[d]}\tilde{\lambda}_{i}\|u_{i}-\tilde{u}_{i}\|,\label{eq: bound on operator norm difference}
\end{align}
where in (\ref{eq: bound on operator norm difference}): The first
term is bounded as 
\[
\|U\Lambda U^{T}-U\tilde{\Lambda}U^{T}\|_{\text{op}}\leq\|\Lambda-\tilde{\Lambda}\|_{\text{op}}=\max_{i\in[d]}|\lambda_{i}-\tilde{\lambda}_{i}|.
\]
The second term is bounded as 
\[
\|U\tilde{\Lambda}U^{T}-\tilde{U}\tilde{\Lambda}\tilde{U}^{T}\|_{\text{op}}=\left\Vert \sum_{i=1}^{d}\tilde{\lambda}_{i}\left(u_{i}u_{i}^{T}-\tilde{u}_{i}\tilde{u}_{i}^{T}\right)\right\Vert _{\text{op}}\leq\sum_{i=1}^{d}\tilde{\lambda}_{i}\left\Vert \left(u_{i}u_{i}^{T}-\tilde{u}_{i}\tilde{u}_{i}^{T}\right)\right\Vert _{\text{op}}\leq\sum_{i=1}^{d}2\tilde{\lambda}_{i}\|u_{i}-\tilde{u}_{i}\|
\]
since for any $v\in\mathbb{S}^{d-1}$
\[
\left|v^{T}\left(u_{i}u_{i}^{T}-\tilde{u}_{i}\tilde{u}_{i}^{T}\right)v\right|=\left|u_{i}^{T}vv^{T}u_{i}-\tilde{u}_{i}^{T}vv^{T}\tilde{u}_{i}\right|=\left|\|u_{i}\|_{vv^{T}}^{2}-\|\tilde{u}_{i}\|_{vv^{T}}^{2}\right|\leq2\|u_{i}-\tilde{u}_{i}\|,
\]
with the last inequality follows as in (\ref{eq: bounding quadratic diff}).
Now, let $\epsilon=\frac{\gamma_{s}}{4dr_{s}}$ and let ${\cal U}$
be an $\epsilon$-net in the Euclidean distance for the unit sphere
$\mathbb{S}^{d-1}$ whose size is less than $|{\cal C}|\leq(\frac{3}{\epsilon})^{d}$
(whose existence is assured from \cite[Corollary 4.2.13]{vershynin2018high}).
Further, let $\epsilon_{0}=\frac{\gamma_{s}}{2}$, and let ${\cal L}$
be a proper $\epsilon_{0}$-net in the in the $\ell_{1}$ norm for
$[0,r_{s}]$ whose size is $|{\cal {\cal L}}|\leq\left(\frac{r_{s}}{\epsilon_{0}}\right).$
By (\ref{eq: bound on operator norm difference}), the set 
\[
\left\{ U\Lambda U^{T}\colon U^{T}U=I_{d},\;U=[u_{1,}\ldots,u_{d}],\;\Lambda=\diag\text{(\ensuremath{\lambda_{1},\ldots,\lambda_{d}})},\;u_{i}\in{\cal U\subset\mathbb{R}}^{d},\;\lambda_{i}\in{\cal L},\;\forall i\in[d]\right\} 
\]
 is a $\gamma_{s}$-cover of ${\cal S}$ whose size is $(|{\cal U}|\cdot|{\cal L}|)^{d}$
(where the $d$th power follows from allowing an independent choice
for $i\in[d]$), which is less the r.h.s. of (\ref{eq: covering number bound of S}). 
\end{IEEEproof}
We say that $\mathcal{K}\subset\mathbb{R}^{n}$ is a $\gamma$-cover
of $\mathcal{L}\subset\mathbb{R}^{n}$ under the metric $\|\cdot\|_{p}$,
$p\in[1,\infty]$, if 
\[
\sup_{l^{n}\in\mathcal{L}}\min_{k^{n}\in\mathcal{K}}\frac{1}{n}\sum_{i=1}^{n}|l_{i}-k_{i}|^{p}\leq\gamma^{p},
\]
where $l^{n}=(l_{1},\ldots,l_{n})$ and $k^{n}=(k_{1},\ldots,k_{n})$.
The $\gamma$-\emph{covering number of }$\mathcal{L}\subset\mathbb{R}^{n}$
under the metric $\|\cdot\|_{p}$ is denoted by 
\[
\N_{p}(\gamma,\mathcal{L})\dfn\min\left\{ |\mathcal{K}|\colon\mathcal{K}\text{ is a }\gamma\text{-cover of }\text{\ensuremath{\mathcal{L}}}\right\} .
\]
The next bound on the empirical Rademacher complexity (\ref{eq: empircal Rademacher complexity})
is well-known (e.g., \cite[Thm. 12.4]{rakhlin2012statistical}).
\begin{lem}[Dudley's entropy integral]
\label{lem: entropy integral} For ${\cal L}\subset\mathbb{R}^{n}$
\[
\Rad(\mathcal{L})\leq\inf_{\alpha\geq0}\left\{ 4\alpha+\frac{12}{\sqrt{n}}\int_{\alpha}^{1}\sqrt{\log\N_{2}\left(\gamma,\mathcal{L}\right)}\cdot\d\gamma\right\} .
\]
\end{lem}
With this bound we may obtain the following bound on the empirical
Rademacher complexity for the loss class induced by the surrogate
error probability loss function:
\begin{lem}
\label{lem: bounding Rad complexity} Let $\boldsymbol{z}=(z_{1},\ldots,z_{n})\in\mathbb{R}^{n}$
be given, and consider the loss class
\begin{equation}
\overline{\mathcal{L}}_{\boldsymbol{z}}\dfn\left\{ \overline{\ell}(C,S,\boldsymbol{z})\in\mathbb{R}_{+}^{n}\colon C\in{\cal C},\;S\in{\cal S}\right\} \subset\mathbb{R}_{+}^{n},\label{eq: surrogate loss class}
\end{equation}
where $\overline{\ell}(C,S,\boldsymbol{z})=(\overline{\ell}(C,S,z_{1}),\ldots\overline{\ell}(C,S,z_{n}))$.
Then, assuming $r_{x}\geq1$ and $r_{s}\geq1$:
\[
\Rad(\overline{\mathcal{L}}_{\boldsymbol{z}})\leq28\sqrt{\frac{(d\vee m)(d+1)\log(31\cdot dr_{s}r_{x})}{n}}.
\]
\end{lem}
\begin{IEEEproof}
We will bound the Rademacher complexity of the loss class induced
by the surrogate loss function in (\ref{eq: surrogate loss class}),
using covering arguments and Dudley's entropy integral. Assume that
$\gamma_{x}\in[0,1]$. Let $\tilde{{\cal C}}_{1}$ be a $\gamma_{x}$-net
of ${\cal C}_{1}$ in the Euclidean norm whose size is less than $|\tilde{{\cal C}}_{1}|\leq(\frac{3r_{x}}{\gamma_{x}})^{d}$,
and whose existence is assured from \cite[Corollary 4.2.13]{vershynin2018high},
and also let $\tilde{{\cal C}}=(\tilde{{\cal C}}_{1})^{m}$. In addition,
let ${\cal \tilde{S}}$ be a $\gamma_{s}$-net of ${\cal S}$ in the
operator norm whose size is less than $|\tilde{{\cal S}}|\leq2(\frac{12dr_{s}}{\gamma_{s}})^{d(d+1)}$,
whose existence is assured from Lemma \ref{lem: covering number of inverse covariance matrices}.
Then, by Lemma \ref{lem: Lipschitzness of surrogate loss function},
the set
\[
\tilde{\mathcal{L}}_{\boldsymbol{z}}\dfn\left\{ \overline{\ell}(C,S,\boldsymbol{z})\in\mathbb{R}_{+}^{n}\colon C\in\tilde{{\cal C}},\;S\in{\cal \tilde{S}}\right\} 
\]
is a $\gamma$-cover of $\overline{\mathcal{L}}_{\boldsymbol{z}}$
with 
\begin{equation}
\gamma=8\gamma_{x}r_{s}(r_{x}+1)+\gamma_{s}(\gamma_{x}^{2}+4r_{x})\label{eq: covering radius to minimze}
\end{equation}
whose size is less than 
\[
2\left[\frac{12d(r_{s}+1)}{\gamma_{s}}\right]^{d(d+1)}\cdot\left(\frac{3r_{x}}{\gamma_{x}}\right)^{m(d+1)}\leq\left[\frac{72\cdot d\cdot r_{s}r_{x}}{\gamma_{s}\gamma_{x}}\right]^{(d\vee m)(d+1)}
\]
We may next optimize over $(\gamma_{x},\gamma_{s})$ to achieve the
minimal covering size for any given $\gamma$, or alternatively, to
minimize $\gamma$ in (\ref{eq: covering radius to minimze}) under
the size constraint defined by $\psi\dfn\gamma_{x}\gamma_{s}>0$,
where we assume that $\psi<1$. Substituting $\gamma_{s}=\psi/\gamma_{x}$
in (\ref{eq: covering radius to minimze}) we obtain
\begin{align}
\gamma & \leq\min_{\gamma_{x}>0}\left\{ \gamma_{x}\left(8r_{s}(r_{x}+1)+\psi\right)+\frac{4r_{x}\psi}{\gamma_{x}}\right\} \\
 & =2\sqrt{\left(8r_{s}(r_{x}+1)+\psi\right)4r_{x}\psi}\\
 & \leq20r_{s}r_{x}\sqrt{\psi}
\end{align}
where the equality follows since the minimizer of $a\gamma_{x}+b/\gamma_{x}$
over $\gamma_{x}\geq0$ for $a,b\in\mathbb{R}_{+}$ is at $\gamma_{x}=\sqrt{b/a}$
and the minimal value is $2\sqrt{ab}$, and the (generous) inequality
using the assumptions $r_{x}\geq1$, $r_{s}\geq1$, and $\psi<1$.
Hence, under these assumptions, 
\[
\psi=\gamma_{x}\gamma_{s}\geq\frac{\gamma^{2}}{400r_{s}^{2}r_{x}^{2}}.
\]
Thus, for a given $\gamma>0$, the logarithm of the cover size is
upper bounded as
\[
(d\vee m)(d+1)\log\left[\frac{28800\cdot dr_{s}^{3}r_{x}^{3}}{\gamma^{2}}\right]=a-2(d\vee m)(d+1)\log\gamma
\]
where $a\dfn(d\vee m)(d+1)\log(28800\cdot dr_{s}^{3}r_{x}^{3})$.
By Lemma \ref{lem: entropy integral}, 
\begin{align}
\Rad(\overline{\mathcal{L}}_{\boldsymbol{z}}) & \leq\lim_{\alpha\to0}\frac{12}{\sqrt{n}}\int_{\alpha}^{1}\sqrt{a-2(d\vee m)(d+1)\log\gamma}\cdot\d\gamma\\
 & \leq12\sqrt{\frac{a}{n}}\lim_{\alpha\to0}\int_{\alpha}^{1}1-\frac{2(d\vee m)(d+1)\log\gamma}{2a}\cdot\d\gamma\\
 & =12\sqrt{\frac{a}{n}}\left[1+\frac{(d\vee m)(d+1)}{a}\right],
\end{align}
where the second inequality follows from $\sqrt{1-t}\leq1-t/2$ for
$t\in(-\infty,1]$, and the following equality using a continuity
argument implied by $\lim_{t\to0}t\cdot\log t=0$. The result follows
by inserting back the definition of $a$ and generously bounding using
$r_{x}\geq1$ and $r_{s}\geq1$, and then simplifying. 
\end{IEEEproof}
The proof of Theorem \ref{thm:Generalization NN surrogate} follows
immediately:
\begin{IEEEproof}[Proof of Theorem \ref{thm:Generalization NN surrogate}]
We use Prop. \ref{prop:Rademacher complexity uniform convergence bound}
and note that 
\begin{align}
\left|\overline{\ell}_{j}(C,S,z)\right| & \leq1\vee\max_{j'\in[m],j'\neq j}\left(\|x_{j}-x_{j'}\|_{S}^{2}+\left|2(x_{j}-x_{j'})^{T}Sz\right|\right)\\
 & \leq1\vee\left\{ 4r_{x}^{2}r_{s}+4r_{x}r_{s}\right\} \nfd r
\end{align}
Bounding the Rademacher complexity using Lemma \ref{lem: bounding Rad complexity}
completes the proof. 
\end{IEEEproof}

\subsection{The Proofs of Theorems \ref{thm:Generalization and empirical error for Gibbs}
and \ref{thm: High probability bound on Gibbs algorithm}\label{subsec:Proof-of-Theorems}}

We will utilize the following lemma regarding the KL divergence.
\begin{lem}
\label{lem: KL for Gibbs}Let $Q$ be a measure and let $l,\tilde{l}\colon{\cal X}\mapsto\mathbb{R}$
be measurable functions on a measurable space ${\cal X}$. Further
let $P(\d x)\propto Q(\d x)\cdot e^{-\beta l(x)}$ (resp. $\tilde{P}(\d x)\propto Q(\d x)\cdot e^{-\beta\tilde{l}(x)}$)
be the Gibbs measures based on the loss functions $l$ (resp. $\tilde{l}$).
If $\|l-\tilde{l}\|_{\infty}\leq\epsilon$ then $\dkl(P||\tilde{P})\leq2\beta\epsilon\cdot\{1\wedge(e^{2\beta\epsilon}-1)\}$. 
\end{lem}
\begin{IEEEproof}
For any $f,g\colon{\cal X}\mapsto\mathbb{R}^{+}$ it holds that $\frac{\E_{Q}\left[f\right]}{\E_{Q}\left[g\right]}\leq\left\Vert \frac{f}{g}\right\Vert _{\infty}$where
$\|\cdot\|_{\infty}$ is the sup norm. Hence, 
\[
\left|\log\frac{P(\d x)}{\tilde{P}(\d x)}\right|=\log\left\{ \frac{e^{-\beta l(x)}}{e^{-\beta\tilde{l}(x)}}\cdot\frac{\E_{Q}\left[e^{-\beta\tilde{l}(x)}\right]}{\E_{Q}\left[e^{-\beta l(x)}\right]}\right\} \leq2\beta\epsilon,
\]
and so it holds that $\dkl(P||\tilde{P})=\E_{P}[\log\frac{P(\d x)}{\tilde{P}(\d x)}]\leq2\beta\epsilon$.
Furthermore, by \cite[Lemma 3.18]{dwork2014algorithmic}, it also
holds that $\dkl(P||\tilde{P})\leq2\beta\epsilon\cdot(e^{2\beta\epsilon}-1)$. 
\end{IEEEproof}
\begin{IEEEproof}[Proof of empirical error bound of Theorem \ref{thm:Generalization and empirical error for Gibbs}
\textendash{} (\ref{eq: bound on the average empirical error})]
 Let $\dtv(\mu||\tilde{\mu})$ denote the total variation distance
between the probability measures $\mu$ and $\tilde{\mu}$. By the
variational representation of the total variation and Pinsker's inequality
\cite[Lemma 2.5]{tsybakov2008introduction}, for any two measures
$\mu,\tilde{\mu}$ and a function $f\colon{\cal X}\mapsto\mathbb{R}$
with $\|f\|_{\infty}\leq1/2$ it holds that 
\begin{equation}
\left|\E_{X\sim\mu}[f(X)]-\E_{X\sim\tilde{\mu}}[f(X)]\right|\leq\dtv(\mu||\tilde{\mu})\leq\sqrt{\frac{\dkl(\mu||\tilde{\mu})}{2\log e}}.\label{eq: bound on average under different measure}
\end{equation}
We further denote by $P_{C_{\mu,1}\cdots C_{\mu,T}}$ (resp. $P_{C_{\boldsymbol{z},1}\cdots C_{\boldsymbol{z},T}}$)
the joint probability measure of $\boldsymbol{C}_{\mu}$ (resp. $\boldsymbol{C}_{\boldsymbol{Z}}$),
and use standard notation for their conditional versions. Furthermore,
we let $\mathfrak{C}$ be the set of all possibly expurgated codebooks
over all $T$ steps, i.e., the initial codebook $C_{0}$, and each
of the ${m_{0} \choose m_{0}-kt}$ codebooks at the $t$th stage,
for any $t\in[T]$. There is a total of 
\[
1+{m_{0} \choose m_{0}-k}+{m_{0} \choose m_{0}-2k}+\cdots{m_{0} \choose m}\trre[\leq,*]T{m_{0} \choose m_{0}/2}\leq\frac{m_{0}^{m_{0}/2+1}}{k}\dfn\alpha_{m_{0},k}
\]
such different codebooks, where $(*)$ follows from the assumption
$m_{0}\ge2m$. For any $\delta\in(0,1)$, define $\epsilon=\sqrt{\frac{1}{2n}\log\frac{\alpha_{m_{0},k}}{\delta}}$
and the event
\[
E\dfn\left\{ \boldsymbol{z}\colon\sup_{C\in\mathfrak{C}}\left|\pe_{\boldsymbol{z}}(C)-\pe_{\mu}(C)\right|<\epsilon\right\} .
\]
Hoeffding's inequality  and the union bound then imply that $\P[E]\geq1-\delta$.
For $\boldsymbol{z}\in E$, the definitions of the Gibbs algorithms
(population and empirical versions) and Lemma \ref{lem: KL for Gibbs}
imply that for any $t\in[T]$ and $C\in\mathfrak{C}$
\begin{equation}
\dkl\left(P_{C_{\boldsymbol{z},t}|C_{\boldsymbol{z},t-1}}(\cdot\mid C)||P_{C_{\mu,t}|C_{\mu,t-1}}(\cdot\mid C)\right)\leq2\beta\epsilon\cdot\left(1\wedge(e^{2\beta\epsilon}-1)\right),\label{eq: bound on KL for Gibbs measures}
\end{equation}
and for $\boldsymbol{z}\in E^{c}$ and any $C\subset(\mathbb{R}^{d})^{m_{t-1}}$
\begin{equation}
\dkl\left(P_{C_{\boldsymbol{z},t}|C_{\boldsymbol{z},t-1}}(\cdot\mid C)||P_{C_{\mu,t}|C_{\mu,t-1}}(\cdot\mid C)\right)\leq2\beta\cdot\left(1\wedge(e^{2\beta}-1)\right).\label{eq: bound on KL for Gibbs measures E complement}
\end{equation}
Condition on any given $\boldsymbol{Z}=\boldsymbol{z}\in E$ and $C_{0}$
(implicitly) and averaging over the randomness of the Gibbs algorithm
\begin{align}
\left|\E\left[\pe_{\mu}(C_{\boldsymbol{z},T})-\pe_{\mu}(C_{\mu,T})\mid\boldsymbol{Z}=\boldsymbol{z}\right]\right| & \trre[\leq,a]\sqrt{\frac{\dkl(P_{C_{\boldsymbol{z},T}}||P_{C_{\mu,T}})}{2\log e}}\\
 & \leq\sqrt{\frac{\dkl(P_{C_{\boldsymbol{z},1}\cdots C_{\boldsymbol{z},T}}||P_{C_{\mu,1}\cdots C_{\mu,T}})}{2\log e}}\\
 & \trre[=,b]\sqrt{\frac{\sum_{t=0}^{T-1}\E\left[\dkl\left(P_{C_{\boldsymbol{z},t}|C_{\boldsymbol{z},t-1}}(\cdot\mid C_{\boldsymbol{z},t-1})||P_{C_{\mu,t}|C_{\mu,t-1}}(\cdot\mid C_{\boldsymbol{z},t-1})\right)\right]}{2\log e}}\\
 & \trre[\leq,c]\sqrt{\frac{2T\beta\epsilon\cdot\left(1\wedge(e^{2\beta\epsilon}-1)\right)}{2\log e}},
\end{align}
where $(a)$ follows from (\ref{eq: bound on average under different measure})
(which holds since $\pe_{\mu}(C)\in[0,1]$), $(b)$ follows from the
chain rule and the Markov property ($P_{C_{\boldsymbol{z},1}\cdots C_{\boldsymbol{z},T}}=P_{C_{\boldsymbol{z},1}}\otimes P_{C_{\boldsymbol{z},2}\mid C_{\boldsymbol{z},1}}\cdots\otimes P_{C_{\boldsymbol{z},T}\mid C_{\boldsymbol{z},T-1}}$)
of $\boldsymbol{C}_{\boldsymbol{z}}$ conditioned on $\boldsymbol{z}$
and a similar property which holds for $\boldsymbol{C}_{\mu}$, and
$(c)$ follows from utilizing (\ref{eq: bound on KL for Gibbs measures}).
A similar bound holds for $\boldsymbol{z}\in E^{c}$ with $\epsilon$
replaced by $1$. Then, 
\begin{align}
\E\left[\pe_{\mu}(C_{\boldsymbol{z},T})-\pe_{\mu}(C_{\mu,T})\right] & =\P[\boldsymbol{Z}\in E]\cdot\E\left[\left(\pe_{\mu}(C_{\boldsymbol{z},T})-\pe_{\mu}(C_{\mu,T})\right)\mid\boldsymbol{Z}\in E\right]\nonumber \\
 & \hphantom{=}+\P[\boldsymbol{Z}\in E^{c}]\cdot\E\left[\left(\pe_{\mu}(C_{\boldsymbol{z},T})-\pe_{\mu}(C_{\mu,T})\right)\mid\boldsymbol{Z}\in E^{c}\right]\\
 & \leq\sqrt{\frac{T\beta}{\log e}}\cdot\left(\sqrt{\epsilon(e^{2\beta\epsilon}-1})+\delta\right).
\end{align}
The proof of the bound is completed by choosing $\delta=\frac{1}{n}$,
the assumption in (\ref{eq: first assumption on beta}) implies that
$2\beta\epsilon\leq1$ and using $e^{x}-1\leq(e-1)x$ for $x\in[0,1]$
we obtain
\begin{align}
\E\left[\pe_{\mu}(C_{\boldsymbol{z},T})-\pe_{\mu}(C_{\mu,T})\right] & \leq\sqrt{\frac{T\beta}{\log e}}\cdot\left(\sqrt{(e-1)\beta\frac{\log(n\alpha_{m_{0},k})}{n}}+\frac{1}{n}\right)\\
 & \leq\sqrt{\frac{4(e-1)}{\log e}\cdot\frac{T\beta^{2}\log(n\alpha_{m_{0},k})}{n}}.
\end{align}
\end{IEEEproof}
We next prove the bounds on the average generalization error. To this
end we note that a randomized learning algorithm $A\colon\boldsymbol{Z}\mapsto{\cal A}$
is conditional probability distribution $P_{A\mid\boldsymbol{Z}}$,
where ${\cal A}$ is the set of possible outputs which is assumed
to be a measurable space. As was shown in \cite{raginsky2016information,xu2017information,raginsky2020information},
the generalization error of learning algorithms can be controlled
by the mutual information between the output hypothesis $A\in{\cal A}$
and the input data, and such a bound is used here. 
\begin{IEEEproof}[Proof of generalization error bound of Theorem \ref{thm:Generalization and empirical error for Gibbs}
\textendash{} (\ref{eq: bound on the average generalization bound})]
 For brevity, let $C_{\boldsymbol{Z},t}$ be denoted here as $C_{t}$.
We use standard information-theoretic notation \cite{cover2012elements}
for entropy and mutual information. Recall that the \emph{erasure
mutual information }\cite{verdu2008information} between $U$ and
$\boldsymbol{V}=(V_{1},\ldots,V_{n})$ is defined as
\[
I^{-}(U;\boldsymbol{V})\dfn\sum_{i=1}^{n}I(U;V_{i}\mid\boldsymbol{V}^{-i})
\]
where $\boldsymbol{V}^{-i}=(V_{1},\ldots,V_{i-1},V_{i+1},\ldots,V_{n})$.
To bound the generalization error we analyze the information-theoretic
stability properties of the Gibbs algorithm, and specifically bound
the mutual information $I(C_{T};\boldsymbol{Z})$. First note the
the expurgation algorithm can be described as an adaptive composition
of algorithms. Specifically, at step $t$, the output of the algorithm
$C_{t}$ is a function of $C_{t-1}$ and $\boldsymbol{Z}$ and so
the Markov relation $(C_{1},\ldots,C_{t-1})-(C_{t},\boldsymbol{Z})-C_{t+1}$
holds. We then have
\begin{align}
I(C_{T};\boldsymbol{Z}) & \trre[\leq,a]I^{^{-}}(C_{T};\boldsymbol{Z})\\
 & =\sum_{i=1}^{n}I(C_{T};Z_{i}\mid\boldsymbol{Z}^{-i})\\
 & \trre[\leq,b]\sum_{i=1}^{n}I(C_{1},\ldots,C_{T};Z_{i}\mid\boldsymbol{Z}^{-i})\\
 & =\sum_{i=1}^{n}\sum_{t=1}^{T}I(C_{t};Z_{i}\mid\boldsymbol{Z}^{-i},C_{t-1},,\ldots,C_{1})\\
 & =\sum_{i=1}^{n}\sum_{t=1}^{T}H(C_{t}\mid\boldsymbol{Z}^{-i},C_{t-1},\ldots,C_{1})-H(C_{t}\mid\boldsymbol{Z},C_{t-1},\ldots,C_{1})\\
 & \trre[\leq,c]\sum_{i=1}^{n}\sum_{t=1}^{T}H(C_{t}\mid\boldsymbol{Z}^{-i},C_{t-1})-H(C_{t}\mid\boldsymbol{Z},C_{t-1})\\
 & =\sum_{t=1}^{T}I^{^{-}}(C_{t};\boldsymbol{Z}\mid C_{t-1})\label{eq: upper bound on MI}
\end{align}
where $(a)$ follows since $\boldsymbol{Z}$ is comprised of $n$
independent random variables and \cite[Prop. 1]{raginsky2020information},
$(b)$ follows from the chain rule and non-negativity of mutual information,
and $(c)$ follows from the Markov property and since conditioning
reduces entropy. We thus upper bound the erasure mutual information
$I^{^{-}}(C_{t};\boldsymbol{Z}\mid C_{t-1})$. As was shown in the
proof of Lemma \ref{lem: KL for Gibbs}, for any $\boldsymbol{z},\boldsymbol{z}'\in(\mathbb{R}^{d})^{n}$
such that $\Hamd(\boldsymbol{z},\boldsymbol{z}')=1$, 
\[
\left|\log\frac{\P\left[C_{t+1}\mid\boldsymbol{Z},C_{t}\right]}{\P\left[C_{t+1}\mid\boldsymbol{Z}',C_{t}\right]}\right|\leq\frac{2\beta}{n}.
\]
Derivation similar to the proof of \cite[Thm. 4]{raginsky2016information}
(see also \cite[Thm. 7]{raginsky2020information}) then implies that
$I^{-}(C_{t};\boldsymbol{Z}\mid C_{t-1})\leq2\beta\wedge\frac{\beta^{2}}{2n}.$
As this holds for all $t\in[T]$, (\ref{eq: upper bound on MI}) thus
implies that $I(C_{T};\boldsymbol{Z})\leq T\cdot(2\beta\wedge\frac{\beta^{2}}{2n})$
and the Gibbs algorithm is stable in the mutual information. Since
the loss function is the error probability that is bounded to $[0,1]$
and so is $\frac{1}{4}$-sub-Gaussian, it follows from \cite[Thm. 1]{xu2017information}
that 
\[
\E\left[\pe_{\mu}(C_{\boldsymbol{Z},T})-\pe_{\boldsymbol{Z}}(C_{\boldsymbol{Z},T})\right]\leq\sqrt{T\left(\frac{\beta}{n}\wedge\frac{\beta^{2}}{4n^{2}}\right)}.
\]
\end{IEEEproof}
To prove the high probability bound of Theorem \ref{thm: High probability bound on Gibbs algorithm}
we recall that a learning algorithm is termed\emph{ $(\varepsilon,\eta)$-differentially-private}
\cite{dwork2014algorithmic} if for any measurable set $F\subseteq{\cal A}$
\[
\Hamd(\boldsymbol{z},\boldsymbol{z}')\leq1\Rightarrow P_{A\mid\boldsymbol{Z}=\boldsymbol{z}}(F)\leq e^{\varepsilon}P_{A\mid\boldsymbol{Z}=\boldsymbol{z}'}(F)+\eta.
\]
Let $C_{0}\subset{\cal C}$ be a super-codebook of size $m_{0}$,
and let ${\cal A}_{t}$ be the set of its subsets of size $m_{t}=m_{0}-kt$,
and $T=\frac{m_{0}-m}{k}$ for some given $k$ (assuming $T$ is integer).
For brevity, we suppress $C_{0}$ from the notation, as it is assumed
that $C_{0}$ is fixed in advance, and in accordance, the high probability
bound of Theorem \ref{thm: High probability bound on Gibbs algorithm}
only refers to the random draw of the noise samples $\boldsymbol{Z}$.
The Gibbs algorithm $C_{\text{G}}\colon(\mathbb{R}^{d})^{n}\mapsto{\cal A}_{T}$
is defined by the sequence of algorithms $\{C_{t}\}_{t=1}^{T}$ in
(\ref{eq: Gibbs algorithm}), as 
\[
P_{C_{\text{G}}\mid\boldsymbol{z},C_{0}}=P_{C_{1}\mid\boldsymbol{z}}\otimes P_{C_{2}\mid C_{1},\boldsymbol{z}}\otimes P_{C_{3}\mid C_{2},\boldsymbol{z}}\otimes\cdots\otimes P_{C_{T}\mid C_{T-1},\boldsymbol{z}}.
\]

\begin{lem}
\label{lem: Gibbs algorithm is differntialy stable}$C_{G}$ is a
$(\frac{T\beta}{n},0)$-differentially private algorithm. Furthermore,
if $\frac{2\beta}{n}\leq1$, then for any $\eta>0$, $C_{G}$ is an
$(\varepsilon,\eta)$-differentially private algorithm, where 
\[
\varepsilon=\sqrt{2T\log\left(\frac{1}{\eta}\right)}\cdot\frac{\beta}{n}+4(e-1)\frac{T\beta^{2}}{n^{2}}.
\]
\end{lem}
\begin{IEEEproof}
For any codebook $C$ and $\boldsymbol{z},\boldsymbol{z}'\in(\mathbb{R}^{d})^{n}$
such that $\Hamd(\boldsymbol{z},\boldsymbol{z}')=1$ it holds that
\[
\left|\pe_{\boldsymbol{z}}(C|j)-\pe_{\boldsymbol{z}'}(C|j)\right|\leq\frac{1}{n}
\]
and so also 
\[
\left|\pe_{\boldsymbol{z}}(C)-\pe_{\boldsymbol{z}'}(C)\right|=\left|\frac{1}{m}\sum_{j\in[m]}\pe_{\boldsymbol{z}}(C\mid j)-\pe_{\boldsymbol{z}'}(C\mid j)\right|\leq\frac{1}{n}.
\]
This bound along with the definition of the Gibbs algorithm (\ref{eq: Gibbs algorithm})
imply that (as in the proof of Lemma \ref{lem: KL for Gibbs}) 
\[
\left|\log\frac{\P\left[C_{t+1}\mid\boldsymbol{z},C_{t}\right]}{\P\left[C_{t+1}\mid\boldsymbol{z}',C_{t}\right]}\right|\leq\frac{2\beta}{n}.
\]
Thus, the algorithm $P_{C_{t+1}\mid\boldsymbol{z},C_{t}}$ is $(\frac{2\beta}{n},0)$-differentially
private for all $t\in[T]$. The algorithm $C_{\text{G}}$ is an \emph{adaptive
composition} \cite[Sec. 2.1]{dwork2015generalization} of the algorithms
$(C_{1},C_{2},\ldots,C_{T})$, for which, in general, $C_{t}$ depends
on the data $\boldsymbol{z}$ and the previous outputs $(C_{1},\ldots,C_{t-1})$
(in this case it only depends on $C_{t-1}$ in a nontrivial way).
By the simple composition theorem (e.g., \cite[Thm. 3]{dwork2015generalization}),
$C_{\text{G}}$ is $(\frac{2\beta T}{n},0)$-differentially private.
Furthermore, under the assumption $\frac{2\beta}{n}<1$, the advanced
composition theorem \cite[Thm. 4]{dwork2015generalization} implies
that for any chosen $\eta>0$, $C_{\text{G}}$ is $(\varepsilon,\eta)$-differentially
private with 
\[
\varepsilon=\sqrt{2T\log\left(\frac{1}{\eta'}\right)}\cdot\frac{\beta}{n}+T\frac{2\beta}{n}(e^{2\beta/n}-1)\leq\sqrt{2T\log\left(\frac{1}{\eta'}\right)}\cdot\frac{\beta}{n}+4(e-1)\frac{T\beta^{2}}{n^{2}}
\]
where the inequality follows from $e^{x}-1\leq(e-1)x$ for $x\in[0,1]$. 
\end{IEEEproof}
We may now turn to the proof of Theorem \ref{thm: High probability bound on Gibbs algorithm}:
\begin{IEEEproof}[Proof of Theorem \ref{thm: High probability bound on Gibbs algorithm}]
If $P_{C\mid\boldsymbol{Z}}$ is $(\varepsilon,\eta)$-differentially
private then for any given $\boldsymbol{z},\boldsymbol{z}'\in(\mathbb{R}^{d})^{n}$
such that $\Hamd(\boldsymbol{z},\boldsymbol{z}')=1$ it can be verified
by standard approximations (or, e.g., \cite[Lemma 6]{dwork2015preserving})
that for any $\tilde{z}\in\mathbb{R}^{d}$
\[
\left|\qe_{\tilde{z}}(C(\boldsymbol{z}))-\qe_{\tilde{z}}(C(\boldsymbol{z}'))\right|\leq e^{\varepsilon}-1+\eta.
\]
Thus, the algorithm $P_{C|\boldsymbol{Z}}$ is $\gamma$-uniformly
stable w.r.t. the loss function $\qe_{\tilde{z}}(\cdot)\in[0,1]$,
with $\gamma=e^{\varepsilon}-1+\eta$. Then, \cite[Thm 1.1]{feldman2019high}
implies that there exists an absolute constant $c>0$ such that 
\begin{equation}
\P\left[\qe_{\mu}(C_{\boldsymbol{Z},T})-\qe_{\boldsymbol{Z}}(C_{\boldsymbol{Z},T})>c\left(\gamma+\frac{1}{\sqrt{n}}\right)\log(n)\cdot\log\frac{n}{\delta}\right]\leq\delta\label{eq: high probability bound for uniformly stable}
\end{equation}
where $\qe_{\mu}(C_{\boldsymbol{Z},T})\dfn\E[\qe_{\tilde{Z}}(C(\boldsymbol{Z}))]$
where $\tilde{Z}\sim\mu$ and independent of $\boldsymbol{Z}$, and
where $\qe_{\boldsymbol{Z}}(C_{\boldsymbol{Z},T})\dfn\frac{1}{n}\sum_{i=1}^{n}\qe_{\boldsymbol{Z}_{i}}(C(\boldsymbol{Z}))$.
By Lemma \ref{lem: Gibbs algorithm is differntialy stable}, for the
Gibbs algorithm $C_{\text{G}}$, a valid uniform stability parameter
is 
\[
\gamma=\exp\left[\sqrt{2T\log\left(\frac{1}{\eta}\right)}\cdot\frac{\beta}{n}+4(e-1)\frac{T\beta^{2}}{n^{2}}\right]-1+\eta
\]
for any $\eta>0$. Under the assumption $\frac{T\beta}{n}\to0$ as
$n\to\infty$, there exists $n_{0}$ such that for all $n>n_{0}$
\[
\gamma\leq(e-1)\sqrt{2T\log\left(\frac{1}{\eta}\right)}\cdot\frac{\beta}{n}+4(e-1)^{2}\frac{T\beta^{2}}{n^{2}}+\eta.
\]
Choosing $\eta=\frac{\sqrt{T}\beta}{n}$, there exists $n_{1}$ such
that for all $n>n_{1}$
\[
\gamma\leq\sqrt{18\log\left(\frac{n}{\beta\sqrt{T}}\right)}\cdot\frac{\sqrt{T}\beta}{n}.
\]
Inserting into (\ref{eq: high probability bound for uniformly stable})
completes the proof.
\end{IEEEproof}

\subsection{The Proofs of Theorems \ref{thm: learning input distributions} and
\ref{thm: uniform convergence for learning weights}}

To prove Theorem \ref{thm: learning input distributions}, we need
a few supporting lemmas. Note that $\|Z\|$ has a density $\mu_{\|Z\|}$,
and let $\overline{\alpha}_{\mu_{\|Z\|}}\dfn\inf\{\alpha\colon\mu_{\|Z\|}([0,\alpha])=1\}\in(0,\infty]$.
For any given $\alpha<\overline{\alpha}_{\mu_{\|Z\|}}$, let $U$
be the following ``compressed norm'' version of the noise to maximal
norm $\alpha$, concretely
\begin{equation}
U_{\alpha}=\begin{cases}
Z, & \|Z\|\leq\alpha\\
\frac{Z}{\|Z\|}\cdot V & \text{otherwise}
\end{cases}\label{eq: truncated random variable with density}
\end{equation}
where $Z\ind V\sim\text{Uniform}[0,\alpha]$. The Wasserstein distance
from the density $\mu_{U_{\alpha}}$ to $\mu_{Z}$ satisfies the following
bound:
\begin{lem}
\label{lem: Wasserstein loss due to truncation}Assume that $\|Z\|$
is $\sqrt{d}\sigma_{Z}$-sub-Gaussian. Then, for any $\alpha\in(0,\overline{\alpha}_{\mu_{\|Z\|}})$
\[
W_{2}(\mu_{Z},\mu_{U_{\alpha}})\leq\sqrt{2}(\alpha+\sqrt{d}\sigma_{Z})\cdot e^{-\alpha^{2}/2d\sigma_{Z}^{2}}.
\]
\end{lem}
\begin{IEEEproof}
Let $(Z^{*},U^{*})$ be the coupling defined by (\ref{eq: truncated random variable with density}).
Then, 
\begin{align}
W_{2}^{2}(\mu_{Z},\mu_{U}) & \leq\E\|U^{*}-Z^{*}\|^{2}\\
 & \leq\E\left[\|Z\|^{2}\cdot\I\{\|Z\|>\alpha\}\right]\\
 & \trre[=,a]\int_{0}^{\infty}\P\left[\|Z\|^{2}\cdot\I\{\|Z\|>\alpha\}>t\right]\d t\\
 & =\int_{0}^{\infty}\P\left[\|Z\|>\alpha\vee\sqrt{t}\right]\d t\\
 & =\int_{0}^{\alpha^{2}}\P\left[\|Z\|>\alpha\right]\d t+\int_{\alpha^{2}}^{\infty}\P\left[\|Z\|>\sqrt{t}\right]\d t\\
 & \trre[\leq,b]\alpha^{2}e^{-\alpha^{2}/d\sigma_{Z}^{2}}+2\int_{\alpha^{2}}^{\infty}e^{-t/d\sigma_{Z}^{2}}\d t\\
 & =\alpha^{2}e^{-\alpha^{2}/d\sigma_{Z}^{2}}+2d\sigma_{Z}^{2}e^{-\alpha^{2}/d\sigma_{Z}^{2}},
\end{align}
where $(a)$ follows from the integral identity (e.g., \cite[Lemma 1.2.1]{vershynin2018high}),
and $(b)$ follows from the assumption that $\|Z\|$ is $(\sqrt{d}\sigma_{Z})$-sub-Gaussian.
\end{IEEEproof}
We next bound the Wasserstein distance $W_{2}(\mu_{Z},\mu_{\hat{Z}_{n}})$
where $\mu_{\hat{Z}_{n}}=\sum_{i=1}^{n}\frac{1}{n}\delta_{Z_{i}}$
is the empirical measure of the noise samples $\boldsymbol{Z}$. For
the case in which $\|Z\|\leq\alpha$ almost surely, \cite[proof of Thm. 3.1]{lee2019learning}
obtained a high probability bound on $W_{1}(\mu_{Z},\mu_{\hat{Z}_{n}})$
utilizing an upper bound on its expected value from \cite{dereich2013constructive}
and McDiarmid\textquoteright s inequality. We use next similar reasoning
and a truncation argument for the more general case which only assumes
sub-Gaussian $\|Z\|$, and also bound the second-order distance rather
than the first-order.
\begin{lem}
\label{lem: Wasserstein distance of noise}Let 
\[
f_{d}(n)\dfn\begin{cases}
n^{-1/4}, & d<4\\
n^{-1/4}\cdot\log n, & d=4\\
n^{-1/d}, & d>4
\end{cases}.
\]
Assume that $\|Z\|$ is $\sqrt{d}\sigma_{Z}$-sub-Gaussian. For any
given $\delta\in(0,1)$, there exists an constant $c_{d}>0$ which
only depends on $d$ and $n_{0}(\sigma_{Z},d)$ such that for any
$n\geq n_{0}$ 
\[
W_{2}(\mu_{Z},\mu_{\hat{Z}_{n}})\leq c_{d}'\sigma_{Z}\log\frac{n}{\delta}\cdot f_{d}(n)\leq c_{d}'\sigma_{Z}\log^{2}\left(\frac{n}{\delta}\right)\cdot n^{-1/(d\vee4)}
\]
with probability larger than $1-\delta$.
\end{lem}
\begin{IEEEproof}
For a given $\alpha>0$, let $U_{\alpha}$ be as in (\ref{eq: truncated random variable with density})
and let $\mu_{U}$ denote its probability measure, where for brevity,
we omit henceforth in the proof the subscript $\alpha$. Also let
$\boldsymbol{U}=(U_{1},\ldots,U_{n})\stackrel{\tiny\mathrm{i.i.d.}}{\sim}\mu_{U}$
and let $\mu_{\hat{U}_{n}}\dfn\frac{1}{n}\sum_{i=1}^{n}\delta_{U_{i}}$
be its empirical measure. 

We first derive a high probability bound on $W_{2}(\mu_{U},\mu_{\hat{U}_{n}})$.
To this end, we utilize the fact that $\|U\|\leq\alpha$ almost surely,
and use a bound on the average Wasserstein distance \cite{fournier2015rate}
as follows:
\begin{align}
\E\left[W_{2}(\mu_{U},\mu_{\hat{U}_{n}})\right] & \trre[\leq,a]\sqrt{\E\left[W_{2}^{2}(\mu_{U},\mu_{\hat{U}_{n}})\right]}\\
 & =\trre[\leq,b]c_{d}\cdot\alpha\begin{cases}
n^{-1/4}, & d<4\\
n^{-1/4}\cdot\log n, & d=4\\
n^{-1/d}, & d>4
\end{cases}\\
 & =c_{d}\alpha\cdot f_{d}(n),\label{eq: average Wasserstein bound}
\end{align}
where $(a)$ follows from Jensen's inequality, and $(b)$ follows
directly from \cite[Thm. 1]{fournier2015rate} by taking, in the notation
there, $p=2$, $q\to\infty$, and $c_{d}$ is a constant which only
depends on $d$.

Now, $\boldsymbol{U}\mapsto W_{2}(\mu_{U},\mu_{\hat{U}_{n}})$ satisfies
a bounded difference inequality with parameter $\frac{2\sqrt{d}\alpha}{n}$,
just as was noticed in \cite[proof of Thm. 3.1]{lee2019learning}
for the first-order Wasserstein distance. Hence, by McDiarmid\textquoteright s
inequality \cite[Thm. 6.2]{boucheron2013concentration}
\begin{equation}
\P\left[W_{2}(\mu_{U},\mu_{\hat{U}_{n}})-\E\left[W_{2}(\mu_{U},\mu_{\hat{U}_{n}})\right]>\alpha\sqrt{\frac{2d}{n}\log\frac{2}{\delta}}\right]\leq\frac{\delta}{2},\label{eq: McDiaramid on Wasserstein}
\end{equation}
and so (\ref{eq: average Wasserstein bound}) and (\ref{eq: McDiaramid on Wasserstein})
imply that with probability larger than $1-\delta/2$ 
\begin{equation}
W_{2}(\mu_{U},\mu_{\hat{U}_{n}})\leq c_{d}\alpha\cdot f_{d}(n)+\alpha\sqrt{\frac{2d}{n}\log\frac{2}{\delta}}.\label{eq: bound on Wasserstein between truncated and empirical}
\end{equation}
We now choose $\alpha>0$ to be large enough so that $W_{2}(\mu_{\hat{U}_{n}},\mu_{\hat{Z}_{n}})=0$
with probability larger than $1-\delta/2$. To this end, consider
the coupling $(Z_{i},U_{i})$ defined by (\ref{eq: truncated random variable with density})
for all $i\in[n]$. Then, 
\[
W_{2}(\mu_{\hat{U}_{n}},\mu_{\hat{Z}_{n}})\leq\sqrt{\frac{1}{n}\sum_{i=1}^{n}\|Z_{i}-U_{i}\|^{2}}.
\]
So, if $Z_{i}=U_{i}$ for all $i\in[n]$ then this upper bound is
zero. Since $\P[\|Z\|>\sqrt{d}\alpha]\leq\exp[-\frac{c_{0}\alpha^{2}}{\sigma_{Z}^{2}}]$
for some absolute constant $c_{0}>0$, it holds for $\alpha=\frac{\sigma_{Z}}{\sqrt{c_{0}}}\sqrt{d\log\frac{2n}{\delta}}$
that
\begin{equation}
\P\left[\bigcap_{i=1}^{n}\{Z_{i}=U_{i}\}\right]=\P\left[\bigcap_{i=1}^{n}\left\{ \|Z_{i}\|\leq\alpha\right\} \right]\geq\left[1-e^{-\frac{c_{0}\alpha^{2}}{d\sigma_{Z}^{2}}}\right]^{n}\geq1-\frac{\delta}{2}\label{eq: no truncation with high probability}
\end{equation}
(the last inequality can be verified using the basic inequality $x^{r}\leq1+r(x-1)$
for $x=1-\delta/2>0$ and $r=\frac{1}{n}\in[0,1]$). 

Consequently, combining the triangle inequality and Lemma \ref{lem: Wasserstein loss due to truncation},
(\ref{eq: bound on Wasserstein between truncated and empirical}),
and (\ref{eq: no truncation with high probability}), that there exists
$n_{0}(\sigma_{Z},d)$ such that for all $n\geq n_{0}$ 
\begin{align}
W_{2}(\mu_{Z},\mu_{\hat{Z}_{n}}) & \leq W_{2}(\mu_{Z},\mu_{U_{\alpha}})+W_{2}(\mu_{U_{\alpha}},\mu_{\hat{U}_{\alpha,n}})+W_{2}(\mu_{\hat{U}_{n}},\mu_{\hat{Z}_{n}})\\
 & =\frac{\sigma_{Z}\sqrt{d\delta}}{\sqrt{2n}}+c_{d}\frac{\sigma_{Z}}{\sqrt{c_{0}}}\sqrt{d\log\frac{2n}{\delta}}\cdot f_{d}(n)+\frac{\sigma_{Z}}{\sqrt{c_{0}}}\sqrt{\frac{2d}{n}\cdot\log\left(\frac{2n}{\delta}\right)\cdot\log\left(\frac{2}{\delta}\right)}\\
 & \leq c_{d}'\sigma_{Z}\cdot\log\left(\frac{n}{\delta}\right)\cdot f_{d}(n)
\end{align}
with probability larger than $1-\delta$, where $c_{d}'>0$ is a constant
which only depends on $d$.
\end{IEEEproof}
We next bound the change in the entropy of the output $Y=X+Z$ when
the noise is replaced by its truncated version $U_{\alpha}$. To this
end we remind the reader the following simple bounds on the differential
entropy of a mixture distribution (see, e.g., \cite{moshksar2016arbitrarily}). 
\begin{lem}
\label{lem: bounds on mixture entropy}Let $\{\nu_{j}\}_{j\in[m]}$
be a set of density functions such that $|h(\nu_{j})|<\infty$ for
all $j\in[m]$, and consider the mixture distribution $f=\sum_{j=1}^{m}a_{j}\nu_{j}$
where $\boldsymbol{a}\in\mathbb{A}^{m-1}$ is a probability vector.
Then, 
\[
\sum_{j=1}^{m}a_{j}h(\nu_{j})\leq h(f)\leq\sum_{j=1}^{m}a_{j}h(\nu_{j})+H(\boldsymbol{a})
\]
where $H(\boldsymbol{a})\dfn-\sum_{j=1}^{m}a_{j}\log a_{j}$ is the
(discrete) entropy of $\boldsymbol{a}$. 
\end{lem}
\begin{IEEEproof}
Let $J\in[m]$ be a random variable such that $\P[J=j]=a_{j}$, and
$X_{j}\sim\nu_{j}$, where $(J,X_{1},\ldots,X_{m})$ are pairwise
independent. Then the left inequality follows from $h(f)=h(X_{J})\geq h(X_{J}\mid J)$,
and the right inequality from $H(J)\geq I(X_{J};J)=h(X_{J})-h(X_{J}\mid J)$.
\end{IEEEproof}
\begin{lem}
\label{lem: entropy loss due to truncation}Suppose that $\E\|X^{2}\|\leq d\sigma_{X}^{2}$,
$|h(X)|\leq\eta_{X}$, and $\|Z\|$ is $\sqrt{d}\sigma_{Z}$-sub-Gaussian.
Then, for any $\alpha\in(0,\overline{\alpha}_{\mu_{\|Z\|}})$
\begin{equation}
\left|h(X+U_{\alpha})-h(X+Z)\right|\leq(2q\eta_{X})\vee\left(qd\log\left[4\pi e\left((\sigma_{X}^{2}+(d^{-1}\alpha^{2})\vee(cq^{-1}\sigma_{Z}^{2})\right)\right]\right)+\binent(q)\label{eq: entropy diff due to truncation}
\end{equation}
where $q=\P[\|Z\|>\alpha]$, $c>0$ is an absolute constant, and $\binent(t)\dfn-t\log t-(1-t)\log(1-t)$
for $t\in(0,1)$ is the binary entropy function. 
\end{lem}
\begin{IEEEproof}
Consider the set $S=\{Z\in\mathbb{R}^{d}\colon\|Z\|<\alpha\}$. Let
$\mu_{Z\mid S}$ be the measure $\mu_{Z}$ conditioned on the set
$S$, i.e., the measure uniquely defined by 
\[
\mu_{Z\mid S}(B\cap S)=\frac{\mu_{Z}(B\cap S)}{\mu_{Z}(S)}
\]
for any Borel set of $\mathbb{R}^{d}$. Then, $\mu_{X+Z}=(1-q)\cdot\mu_{Z\mid S}*\mu_{X}+q\cdot\mu_{Z\mid S^{c}}*\mu_{X}$
where $q=\mu_{Z}(S^{c})$. Similarly, $\mu_{X+U}=(1-q)\cdot\mu_{Z|S}*\mu_{X}+q\cdot\nu*\mu_{X}$
where $\nu$ is defined via the truncation operation (\ref{eq: truncated random variable with density}).
Thus, both $\mu_{X+Z}$ and $\mu_{X+U}$ are mixtures of two components,
and so Lemma \ref{lem: bounds on mixture entropy} implies that
\begin{align}
\left|h(X+U_{\alpha})-h(X+Z)\right| & \leq q\cdot\left|h(\nu*\mu_{X})-h(\nu*\mu_{X})\right|+\binent(q)\\
 & \leq q\cdot\left|h(\nu*\mu_{X})\right|+q\cdot\left|h(\mu_{Z|S^{c}}*\mu_{X})\right|+\binent(q).
\end{align}
We bound each of the entropies in the last display in a similar fashion.
Let $V\sim\nu$ such that $X\ind V$. Then, since Gaussian random
vector whose covariance matrix is proportional to the identity matrix
maximizes differential entropy under variance constraint (as easily
follows, e.g., from Hadamard\textquoteright s inequality \cite[Thm. 17.9.2]{cover2012elements})
\begin{align}
h(\nu*\mu_{X}) & =h(X+V)\\
 & \leq\frac{d}{2}\log(2\pi e\E[\tfrac{1}{d}\|X+V\|^{2}])\\
 & \leq\frac{d}{2}\log\left(4\pi e\E[\tfrac{1}{d}\|X\|^{2}+\tfrac{1}{d}\|V\|^{2}]\right)\\
 & \leq\frac{d}{2}\log(4\pi e(\sigma_{X}^{2}+d^{-1}\alpha^{2}]).
\end{align}
Since also $h(X+V)\geq h(X+V\mid V)=h(X\mid V)=h(X)\geq-\eta_{X}$
we obtain 
\[
\left|h(\nu*\mu_{X})\right|\leq\eta_{X}\vee\frac{d}{2}\log(4\pi e(\sigma_{X}^{2}+d^{-1}\alpha^{2}]).
\]
Note that conditioned on $\|Z\|>\alpha$, the density of $Z$ is $\mu_{Z|S^{c}}$.
Now, using the assumption that $\|Z\|$ is $(\sqrt{d}\sigma_{Z})$-sub-Gaussian,
there exists an absolute constant $c>0$ such that 
\[
\E\left[\|Z\|^{2}\mid\|Z\|>\alpha\right]=\frac{\E\left[\|Z\|^{2}\cdot\I\{\|Z\|>\alpha\}\right]}{q}\leq\frac{\E\left[\|Z\|^{2}\right]}{q}\leq\frac{cd\sigma_{Z}^{2}}{q}.
\]
Thus, we obtain, similarly to the bound on $|h(\nu*\mu_{X})|$ that
\[
\left|h(\nu*\mu_{Z|S^{c}})\right|\leq\eta_{X}\vee\frac{d}{2}\log\left(4\pi e(\sigma_{X}^{2}+cq^{-1}\sigma_{Z}^{2})\right).
\]
\end{IEEEproof}
We may now prove Theorem \ref{thm: learning input distributions}.
\begin{IEEEproof}[Proof of Theorem \ref{thm: learning input distributions}]
Let $Z_{1},\ldots,Z_{n}$ be $n$ independent samples from $\mu_{Z}$.
By the assumption that $\|Z\|$ is $\sqrt{d}\sigma_{Z}$-sub-Gaussian,
there exists a minimal absolute constant $c>0$ such that if $\alpha_{n}=c_{\alpha}\sigma_{Z}\sqrt{2d\log\frac{n^{2}}{2}}$
then $q\equiv q_{n}\dfn\P[\|Z\|>\alpha_{n}]\leq\tfrac{1}{2n^{2}}$.
If $c_{\alpha}<1/\sqrt{2}$ we will increase it in the definition
of $\alpha_{n}$ to some arbitrary $c_{\alpha}>\frac{1}{\sqrt{2}}$.
Then, by the union bound, $\max_{1\leq i\leq n}\|Z_{i}\|\leq\alpha_{n}$
with probability larger than $1-\tfrac{1}{2n}$. We will assume from
this point onward that this event holds. 

For any $X\sim\mu_{X}\in{\cal P}^{*}$ it holds that 
\begin{equation}
\left|h(X+Z)-h(X+\hat{Z}_{n})\right|\leq\left|h(X+Z)-h(X+U_{\alpha_{n}})\right|+\left|h(X+U_{\alpha_{n}})-h(X+\hat{Z}_{n})\right|\nfd G_{1}+G_{2}.\label{eq: entropy difference with triangle}
\end{equation}

\uline{Analysis of \mbox{$G_{1}$}:} Since $q_{n}\leq\tfrac{1}{2n^{2}}\leq\frac{1}{2}$
and as $-\log(1-t)\leq t$ for $t\in(0,1/2)$, we may bound 
\[
\binent(q_{n})\leq\binent\left(\frac{1}{2n^{2}}\right)=\frac{\log(2n^{2})}{2n^{2}}-\left(1-\frac{1}{2n^{2}}\right)\log\left(1-\frac{1}{2n^{2}}\right)\leq\frac{2\log n+\log(2e)}{2n^{2}}.
\]
Substituting this bound in Lemma \ref{lem: entropy loss due to truncation},
we deduce that there exists $n_{0}\in\mathbb{N}$ and $C_{0}>0$,
both which depend on $(d,\sigma_{X}^{2},\eta_{X},\sigma_{Z}^{2})$,
such that \textbf{
\begin{equation}
G_{1}\leq C_{0}\frac{\log n}{n^{2}}.\label{eq: entropy difference between noise and truncated version proof}
\end{equation}
}Indeed, to evaluate the asymptotic order of the term inside the parenthesis
in (\ref{eq: entropy diff due to truncation}), we note that $q\mapsto q\log q^{-1}$
is monotonic increasing on $[0,\frac{1}{4}]$ and so $\max_{0\leq q\leq n^{-2}}q\log q^{-1}=2\frac{\log n}{n^{2}}$
for $n\geq2$. 

\uline{Analysis of \mbox{$G_{2}$}:}\textbf{ }By definition $\|U_{\alpha_{n}}\|\leq\alpha_{n}$,
and under the high probability assumption, $\|\hat{Z}_{n}\|\leq\alpha_{n}$
holds too. It was shown in \cite[Prop. 3]{polyanskiy2016wasserstein}
that if $V=B+W$ where $B\ind W$, $\|B\|<\sigma^{2}d$ with probability
$1$, and $W$ is $(\psi_{1},\psi_{2})$-regular, then $\mu_{V}$
is $(\psi_{1},\psi_{2}+\psi_{1}\sigma\sqrt{d})$-regular. Since $\mu_{X}$
is $(\psi_{1},\psi_{2})$-regular by assumption, it holds that both
$X+U_{\alpha_{n}}$ and $X+\hat{Z}_{n}$ are $(\psi_{1},\tilde{\psi}_{2})$-regular
with $\tilde{\psi}_{2}\dfn\psi_{2}+\psi_{1}c_{\alpha}\sigma_{Z}\sqrt{4d\log n}.$
Then, by \cite[Prop. 1]{polyanskiy2016wasserstein}, it holds with
probability larger than $1-\frac{1}{n}$ that 
\begin{equation}
G_{2}\leq\left(\psi_{1}\sqrt{\E[\|X+\hat{Z}_{n}\|^{2}]}+\psi_{1}\sqrt{\E[\|X+U_{\alpha_{n}}\|^{2}]}+\tilde{\psi}_{2}\right)\cdot W_{2}(\mu_{X+\hat{Z}_{n}},\mu_{X+U_{\alpha_{n}}}).\label{eq: Wasserstein distance bound on entropy difference}
\end{equation}
To bounds the r.h.s. of (\ref{eq: Wasserstein distance bound on entropy difference}),
we note that with probability larger than $1-\frac{1}{2n}$ there
exists a constant $C_{1}(d,\sigma_{Z})$ such that
\begin{align}
W_{2}(\mu_{X+\hat{Z}_{n}},\mu_{X+U_{\alpha_{n}}}) & \trre[\leq,a]W_{2}(\mu_{\hat{Z}_{n}},\mu_{U_{\alpha_{n}}})\\
 & \trre[\leq,b]W_{2}(\mu_{\hat{Z}_{n}},\mu_{Z})+W_{2}(\mu_{Z},\mu_{U_{\alpha_{n}}})\\
 & \trre[\leq,c]c_{d}'\sigma_{Z}\log^{2}(4n^{2})\cdot n^{-1/(d\vee4)}+\sqrt{2}(\alpha_{n}+\sqrt{d}\sigma_{Z})e^{-\alpha_{n}^{2}/2d\sigma_{Z}^{2}}\\
 & \trre[\leq,d]C_{1}(d,\sigma_{Z})\frac{\log^{2}n}{n^{1/(d\vee4)}},\label{eq: Wasserstein between X+emprical and X + truncated noise}
\end{align}
where $(a)$ follows since Wasserstein distance is non-increasing
under a convolution operation,\footnote{Namely, if $X\ind Z_{1}$ and $X\ind Z_{2}$ and let $U=X+Z_{1}$
and $V=X+Z_{2}$. Then, $W_{p}(\mu_{U},\mu_{V})\leq W_{p}(\mu_{Z_{2}},\mu_{Z_{1}})$.} $(b)$ follows from the triangle inequality, $(c)$ follows from
Lemmas \ref{lem: Wasserstein loss due to truncation} and \ref{lem: Wasserstein distance of noise},
and $(d)$ utilizes the assumptions $c_{\alpha}>\frac{1}{\sqrt{2}}$
and $\sigma_{Z}=\Omega(n^{-(d-2)/(4d)})$ to show that $\sqrt{2}(\alpha_{n}+\sqrt{d}\sigma_{Z})e^{-\alpha_{n}^{2}/2d\sigma_{Z}^{2}}=o(n^{-1/d})$. 

Substituting the bound (\ref{eq: Wasserstein between X+emprical and X + truncated noise}),
as well as $\E[\|X+\hat{Z}_{n}\|^{2}]\leq2d\sigma_{X}^{2}+2\alpha_{n}^{2}$
and $\E[\|X+U_{\alpha_{n}}\|^{2}]\leq2d\sigma_{X}^{2}+2\alpha_{n}^{2}$,
in (\ref{eq: Wasserstein distance bound on entropy difference}),
and then using (\ref{eq: entropy difference between noise and truncated version proof})
and (\ref{eq: Wasserstein between X+emprical and X + truncated noise})
back in (\ref{eq: entropy difference with triangle}) implies that
there exists a constant $C_{2}>0$ which depends on $(d,\sigma_{X},\eta_{X},\psi_{1},\psi_{2},\sigma_{Z})$
such that 
\[
\sup_{X\colon\mu_{X}\in{\cal P}^{*}}\left|h(X+Z)-h(X+\hat{Z}_{n})\right|\leq C_{2}\frac{\log^{2}n}{n^{1/(d\vee4)}}
\]
with probability larger than $1-\frac{1}{n}$, and completes the proof.
\end{IEEEproof}
We now turn to prove Theorem \ref{thm: uniform convergence for learning weights}.
The next lemma bounds the difference in the differential entropy of
two mixtures which have the same component densities but perhaps different
mixing weights. The bound is given in terms of the total variation
and chi-square distance between the weights of the two densities,
and also depends on the maximum second-order R\'{e}nyi entropy of
the component densities. 
\begin{lem}
\label{lem: Entropy difference wrt TV and Chi2}Let $\{\nu_{j}\}_{j\in[m]}$
be a set of density functions such that $|h_{2}(\nu_{j})|\leq A_{\nu,2}$
for all $j\in[m]$ (where $h_{2}(\cdot)$ is the second-order differential
R\'{e}nyi entropy). Consider the mixture distributions $f=\sum_{j=1}^{m}a_{j}\nu_{j}$
and $g=\sum_{j=1}^{m}b_{j}\nu_{j}$ where $\boldsymbol{a},\boldsymbol{b}\in\mathbb{A}^{m-1}$
are probability vectors. If $\min_{j\in[m]}b_{j}\geq\epsilon$ then
\[
\left|h(f)-h(g)\right|\leq\Hamd(\boldsymbol{a},\boldsymbol{b})\cdot\dchis(\boldsymbol{a},\boldsymbol{b})+2(\log\tfrac{1}{\epsilon}+A_{\nu,2})\cdot\dtv(\boldsymbol{a},\boldsymbol{b}).
\]
\end{lem}
\begin{IEEEproof}
By the triangle inequality and the fact the chi-square divergence
dominates the KL divergence (e.g. \cite[Lemma 2.7]{tsybakov2008introduction})
\begin{align}
\left|h(f)-h(g)\right| & \leq\left|-\int f\log f+\int f\log g\right|+\left|-\int f\log g+\int g\log g\right|\\
 & =\dkl(f,g)+\left|\int(g-f)\log g\right|\\
 & \leq\dchis(f,g)+\left|\int(g-f)\log g\right|.
\end{align}
We obtain the desired bound by bounding each of the terms in the last
equation separately. For the first term, using Cauchy\textendash Schwarz
inequality,
\begin{align}
\dchis(f,g) & =\int\frac{\left(\sum_{j\colon a_{j}\neq b_{j}}(a_{j}-b_{j})\nu_{j}\right)^{2}}{\sum b_{j}\nu_{j}}\\
 & \leq\int\frac{\left(\sum_{j\colon a_{j}\neq b_{j}}\frac{(a_{j}-b_{j})}{\sqrt{b_{j}}}\sqrt{b_{j}}\nu_{j}\right)^{2}}{\sum_{j\colon a_{j}\neq b_{j}}b_{j}\nu_{j}}\\
 & \leq\int\sum_{j\colon a_{j}\neq b_{j}}\frac{(a_{j}-b_{j})^{2}}{b_{j}}\cdot\frac{\sum_{j\colon a_{j}\neq b_{j}}b_{j}\nu_{j}^{2}}{\sum_{j\colon a_{j}\neq b_{j}}b_{j}\nu_{j}}\\
 & =\dchis(\boldsymbol{a},\boldsymbol{b})\int\cdot\frac{\sum_{j\colon a_{j}\neq b_{j}}b_{j}\nu_{j}^{2}}{\sum_{j\colon a_{j}\neq b_{j}}b_{j}\nu_{j}}\\
 & =\dchis(\boldsymbol{a},\boldsymbol{b})\cdot\int\max_{j\in[m]\colon a_{j}\neq b_{j}}\nu_{j}\\
 & \leq\dchis(\boldsymbol{a},\boldsymbol{b})\cdot\sum_{j\in[m]\colon a_{j}\neq b_{j}}\int\nu_{j}\\
 & =\dchis(\boldsymbol{a},\boldsymbol{b})\cdot\Hamd(\boldsymbol{a},\boldsymbol{b}).
\end{align}
For the second term, it holds by Jensen and Cauchy\textendash Schwarz
inequalities that 
\[
\int\nu_{j}\log\left(\sum_{l}b_{l}\nu_{l}\right)\leq\log\left(\sum_{l}b_{l}\int\nu_{j}\nu_{l}\right)\leq\log\left(\max_{l}\int\nu_{j}\nu_{l}\right)\leq\max_{l}\frac{1}{2}\log\left(\int\nu_{j}^{2}\int\nu_{l}^{2}\right)\leq A_{\nu,2}.
\]
Since R\'{e}nyi entropies are decreasing functions of their order\footnote{For example, \cite[Eq. (3) and Thm. 3]{van2014renyi} state this for
densities with finite support and proves this by a related property
for the R\'{e}nyi divergence, but the result holds for general densities
and can be proved similarly to the discrete case.} it holds that $h(\nu_{j})\geq h_{2}(\nu_{j})$ and so 
\[
\int\nu_{j}\log\left(\sum_{l}b_{l}\nu_{l}\right)\geq\int\nu_{j}\log\left(b_{j}\nu_{j}\right)=-\log\frac{1}{\epsilon}-h(\nu_{j})\geq-\log\frac{1}{\epsilon}-A_{\nu,2}.
\]
The upper bound then follows from combining the above two bounds and
\[
\left|\int(g-f)\log g\right|=\left|\sum_{j}(b_{j}-a_{j})\int\nu_{j}\log\left(\sum_{l}b_{l}\nu_{l}\right)\right|\leq\sum_{j}|b_{j}-a_{j}|\cdot\left|\int\nu_{j}\log\left(\sum_{l}b_{l}\nu_{l}\right)\right|.
\]
\end{IEEEproof}
The next lemma provides an upper bound on the second-order R\'{e}nyi
entropy of $\tilde{Z}_{n}$.
\begin{lem}
\label{lem: Renyi entropy of KDE}Suppose that $\tilde{Z}_{n}\sim\frac{1}{\theta^{d}}\sum_{i=1}^{n}a_{i}\kappa_{\theta,Z_{i}}\dfn\mu_{\tilde{Z}_{n}}$,
where $|h_{2}(\kappa)|\leq A_{\kappa,2}$, $a_{i}\geq\epsilon>0$
for all $i\in[n]$, and $\theta<1$. Then, 
\[
|h_{2}(\tilde{Z}_{n})|\leq A_{\tilde{Z},2}\dfn A_{\kappa,2}+d\log\tfrac{1}{\theta}+2\log\tfrac{1}{\epsilon}.
\]
\end{lem}
\begin{IEEEproof}
By Jensen's inequality 
\[
\int\mu_{\tilde{Z}_{n}}^{2}=\int\left(\frac{1}{\theta^{d}}\sum_{i=1}^{n}a_{i}\kappa_{\theta,Z_{i}}\right)^{2}\leq\frac{1}{\theta^{2d}}\sum_{i=1}^{n}a_{i}\int\kappa_{\theta,Z_{i}}^{2}\leq\frac{1}{\theta^{2d}}\int\kappa^{2}\left(\frac{z}{\theta}\right)\d z=\frac{1}{\theta^{d}}\int\kappa^{2}\leq\frac{e^{A_{\kappa,2}}}{\theta^{d}}.
\]
Furthermore,
\[
\int\mu_{\tilde{Z}_{n}}^{2}=\int\left(\frac{1}{\theta^{d}}\sum_{i=1}^{n}a_{i}\kappa_{\theta,Z_{n}}\right)^{2}\geq\frac{\epsilon^{2}}{\theta^{2d}}\int\left(\max_{i\in[n]}\kappa_{\theta,Z_{i}}\right)^{2}\geq\frac{\epsilon^{2}}{\theta^{2d}}\max_{i\in[n]}\int\left(\kappa_{\theta,Z_{i}}\right)^{2}\geq\frac{\epsilon^{2}e^{-A_{\kappa.2}}}{\theta^{d}}.
\]
The results then follows by combining both bounds.
\end{IEEEproof}
The next lemma states that the entropy difference $|h(X_{\boldsymbol{a}}+Z)-h(X_{\boldsymbol{a}}+\tilde{Z}_{n})|$
concentrates fast around its mean value.
\begin{lem}
\label{lem: concentration of entropy difference}Let $\epsilon>0$
be given. Let $\Gamma(\boldsymbol{Z})\dfn|h(X_{\boldsymbol{a}}+Z)-h(X_{\boldsymbol{a}}+\tilde{Z}_{n})|$
where $X_{\boldsymbol{a}}\sim\sum_{j=1}^{m}a_{j}\delta_{x_{j}}$ and
$a_{j}\geq\epsilon>0$ for all $j\in[m]$, and $\tilde{Z}_{n}\sim\frac{1}{n\theta^{d}}\sum_{i=1}^{n}\kappa_{\theta,Z_{i}}$
where $|h_{2}(\kappa)|\leq A_{\kappa,2}$. Then, for any $\delta_{1}\in(0,1)$
\[
\Gamma(\boldsymbol{Z})\leq\E\Gamma(\boldsymbol{Z})+\sqrt{\frac{(6+2\log n+A_{\kappa_{C}})^{2}}{2n}\log\frac{1}{\delta_{1}}}
\]
 with probability larger than $1-\delta_{1}$, where 
\begin{equation}
A_{\kappa_{C}}\dfn A_{\kappa,2}+d\log\tfrac{1}{\theta}+2\log\tfrac{1}{\epsilon}.\label{eq: Codebook Renyi entropy bound}
\end{equation}
\end{lem}
\begin{IEEEproof}
Let $\boldsymbol{z}^{(0)},\boldsymbol{z}^{(1)}\in\mathbb{R}^{d}$
be such that $\Hamd(\boldsymbol{z}^{(0)},\boldsymbol{z}^{(1)})\leq1$,
and further let $\tilde{Z}_{n}^{(l)}\sim\frac{1}{n\theta^{d}}\sum_{i=1}^{n}\kappa_{\theta,z_{i}^{(l)}}$
for $l=0,1$ be the corresponding KDEs of $\mu_{Z}$, and assume w.l.o.g.
that $z_{i}^{(0)}=z_{i}^{(1)}$ for all $i\in[n-1]$. Let $\overline{z}=(z_{1}^{(0)},z_{2}^{(0)},\ldots,z_{n}^{(0)},z_{n}^{(1)})\in\mathbb{R}^{n+1}$
and denote
\[
\boldsymbol{q}^{(0)}=\left(\tfrac{1}{n},\ldots,\tfrac{1}{n},\tfrac{1}{n},0\right)\in[0,1]^{n+1}
\]
\[
\boldsymbol{q}^{(1)}=\left(\tfrac{1}{n},\ldots\tfrac{1}{n},0,\tfrac{1}{n}\right)\in[0,1]^{n+1}
\]
\[
\boldsymbol{q}^{(1/2)}=\left(\tfrac{1}{n},\ldots,\tfrac{1}{n},\tfrac{1}{2n},\tfrac{1}{2n}\right)\in[0,1]^{n+1}
\]
such that $\hat{Z}_{n}^{(l)}\sim\sum_{i=1}^{n+1}q_{i}^{(l)}\delta_{\overline{z}_{i}}$
for $l=0,1$. Further denote the ``codebook kernel'' $\kappa_{C}\dfn\frac{1}{\theta^{d}}\sum_{j=1}^{m}a_{j}\kappa_{\theta,x_{j}}$,
which, by Lemma \ref{lem: Renyi entropy of KDE}, satisfies that $|h_{2}(\kappa_{C})|\leq A_{\kappa_{C}}$
for $A_{\kappa_{C}}$ in (\ref{eq: Codebook Renyi entropy bound}).
Then, for $V\eqd X+\theta U\sim\kappa_{C}$ where $U\sim\kappa$ and
$U\ind X$, it holds that
\begin{align}
\left|\Gamma(\boldsymbol{z}^{(0)})-\Gamma(\boldsymbol{z}^{(1)})\right| & \trre[\leq,a]\left|h(X+\tilde{Z}_{n}^{(0)})-h(X+\tilde{Z}_{n}^{(1)})\right|\\
 & =\left|h(V+\hat{Z}_{n}^{(0)})-h(V+\hat{Z}_{n}^{(1)})\right|\\
 & \trre[\leq,b]\left|h(V+\hat{Z}_{n}^{(0)})-h(V+\hat{Z}_{n}^{(1/2)})\right|+\left|h(V+\hat{Z}_{n}^{(1/2)})-h(V+\hat{Z}_{n}^{(1)})\right|\\
 & \trre[\leq,c]\frac{4}{n}+\frac{2}{n}(\log(2n)+A_{\kappa_{C}}),
\end{align}
where $(a)$ follows from the reverse triangle inequality, $(b)$
follows from the triangle inequality, and $(c)$ follows by bounding
the two terms in the same manner using Lemma \ref{lem: Entropy difference wrt TV and Chi2}.
Specifically, for the first term, we set $\epsilon=\tfrac{1}{2n}$
and $\nu_{i}=\kappa_{C,\overline{z}_{i}}$ (which is a shift of $\kappa_{C}$
by $Z_{i}$) and $a_{i}=q_{i}$ for $i\in[n+1]$ and note that $\boldsymbol{q}^{(1/2)}\gg\boldsymbol{q}^{(0)}$,
and $\dtv(\boldsymbol{q}^{(0)},\boldsymbol{q}^{(1/2)})=\frac{1}{2n}$,
$\dchis(\boldsymbol{q}^{(0)},\boldsymbol{q}^{(1/2)})=\frac{1}{n}$,
and $\Hamd(\boldsymbol{q}^{(0)},\boldsymbol{q}^{(1/2)})=2$.

Now, the function $\Gamma(z)$ satisfies $\tfrac{1}{n}(6+2\log n+A_{\kappa_{C}})$-bounded-difference
property, and so the stated result follows from McDiarmid's inequality
(bounded differences inequality) \cite[Thm. 6.2]{boucheron2013concentration}
which implies that for any $t\geq0$
\[
\P\left[\Gamma(\boldsymbol{Z})>\E\Gamma(\boldsymbol{Z})+t\right]\leq\exp\left[-\frac{2nt^{2}}{(6+2\log n+A_{\kappa_{C}})^{2}}\right].
\]
\end{IEEEproof}
We may now prove the theorem.
\begin{IEEEproof}[Proof of Theorem \ref{thm: uniform convergence for learning weights}]
Let $\epsilon>0$ be given such that $m^{2}\epsilon<1$ and consider
an $\epsilon$-net for $[0,1]$ given by ${\cal F}=\{\epsilon,2\epsilon,\ldots,1\}$
where we assume that $\epsilon^{-1}$ is integer. In addition, consider
a net ${\cal A}_{\epsilon}^{m-1}\subset\mathbb{A}^{m-1}$ constructed
by a quantization of $\boldsymbol{a}\in\mathbb{A}^{m-1}$ to $\boldsymbol{b}\in{\cal A}_{\epsilon}^{m-1}$
in the following way. Assume for the sake of notational simplicity
that $\boldsymbol{a}$ is ordered such that $a_{1}\leq a_{2}\leq\cdots\leq a_{m}$.
The first $m-1$ coordinates of $\boldsymbol{a}$ are rounded upwards
so that $b_{j}=\min\{b\in{\cal F}\colon b\geq a_{j}\}$ for any $j\in[m-1]$,
and $b_{m}=1-\sum_{j=1}^{m}\overline{a}_{j}\in{\cal F}$. So, by construction
$b_{j}\geq\epsilon$ for all $j\in[m-1]$, and since by the ordering
assumption $a_{m}\geq\tfrac{1}{m}$ must hold, the condition $m^{2}\epsilon<1$
and the quantization definition imply that $b_{m}>\epsilon$ also
holds. The number of possible probability vectors $\boldsymbol{b}$
obtained in this way is at most $|{\cal A}_{\epsilon}^{m-1}|\leq\left(\frac{1}{\epsilon}\right)^{m}$.
Furthermore, if $\boldsymbol{a}$ is mapped to $\boldsymbol{b}\in{\cal A}_{\epsilon}^{m-1}$
then 
\[
\dtv(\boldsymbol{a},\boldsymbol{b})=\frac{1}{2}\sum_{j=1}^{m}\left|a_{j}-b_{j}\right|\leq m\epsilon,
\]
and 
\[
\dchis(\boldsymbol{a},\boldsymbol{b})=\sum_{j=1}^{m}\frac{(a_{j}-b_{j})^{2}}{b_{j}}\leq\sum_{j=1}^{m-1}\frac{\epsilon^{2}}{\epsilon}+\frac{m^{2}\epsilon^{2}}{\epsilon}\leq2m^{2}\epsilon.
\]
Hence, for any $\boldsymbol{a}\in{\cal A}$ there exists $\boldsymbol{b}\in{\cal A}_{\epsilon}^{m-1}$
such that $\dtv(\boldsymbol{a},\boldsymbol{b})\leq m\epsilon$ and
$\dchis(\boldsymbol{a},\boldsymbol{b})\leq2m^{2}\epsilon.$\footnote{In other words, ${\cal A}_{\epsilon}^{m-1}$ is an $m\epsilon$-cover
of $\mathbb{A}^{m-1}$ in the total variation distance and an $(2m^{2}\epsilon)$-''cover''
of $\mathbb{A}^{m-1}$ in the chi-square divergence (the latter statement,
however, is not rigorous since the chi-square divergence is not symmetric
and thus not a metric).} 

Let $\boldsymbol{a}\in\mathbb{A}^{m-1}$ be mapped to $\boldsymbol{b}\in{\cal A}_{\epsilon}^{m-1}$.
Then, 
\begin{align}
 & \left|h(X_{\boldsymbol{a}}+Z)-h(X_{\boldsymbol{a}}+\tilde{Z}_{n})\right|\nonumber \\
 & \leq\left|h(X_{\boldsymbol{a}}+Z)-h(X_{\boldsymbol{b}}+Z)\right|+\left|h(X_{\boldsymbol{b}}+Z)-h(X_{\boldsymbol{b}}+\tilde{Z}_{n})\right|+\left|h(X_{\boldsymbol{a}}+\tilde{Z}_{n})-h(X_{\boldsymbol{b}}+\tilde{Z}_{n})\right|\\
 & \dfn G_{1}+G_{2}+G_{3}.
\end{align}

\uline{Analysis of \mbox{$G_{1}$}:} By the assumption on the R\'{e}nyi
entropy of $\mu_{Z}$ and Lemma \ref{lem: Entropy difference wrt TV and Chi2}
with $\nu_{j}(z)=\mu_{Z}(z-x_{j})$ it holds that 
\[
G_{1}\leq2m^{3}\epsilon+2(\log\tfrac{1}{\epsilon}+A_{Z,2})m\epsilon.
\]

\uline{Analysis of \mbox{$G_{2}$}:} Take $\delta_{1}=\frac{\delta}{|{\cal A}_{\epsilon}^{m-1}|}$,
and denote $\Gamma_{\boldsymbol{b}}(\boldsymbol{Z})\dfn|h(X_{\boldsymbol{b}}+Z)-h(X_{\boldsymbol{b}}+\tilde{Z}_{n})|$.
Since $b_{j}\geq\epsilon$ for all $j\in[m]$, Lemma \ref{lem: concentration of entropy difference}
and the union bound assure that 
\[
\P\left[\bigcap_{\boldsymbol{b}\in{\cal A}_{\epsilon}^{m-1}}\left\{ \Gamma_{\boldsymbol{b}}(\boldsymbol{Z})<\E\Gamma_{\boldsymbol{b}}(\boldsymbol{Z})+\sqrt{\frac{(6+2\log n+A_{\kappa_{C}})^{2}}{2n}\log\frac{1}{\delta_{1}}}\right\} \right]\geq1-\delta,
\]
where $A_{\kappa_{C}}$ is as in (\ref{eq: Codebook Renyi entropy bound}).
Thus, given that this event holds, for any $\boldsymbol{b}\in{\cal A}_{\epsilon}^{m-1}$
\[
G_{2}\leq\E\left[\left|h(X_{\boldsymbol{b}}+Z)-h(X_{\boldsymbol{b}}+\tilde{Z}_{n})\right|\right]+(6+2\log n+A_{\kappa_{C}})\sqrt{\frac{1}{2n}\log\frac{1}{\delta}+\frac{m}{2n}\log\frac{2}{\epsilon}}.
\]

\uline{Analysis of \mbox{$G_{3}$}:} It holds that $\tilde{Z}_{n}\sim\frac{1}{n\theta^{d}}\sum_{i=1}^{n}\kappa_{\theta,Z_{i}}\dfn\mu_{\tilde{Z}_{n}}$,
and so by Lemma \ref{lem: Renyi entropy of KDE} 
\[
\left|h_{2}(\mu_{\tilde{Z}_{n}})\right|\leq A_{\kappa}+2\log n+2d\log\frac{1}{\theta}\nfd A_{\tilde{Z},2}.
\]
Lemma \ref{lem: Entropy difference wrt TV and Chi2} with $\nu_{j}(z)=\mu_{\tilde{Z}_{n}}(z-x_{j})$
implies that 
\[
G_{3}\leq2m^{3}\epsilon+2\left(\log\tfrac{1}{\epsilon}+A_{\kappa,2}+2\log n+2d\log\tfrac{1}{\theta}\right)m\epsilon.
\]

From the bounds on $G_{1},G_{2}$ and $G_{3}$ we deduce that it holds
with probability larger than $1-\delta$ that
\begin{align}
\sup_{\boldsymbol{a}\in\mathbb{A}^{m-1}}\left|h(X_{\boldsymbol{a}}+Z)-h(X_{\boldsymbol{a}}+\tilde{Z}_{n})\right| & \leq\max_{\boldsymbol{b}\in{\cal A}_{\epsilon}^{m-1}}\E\left[\left|h(X_{\boldsymbol{b}}+Z)-h(X_{\boldsymbol{b}}+\tilde{Z}_{n})\right|\right]+\Delta_{0}(\epsilon)\\
 & \leq\sup_{\boldsymbol{a}\in\mathbb{A}^{m-1}}\E\left[\left|h(X_{\boldsymbol{b}}+Z)-h(X_{\boldsymbol{b}}+\tilde{Z}_{n})\right|\right]+\Delta_{0}(\epsilon),
\end{align}
where 
\begin{align}
\Delta_{0}(\epsilon) & =4m^{3}\epsilon+2\left[2\log\tfrac{1}{\epsilon}+2\log n+2d\log\tfrac{1}{\theta}+A_{Z,2}+A_{\kappa,2}\right]m\epsilon\nonumber \\
 & \hphantom{=}+(6+2\log n+A_{\kappa,2}+d\log\tfrac{1}{\theta}+2\log\tfrac{1}{\epsilon})\sqrt{\frac{1}{2n}\log\frac{1}{\delta}+\frac{m}{2n}\log\frac{2}{\epsilon}}.
\end{align}
Choosing $\epsilon=\frac{1}{\lceil nm^{2}\sqrt{m}\rceil}$ and simplifying\footnote{The ceiling operation in the choice of $\epsilon$ has a negligible
effect on the final result.} completes the proof.
\end{IEEEproof}

\section{Experiments Details \label{sec:Experiments-Details}}

We denote by $Q_{\alpha}\dfn\left[\begin{array}{cc}
\cos(\alpha) & \sin(\alpha)\\
-\sin(\alpha) & \cos(\alpha)
\end{array}\right]$ a rotation matrix of angle $\alpha$. 

\begin{table}[H]
\begin{centering}
\begin{tabular}{|c|c|c|c|}
\hline 
{\small{}Notation} & {\small{}Description} & {\small{}Value (Sec. \ref{subsec:Two-dimensional-Gaussian-noise})} & {\small{}Value (Sec. \ref{subsec:Gaussian-Mixture-noise})}\tabularnewline
\hline 
\hline 
{\small{}$d$} & {\small{}channel dimension} & {\small{}$2$} & {\small{}$4$}\tabularnewline
\hline 
{\small{}$m$} & {\small{}codebook cardinality} & {\small{}$32$} & {\small{}$(8,16,32,64)$}\tabularnewline
\hline 
{\small{}$r_{s}$} & {\small{}maximal eigenvector of $S$} & {\small{}$10$} & {\small{}$10$}\tabularnewline
\hline 
{\small{}$\phi_{x}$} & {\small{}input distribution projection parameter} & {\small{}$2$} & {\small{}$2$}\tabularnewline
\hline 
{\small{}$\Gamma$} & {\small{}``gap-to-capacity'' parameter} & {\small{}$10$} & {\small{}$20$}\tabularnewline
\hline 
{\small{}$n$} & {\small{}number of training samples} & {\small{}$2\cdot10^{3}$ } & {\small{}$2\cdot10^{3}$}\tabularnewline
\hline 
\multirow{4}{*}{{\small{}$Z$}} & \multirow{4}{*}{{\small{}noise distribution}} & \multirow{4}{*}{{\small{}$Z\sim N(0,K)$}} & {\small{}$Z\eqd\Psi\left[\alpha\left(\sum_{l=1}^{k}R_{l}\cdot v_{l}+W\right)\right]$}\tabularnewline
 &  &  & {\small{}$W\sim N(0,K_{W})$}\tabularnewline
 &  &  & {\small{}$R_{l}\sim\text{Uniform}\{\pm1\}$, i.i.d.}\tabularnewline
 &  &  & {\small{}$W\ind\{R_{l}\}$}\tabularnewline
\hline 
{\small{}$\phi_{z}$} & {\small{}noise distribution projection parameter} & {\small{}N/A} & {\small{}$2$}\tabularnewline
\hline 
{\small{}$K$ or $K_{W}$} & {\small{}Gaussian noise covariance matrix} & {\small{}$K=10^{-1}\cdot Q_{\alpha}\Lambda Q_{\alpha}^{T},$} & {\small{}$K_{W}=AA^{T}$}\tabularnewline
 &  & {\small{}$\Lambda\dfn\left[\begin{array}{cc}
1 & 0\\
0 & 3
\end{array}\right]$, $\alpha=30^{\circ}$} & {\small{}$A_{i_{1}i_{2}}\sim N(0,1)$ i.i.d.}\tabularnewline
\hline 
{\small{}$\{v_{l}\}_{l=1}^{s}$} & {\small{} interference vectors } & {\small{}N/A} & {\small{}$v_{l}\sim N(0,I_{d})$ i.i.d., $s=5$}\tabularnewline
\hline 
{\small{}$r_{x}$} & {\small{}maximal power constraint} & \multicolumn{2}{c|}{{\small{}$r_{x}=\Gamma\cdot r_{\text{min}}$}}\tabularnewline
\hline 
{\small{}$C^{(0)}=\left\{ X_{j}^{(0)}\right\} _{j=1}^{m}$} & {\small{}initial codebook generation} & \multicolumn{2}{c|}{{\small{}$X_{j}^{(0)}\sim N(0,\frac{r_{x}^{2}}{\chi d}\cdot I_{d})$
i.i.d.}}\tabularnewline
\hline 
{\small{}$\lambda^{(i)}\equiv\lambda$} & {\small{}SGD step sizes \textendash{} codeword update} & \multicolumn{2}{c|}{{\small{}$10^{-1}$ }}\tabularnewline
\hline 
{\small{}$\eta^{(i)}\equiv\eta$ } & {\small{}SGD step sizes \textendash{} covariance update} & \multicolumn{2}{c|}{{\small{}$10^{-1}$}}\tabularnewline
\hline 
{\small{}$\tilde{n}$} & {\small{}number of validation samples} & {\small{}$10^{4}$ } & {\small{}$10^{3}$}\tabularnewline
\hline 
 & {\small{}total number of runs} & {\small{}N/A} & {\small{}$10^{3}$}\tabularnewline
\hline 
 & {\small{}number of runs per distribution} & {\small{}N/A} & {\small{}$10$}\tabularnewline
\hline 
\end{tabular}
\par\end{centering}
~
\centering{}\caption{SGD algorithm experiments parameters \label{tab: SGD Experiments-parameters}}
\end{table}

\begin{table}[H]
\begin{centering}
\begin{tabular}{|c|c|c|}
\hline 
{\small{}Notation} & {\small{}Description} & {\small{}Value (}Sec. \ref{subsec:Gaussian-Mixture-noise-Gibbs})\tabularnewline
\hline 
\hline 
{\small{}$d$} & {\small{}channel dimension} & {\small{}$4$}\tabularnewline
\hline 
{\small{}$m$} & {\small{}codebook cardinality} & $2\leq m\leq64$\tabularnewline
\hline 
{\small{}$\Gamma$} & {\small{}``gap-to-capacity'' parameter} & {\small{}$10$}\tabularnewline
\hline 
{\small{}$n$} & {\small{}number of training samples } & {\small{}$10^{2}$}\tabularnewline
\hline 
{\small{}$Z$} & {\small{}noise distribution} & {\small{}$Z\eqd\sum_{l=1}^{k}R_{l}\cdot v_{l}+W,$}\tabularnewline
 &  & {\small{}$W\sim N(0,K_{W}),\;R_{l}\sim\text{Uniform}\{\pm1\}$}\tabularnewline
\hline 
{\small{}$K_{W}$} & {\small{}Gaussian noise covariance matrix} & {\small{}$K_{W}=AA^{T},\;A_{i_{1}i_{2}}\sim N(0,1)$ i.i.d.}\tabularnewline
\hline 
{\small{}$\{v_{l}\}_{l=1}^{s}$} & {\small{} interference vectors } & {\small{}$v_{l}\sim N(0,I_{d})$ i.i.d., $s=10$}\tabularnewline
\hline 
{\small{}$r_{x}$} & {\small{}maximal power constraint} & {\small{}$r_{x}=\Gamma\cdot r_{\text{min}}$}\tabularnewline
\hline 
{\small{}$C^{(0)}=\left\{ X_{j}^{(0)}\right\} _{j=1}^{m}$} & {\small{}initial codebook generation} & {\small{}$X_{j}^{(0)}\sim N(0,\frac{r_{x}^{2}}{d}\cdot I_{d})$ i.i.d.}\tabularnewline
\hline 
{\small{}$m_{0}$} & {\small{}number of initial codewords} & {\small{}$64$}\tabularnewline
\hline 
{\small{}$k$} & {\small{}number of codewords removed at each stage} & {\small{}$1$}\tabularnewline
\hline 
{\small{}$\beta$} & {\small{}inverse temperature parameter} & {\small{}$(1,10,100)\cdot10^{2}$}\tabularnewline
\hline 
{\small{}$Q(\d x)$} & {\small{}reference measure} & {\small{}Lebesgue measure}\tabularnewline
\hline 
 & {\small{}decoder} & {\small{}minimum distance $S=I_{d}$}\tabularnewline
\hline 
{\small{}$\tilde{n}$} & {\small{}number of validation samples} & {\small{}$10^{4}$ }\tabularnewline
\hline 
 & {\small{}total number of runs} & {\small{}$2.5\cdot10^{3}$}\tabularnewline
\hline 
 & {\small{}number of runs per distribution} & $10$\tabularnewline
\hline 
\end{tabular}
\par\end{centering}
~
\centering{}\caption{Gibbs algorithm experiment parameters \label{tab: Gibbs Experiments-parameters}}
\end{table}

\section{Memoization Implementation of the Gibbs Algorithm \label{sec:Memoization-Implementation-of}}

The main computational task required by Algorithm \ref{alg: Gibbs expurgation}
is an efficient computation of the average error probability of a
codebook. In principle, at each stage of the algorithm, the average
error probability should be computed for any of the candidates codebooks.
A possible efficient implementation may compute these error probabilities
based on a pairwise error array computed for the initial codebook
$C_{0}=\{x_{1},\ldots,x_{m_{0}}\}$ and the given noise samples. Specifically,
consider the array $E\in\{2^{[n]}\}^{m_{0}\times m_{0}}$ where $2^{[n]}$
is the power set of $[n]$, such that the $(j_{1},j_{2})$th entry
of $E$ is given by 
\[
E(j_{1},j_{2})\dfn\left\{ i\colon\|x_{j_{1}}-x_{j_{2}}\|_{S}^{2}+2(x_{j_{1}}-x_{j_{2}})^{T}Sz_{i}<0\right\} .
\]
To wit, this entry $E(j_{1},j_{2})$ is the set of noise samples indices
such that if for a codebook consisting only the codewords $(x_{j_{1}},x_{j_{2}})$,
the noise sample $z_{i}$ will cause a decoding error when $x_{j_{1}}$
is transmitted. In accordance, $\bigcup_{j_{2}\in[m_{0}]}\{E(j_{1},j_{2})\}\subset[n]$
is the set of noise sample indices such that a decoding error occurs
when $x_{j_{1}}$ is transmitted and the codebook is $C_{0}$. The
array $E$ can be computed \emph{once} at initialization of the algorithm,
and then the error probabilities required by Algorithm \ref{alg: Gibbs expurgation}
can be computed based only on this array. At first, for $C_{0}=\{x_{1},\ldots,x_{m_{0}}\}$,
the error probability is given by
\begin{equation}
\pe_{\boldsymbol{z}}(C_{0})=\frac{1}{n}\sum_{i=1}^{n}\frac{1}{m_{0}}\sum_{j_{1}=1}^{m_{0}}\left|\bigcup_{j_{2}\in[m_{0}]}\{E(j_{1},j_{2})\}\right|.\label{eq: error probability using pairwise error matrix}
\end{equation}
The error probability $\pe_{\boldsymbol{z}}(C_{1})$ for $C_{1}=C_{0}\backslash(x_{j_{1}},\ldots,x_{j_{k}})$
from $C_{0}$ can be computed by first removing the rows indexed by
$(j_{1},\ldots,j_{k})$ and the rows indexed by $(j_{1},\ldots,j_{k})$
from the array $E(j_{1},j_{2})$ and then compute as in (\ref{eq: error probability using pairwise error matrix})
with the new array. The main advantage of this approach is that the
array tend to be sparse in the sense that either $|E(j_{1},j_{2})|$
is typically low \textendash{} when transmitting $x_{j}$, there is
only a relatively a small number of noise samples that will cause
an error. This array become sparser with the steps of the algorithm,
and moreover, high signal-to-noise ratio and high dimension lead to
sparser arrays.

\bibliographystyle{plain}
\bibliography{Learning_codes}

\end{document}